# End-of-life Disposal Dynamics Assessment for Lunar Distant Retrograde Orbits

Yongjie Chen[1], Yue Wang[1, 2*], Haichao Gui[1, 2]

[1] *School of Astronautics, Beihang University, Beijing 102206, China*
[2] *Key Laboratory of Spacecraft Design Optimization and Dynamic Simulation Technology, Ministry of Education, Beijing 102206, China*

**Abstract**

The disposal dynamics of cislunar periodic orbits are critical for end-of-life disposal design. Governed by the combined gravitational influences of the Earth, Moon, and Sun, the dynamics are highly sensitive and chaotic, demanding rigorous analysis for both physical insight and practical guidance. This study investigates the disposal dynamics of lunar distant retrograde orbits (DROs), covering both the departure and cislunar escape phases. The departure phase concerns removing the object from the vicinity of the nominal DRO, whereas the escape phase assesses whether the post-disposal trajectory leaves cislunar space and enters a heliocentric orbit. A new departure indicator, based on the dimensionless momentum integral, is proposed to identify the departure epoch. Cislunar escape dynamics are analyzed combining Lagrangian descriptors with semi-analytical equations governing energy evolution. A two-layer clustering method is proposed for classifying escape orbital families based on the Lagrangian descriptor field. The inherent stability of DRO necessitates high-magnitude disposal maneuvers for both departure and escape. The effects of magnitudes and directions of disposal maneuvers, Sun–Earth–Moon geometry, and orbital phase are systematically evaluated, and favorable conditions for disposal design are compared using the Lagrangian descriptor field. The framework is further validated for near rectilinear halo orbit (NRHO) and high-fidelity ephemeris model.

**Keywords:** Cislunar disposal dynamics; Bicircular restricted four-body problem; Distant retrograde orbit; Lagrangian descriptor; Orbital clustering

## 1. Introduction

Cislunar space serves as a critical bridge between Earth and deep space. Recent programs, including NASA's

*Corresponding author.
*E-mail addresses:* chenyongjie@buaa.edu.cn (Y. Chen), ywang@buaa.edu.cn (Y. Wang), hcgui@buaa.edu.cn (H. Gui)

Artemis [1], Lunar Orbital Platform–Gateway (LOP-G) program [2], and CNSA's Chang'e missions [3–5], underscore its growing strategic value [6]. Consequently, human and autonomous robotic presence in this region is expanding rapidly [7]. This escalating activity will inevitably generate cislunar space debris, such as defunct spacecraft and rocket upper stages. Effective end-of-life disposal strategies are therefore vital to ensure cislunar sustainability [8]. The lunar DROs have gained extensively attention as they maintain long-term stability and have broad accessibility to different cislunar regions [8,9,10]. In the future, there will be a large number of spacecraft operating on lunar DROs. However, the stability of DROs makes it difficult to achieve departure and escape from the vicinity of nominal orbits [10]. Thus, it is necessary to analyze disposal dynamics from DROs particularly to enable the end-of-life disposal design.

Generally, the end-of-life disposal should prioritize collision avoidance with operational spacecraft, the Earth, and the Moon. Ideally, abandoned objects should be permanently removed from the cislunar space [11,12]. According to the European Space Agency Space Debris Mitigation Working Group [13], the heliocentric space is also determined as the first preferred disposal destination. To achieve this, the disposal process can be generally divided into two dynamical phases: local departure and cislunar escape [14,15].

The departure phase makes the object deviate from its nominal operational orbit and reduces the likelihood of subsequent returns to the interior cislunar region. As a pioneering project, the Gateway has been proposed as a crewed outpost supporting a variety of exploration activities. Previous studies have primarily investigated disposal from near rectilinear halo orbit (NRHO), driven by the Gateway's operational needs [15] and lunar impact [16]. As a useful and intuitive metric, the momentum integral (MI) [17] has been proposed to reflect the successful departure from the 9:2 NRHO. The successful departure from the NRHO is defined when the MI along the post-disposal trajectory crosses a threshold value [14,15]. The duration from maneuver execution to the MI threshold crossing defines the time-to-depart. However, relying on a constant MI threshold lacks universality across different cislunar periodic orbits. For instance, the Earth-Moon 9:2 NRHO oscillates between lunar distances of approximately 3,100 km and 71,000 km. In contrast, a 2:1 distant retrograde Orbit (DRO) maintains a much larger lunar distance, ranging from roughly 72,000 km to 97,000 km. Therefore, the amplitude of MI is much larger than that of 9:2 NRHO within a 2:1 DRO period. Consequently, the baseline MI amplitude for 2:1 DRO is significantly larger than that of 9:2 NRHO. Applying the existing MI criterion to the disposal analysis for other periodic orbits may not always be effective, necessitating an improved metric for departure assessment.

Escaping into the heliocentric space is a comparatively safe strategy as the final result of disposal. Previous studies about the disposal from NRHO highlight the critical role of solar gravity perturbations (SGP) in facilitating this escape [15,18]. By leveraging the SGP, trajectories can transition from the Earth-Moon dynamical regime into the Sun-Earth governed phase space [19]. This process shares fundamental mechanics with low-energy transfers and ballistic capture [20]. Thus, characterizing these low-energy mechanisms remains critical to cislunar escape research [21,22]. However, the precise influence of SGP on escape trajectories remains highly complex and unintended outcomes are common for the disposal design. For example, pre-launch analyses of the Artemis 1 Interim Cryogenic Propulsion Stage (ICPS) revealed that certain launch windows would result in a quick return to the cislunar region and the bounded motion within two years, rather than a true heliocentric escape [23]. The long-term bounded motions are typical unsuccessful, hazardous disposals. Distinguishing the underlying dynamics of successful escapes from unsuccessful results is therefore essential. Besides, the cislunar escape trajectories are sensitive to the disposal maneuver parameters, thus it is necessary to analyze the disposal design parameters thoroughly [20]. Moreover, there is still a lack of research on the escape process from DROs. Different from the dynamics of libration point orbits with complicated invariant manifolds [24], the sticky region [25] near DROs makes the escape dynamics complicated. This requires methods capable of capturing the physical processes of cislunar escape and the influence of design parameters. The Lagrangian Descriptor (LD) has emerged as a powerful tool for mapping phase space geometry and identifying dynamical separatrices in the complex dynamical systems [26,27]. Given its recent success in analyzing weak-stability-boundary (WSB) transport and ballistic capture [28], the LD method shows immense potential for decoding the complexities of low-energy cislunar escape dynamics.

To address these gaps, this study develops a new framework for analyzing disposal dynamics from cislunar periodic orbits in the bicircular restricted four-body problem (BCRFBP) model. Furthermore, spatial disposal maneuvers are incorporated to fully capture the three-dimensional dynamical behaviors of the disposal process. For the first phase of disposal, the departure dynamics from DRO are analyzed with an improved metric based on MI. Unlike previous methods relying on a constant MI threshold, this approach evaluates relative variations. First, the difference between the post-disposal MI and the nominal periodic orbit's MI, denoted as $\Delta\mathrm{MI}(t)$, is calculated. The $\left|\Delta\mathrm{MI}(t)\right|$ is then normalized by the amplitude of the nominal orbit's MI. A successful departure is confirmed only when this newly defined, normalized metric exceeds a predefined dimensionless threshold. Because it evaluates relative deviation rather than absolute values, this proposed metric is independent of specific orbital geometries, thereby offering significantly broader applicability.

For the second phase of disposal, the cislunar escape dynamics are investigated through the framework combining the interpretation of escape behavior and the dependence of disposal parameters. The LD field is portraited by integrating the modulus of the orbital angular momentum w.r.t. the Earth-Moon barycenter (EMB). Trajectories achieving successful heliocentric escape are characterized by a pronounced increase in angular momentum modulus. In contrast, unsuccessful escape outcomes, such as bounded motion, lack this dynamical signature. Consequently, the constructed LD field can effectively distinguish successful escape from unsuccessful one. Furthermore, to systematically extract the escape results in LD field, a two-layer clustering algorithm is proposed. Utilizing the LD value as a key sorting parameter, this algorithm automatically extracts and categorizes orbital families that share similar escape behaviors with a fixed disposal maneuver magnitude. Finally, the influence of initial solar phase and disposal maneuver magnitude on escape opportunities is comprehensively discussed. It should be noted that this work applies only to objects that are capable of performing disposal maneuvers, not to defunct satellites or debris. The complete phases in the post-disposal orbital dynamics are shown in Fig. 1.

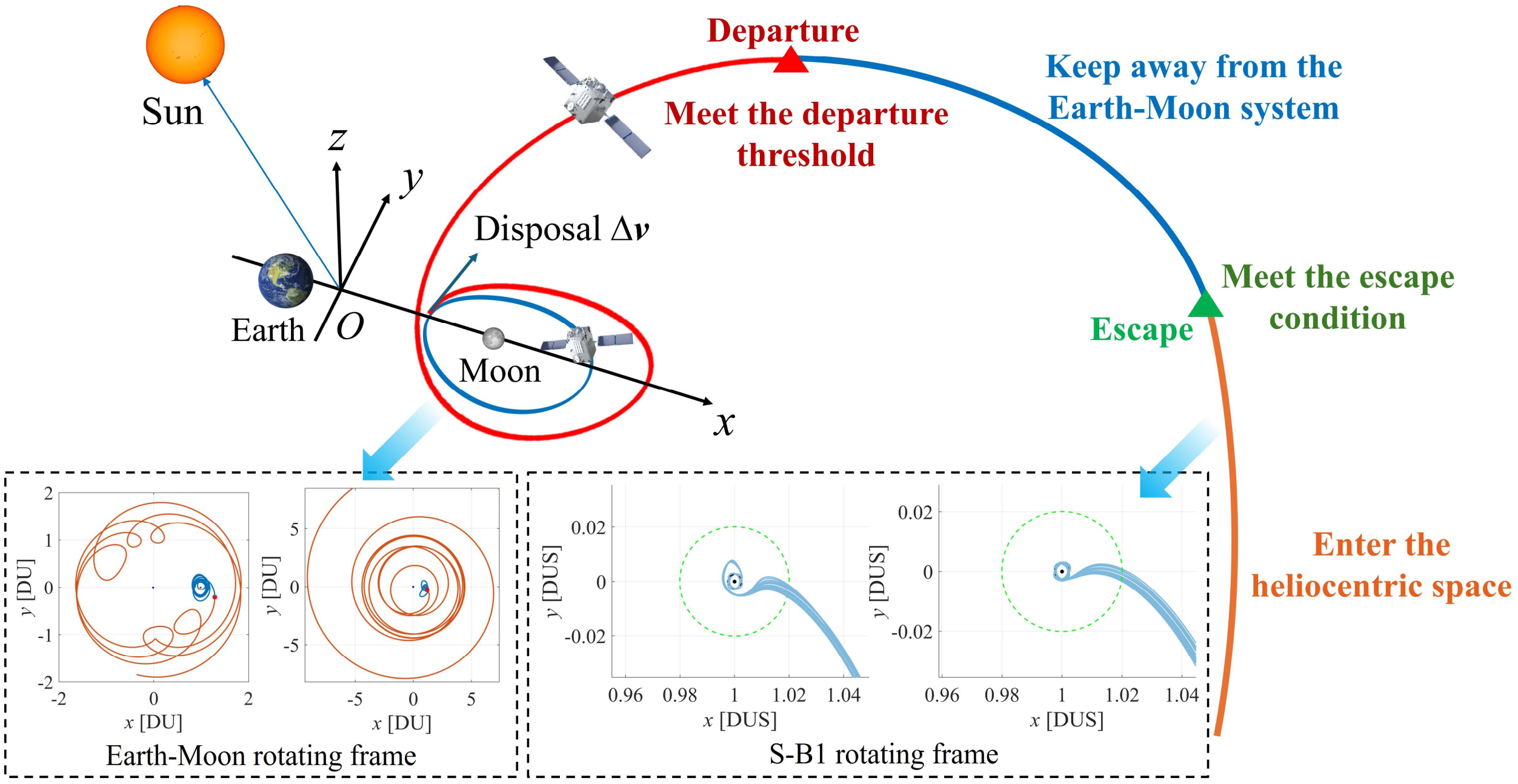

**Fig. 1 The phases in the post-disposal orbital evolution.**

The organization of this paper is described as follows: The orbital dynamical model and the backgrounds on the disposal dynamics are introduced in Section 2. The departure dynamics from 2:1 DRO and 3:1 DRO are analyzed in detail in Section 3. Subsequently, the escape dynamics from 2:1 DRO and 3:1 DRO are revealed in Section 4. Finally, some concluding remarks are presented in Section 5.

## 2. Orbital Dynamical Models

This investigation employs two orbital dynamical models for the computation and analysis of the cislunar escape trajectories. The first model is the CRTBP, which is leveraged to analyze the characteristics of orbital evolution dominated by the Earth and Moon. The BCRFBP model, including the gravitational influence of the Sun, is adopted for the generalized analysis of cislunar escape dynamics. Subsequently, the evolution patterns of mechanical energy and orbital angular momentum in CRTBP are examined to reveal the characteristics of orbital evolution with gravity assistant (GA). Then, the basic principle of LD is introduced, which is used in this study to reveal the low-energy cislunar escape field.

### 2.1 The Orbital Dynamical Models

#### 2.1.1 The CRTBP and BCRFBP Model

Assuming that, $m_1$ and $m_2$ represent the masses of the Earth and Moon, respectively. The mass of the spacecraft $m_3$ can be ignored compared to other two celestial bodies, and thus the spacecraft can be seen as a massless point. Both the Earth and Moon are considered as point masses moving around EMB in circular orbits. The motion of the spacecraft is formulated in the Earth-Moon rotating frame, with the origin located at the EMB, the ***x*** axis pointing from the origin to the Moon, the ***z***-axis paralleling to the angular momentum of the Earth-Moon system, and the ***y*** axis satisfying the right-handed rule, as shown in Fig. 2a. The mass ratio of the system is expressed as $\mu = m_2/(m_1+m_2)$. The state vector of spacecraft $\boldsymbol{X} = [x,\ y,\ z,\ \dot{x},\ \dot{y},\ \dot{z}]^T$, which represents the position and velocity in the rotating frame, is governed by the following equations of motion

$$\ddot{x} - 2\dot{y} = \frac{\partial \Omega_3}{\partial x},\quad \ddot{y} + 2\dot{x} = \frac{\partial \Omega_3}{\partial y},\quad \ddot{z} = \frac{\partial \Omega_3}{\partial z}. \tag{1}$$

where $\Omega_3$ is the effective potential, and is given by

$$\Omega_3(x,y,z) = \frac{1}{2}(x^2+y^2) + \frac{1-\mu}{r_1} + \frac{\mu}{r_2} + \frac{1}{2}\mu(1-\mu). \tag{2}$$

$r_1$ and $r_2$ denote the distances from the spacecraft to the Earth and Moon, respectively,

$$r_1 = \sqrt{(x+\mu)^2 + y^2 + z^2},\ r_2 = \sqrt{(x-1+\mu)^2 + y^2 + z^2}. \tag{3}$$

Without perturbations form other objects, the dynamical model of CRTBP is autonomous and holds an integral of motion, i.e., the Jacobi energy:

$$C = -(\dot{x}^2 + \dot{y}^2 + \dot{z}^2) + 2\Omega_3. \tag{4}$$

In the BCRFBP, the gravitational influence of the Sun is considered. The Sun with the mass $m_S$ is assumed to be in a coplanar circular orbit around the EMB with an angular velocity of $\omega_S$ in the Earth-Moon rotating frame. The solar phase in the Earth-Moon rotating frame is denoted as $\alpha_S = \omega_S t$. The distance between the EMB and the Sun is indicated as $\rho$. Two kinds of BCRFBP models are shown in Fig. 2b and Fig. 2c, respectively. The first model is derived in the Earth-Moon rotating frame, with the state of motion governed by the following equations

$$\ddot{x} - 2\dot{y} = \frac{\partial \Omega_4}{\partial x}, \quad \ddot{y} + 2\dot{x} = \frac{\partial \Omega_4}{\partial y}, \quad \ddot{z} = \frac{\partial \Omega_4}{\partial z}, \tag{5}$$

where $\Omega_4$ is the effective potential considering the solar gravity, and is given by

$$\Omega_4(x, y, z) = \Omega_3 + \Omega_S = \Omega_3 + \frac{m_S}{r_{3S}(t)} - \frac{m_S}{\rho^2}[x\cos(\omega_S t) + y\sin(\omega_S t)]. \tag{6}$$

$r_{3S}$ denotes the distance from the spacecraft to the Sun,

$$r_{3S}(t) = \sqrt{[x - \rho\cos(\omega_S t)]^2 + [y - \rho\sin(\omega_S t)]^2 + z^2}. \tag{7}$$

The second model is constructed in the Sun-EMB (S-B1) rotating frame, as shown in Fig. 2c, which is mainly used to analyze the geometric relationship between solar gravitational influence and the motion of spacecraft.

Due to the time-varying nature of solar gravitational influence, there is no integral of motion in the BCRFBP. However, the instantaneous Jacobi constant of spacecraft in the Earth-Moon system can reflect the changes in orbital energy, which will be analyzed in the following. The physical parameters of the Sun–Earth–Moon system and the nondimensional units are all listed in Table 1. The conception of escape boundary of cislunar region will be introduced in the following sections.

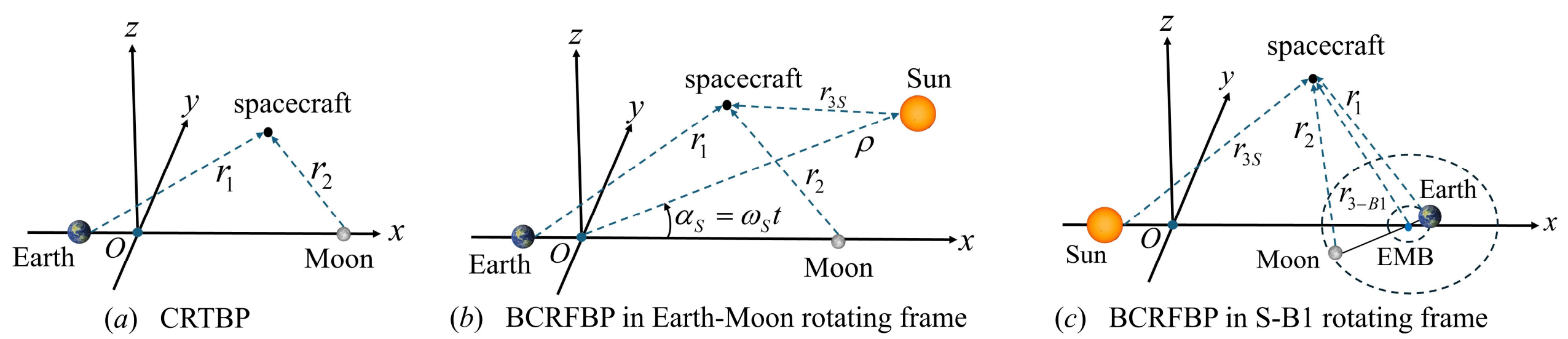

**Fig. 2 The CRTBP and BCRFBP models.**

**Table 1 Physical parameters used in the dynamical model.**

| Symbol | Value | Meaning |
|---|---|---|
| $\mu$ | $1.21506683\times10^{-2}$ | The mass ratio of the Earth-Moon system |
| DU, km | $3.84405000\times10^{5}$ | Distance unit in the Earth-Moon system (The nominal distance between the Earth and Moon) |
| DUS, km | $1.49460947\times10^{8}$ | (The nominal distance between the Sun and EMB) |
| TU, days | 4.34811305 $[T_{EM}/(2\pi)]$ | Time unit in the Earth-Moon system ($T_{EM}$ denotes the period of the Moon surrounding the Earth) |
| VU, km/s | 1.02323281 | Velocity unit (DU/TU) |
| $m_S$ | $3.28900541\times10^{5}$ | Nondimensional mass of the Sun in the Earth-Moon system (unit of mass is $m_1+m_2$) |
| $\omega_S$ | $-9.25195985\times10^{-1}$ | Nondimensional angular velocity of the Sun in the Earth-Moon system |
| $\rho$ | $3.88811143\times10^{2}$ | Nondimensional distance between the Sun and EMB in the Earth-Moon system |
| $R_{\text{lunar SOI}}$, km | 66180.0 (approximately 0.1722 DU) | The lunar SOI range w.r.t. the Moon's center |
| $R_{escape}$, km | $3.002585195\times10^{6}$ | The escape boundary of cislunar region w.r.t. the EMB |

### 2.1.2 The High-Fidelity Ephemeris Model

For applications in mission scenarios where high-fidelity modeling accuracy is required, the ephemeris model is established according to the restricted N-body problem (RNBP) by considering the planetary ephemerides. The Earth is selected as the central body, while the real position of the moon and sun with respect to the Earth is instantaneously obtained by the DE438 file.

The whole equation of the motion of the spacecraft in the Earth-centered J2000 inertial frame is written as

$$\ddot{\boldsymbol{r}}_{qi} = -\frac{G(m_i+m_q)\boldsymbol{r}_{qi}}{\left\|\boldsymbol{r}_{qi}\right\|^3} + G\sum_{\substack{j=1\\ j\neq i,q}}^{n} m_j\left(\frac{\boldsymbol{r}_{ij}}{\left\|\boldsymbol{r}_{ij}\right\|^3} - \frac{\boldsymbol{r}_{qj}}{\left\|\boldsymbol{r}_{qj}\right\|^3}\right) + \frac{(1+P)P_0\cdot r_{\text{S}}^{\,2}}{B}. \tag{8}$$

where $i$, $q$, and $j$ represent the spacecraft, the central body (Earth), and the perturbing bodies, respectively; the last term in Eq. (8) is the acceleration of solar radiation pressure, and the parameters are set as $P$=0.4, $P_0$=0.0046 kg/(m×s$^2$), $B$=21.74 m$^2$/kg, $r_{\text{S}}$ = 1 DUS, which are the same as in prior research [11].

## 2.2 The Evolution of Mechanical Energy and Angular Momentum in CRTBP

The mechanical energy in Keplerian orbit surrounding a celestial body maintains constant, while in the CRTBP, it is modified in the expression and usually changes with the evolution of trajectory. The mechanical energy relative to the EMB can be derived as the following:

$$E = \frac{1}{2}(\dot{x}^2 + \dot{y}^2 + \dot{z}^2) + \frac{1}{2}(x^2 + y^2) - \dot{x}y + \dot{y}x - \frac{1-\mu}{r_1} - \frac{\mu}{r_2}. \tag{9}$$

Besides, the expression of $E$ can establish a connection with Jacobi energy $C$, which is derived as [29]

$$E + \frac{C}{2} = x^2 + y^2 - \dot{x}y + x\dot{y} + \frac{\mu(1-\mu)}{2} = H_z + \frac{\mu(1-\mu)}{2}, \tag{10}$$

where $H_z$ represents the $z$-component orbital angular momentum w.r.t. EMB.

The mechanical energy $E$ in Eq. (9) combines the Earth's and Moon's gravitational potential. Especially, the lunar gravity assistant (LGA) has been proven to cause the lager increase in mechanical energy than the Earth gravity assistant (EGA) [29]. The lunar sphere of influence (SOI), with a radius of 66,180 km measured from the Moon's center, is commonly adopted as the boundary of the lunar encounter region [9,29]. The derivative of $E$ w.r.t. $t$ can be derived as

$$\frac{dE}{dt} = \frac{d\left(E + \frac{C}{2}\right)}{dt} = \frac{dH_z}{dt} = 2x\dot{x} + 2y\dot{y} + x\ddot{y} - y\ddot{x} = x\frac{\partial \Omega_3}{\partial y} - y\frac{\partial \Omega_3}{\partial x} = (1-\mu)\mu y\left(\frac{1}{r_1^3} - \frac{1}{r_2^3}\right). \tag{11}$$

To explain the impact of LGA on the change of orbital mechanical energy, the following part briefly illustrates the relationship between $dE/dt$ and position conditions inside the lunar SOI. The spatial position vector from the Moon to spacecraft is defined with $\theta$ and $\phi$ w.r.t. the Earth-Moon rotating frame, as shown in Fig. 3a, which can be expressed as $\boldsymbol{r}_2 = r_2[\cos\phi\cos\theta,\ \cos\phi\sin\theta,\ \sin\phi]^T$, and $\theta \in [0,\ 360°]$, $\phi \in [-90°,\ 90°]$. The distribution patterns of $dE/dt$ with different distances from the Moon's center are displayed in Fig. 3b~Fig. 3d. The obvious feature is that, in the region where $\theta \in (180°,\ 360°)$ (i.e., $y < 0$), $dE/dt$ remains positive, while in the region where $\theta \in (0,\ 180°)$, $dE/dt$ remains negative. The highest value of $|dE/dt|$ with a constant distance $r_2$ relative to the Moon's center locates at $\theta = 90° \,|\, 270°$ & $\phi = 0$. Overall, when the spacecraft transits in the $y < 0$ region of lunar SOI, $E$ increases, while when it transits in the $y > 0$ region of lunar SOI, $E$ decreases. Therefore, the lunar SOI region with $y < 0$ usually provides favorable conditions for exterior low-energy transfer and cislunar escape.

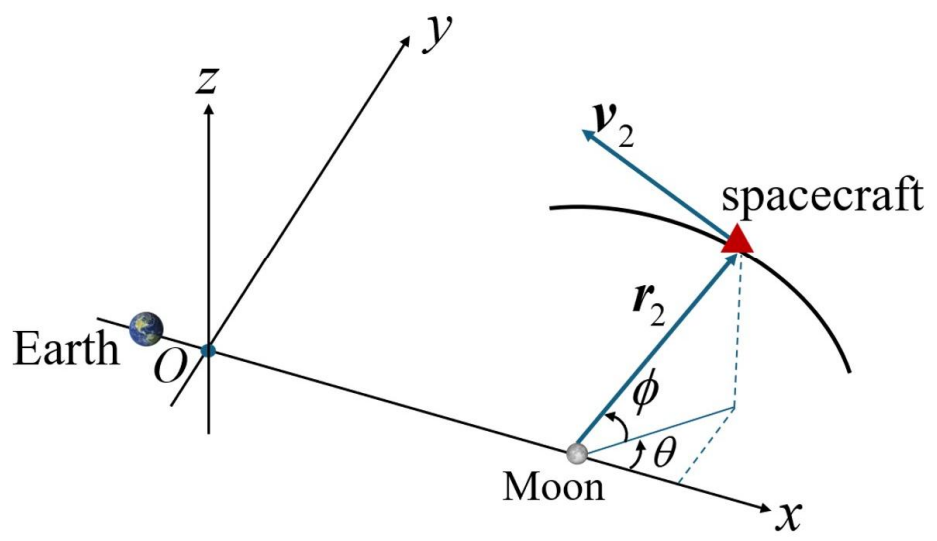

($a$) LGA trajectory of spacecraft within the lunar SOI

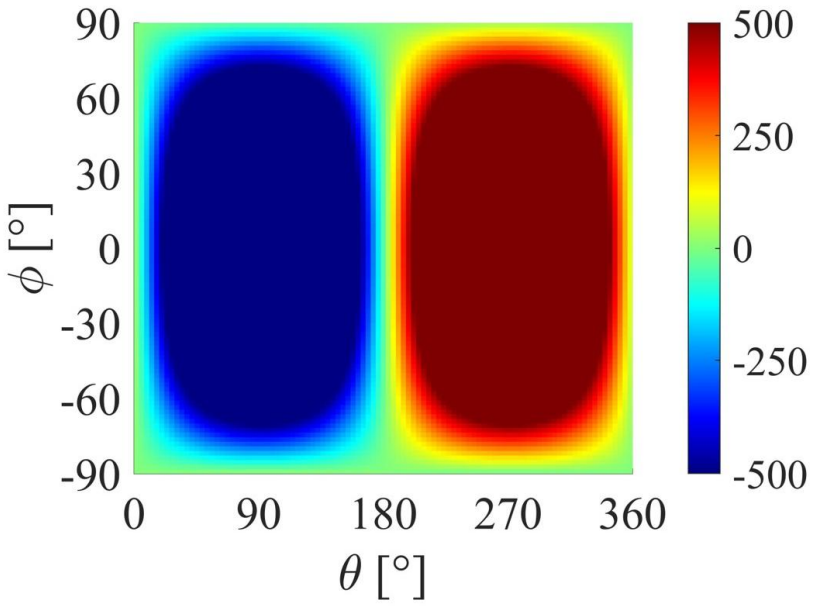

($b$) $r_2 = 0.025$ [DU]

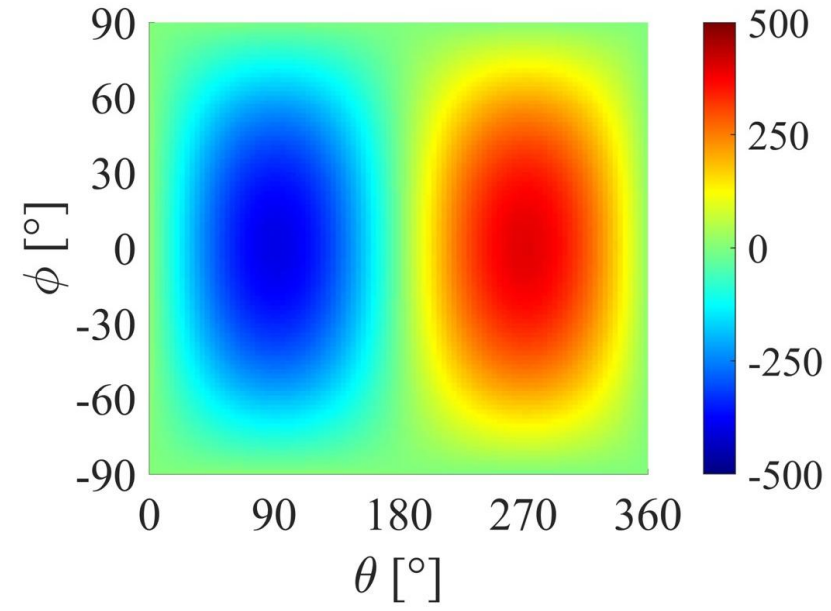

($c$) $r_2 = 0.050$ [DU]

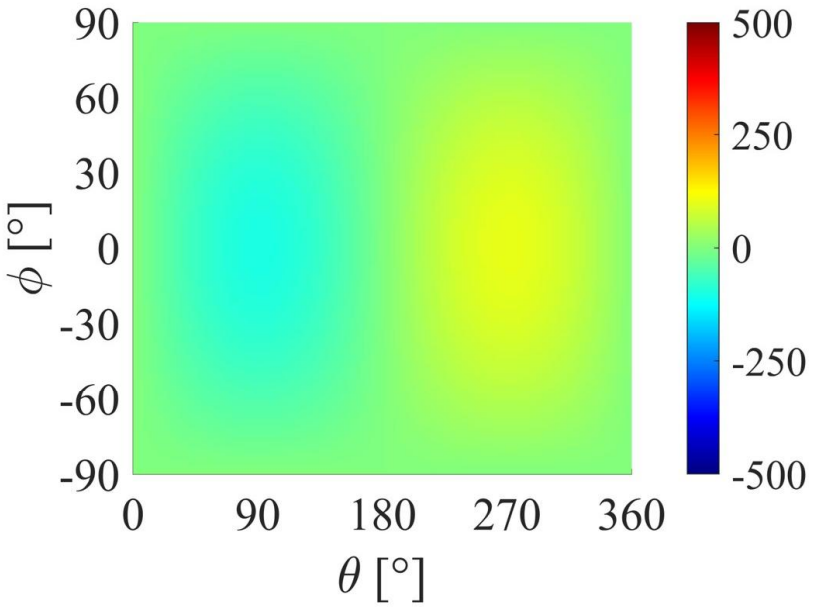

($d$) $r_2 = 0.100$ [DU]

**Fig. 3 The LGA trajectory and distribution patterns of $dE/dt$ with different distances from lunar center.**

### 2.3 The Lagrangian Descriptor

The concept of Lagrangian descriptor (LD) was first introduced in the research about the geometry of ocean flows [30] and was later shown to outperform finite time Lyapunov exponent (FTLE) in revealing phase-space structures in dynamical systems [26]. Unlike FTLE, which quantifies the maximum finite-time exponential separation of nearby trajectories, LD offers a different viewpoint: it characterizes phase-space geometry by accumulating a prescribed dynamical quantity along trajectories. As a result, LD fields can clearly expose structures such as invariant manifolds, resonance islands, and chaotic regions [26,31]. Because of this ability to emphasize geometrical organization directly, LD constitutes the core methodology of this work.

Below is the LD calculation framework presented in this study. Considering a general time-dependent vector field on $\mathbb{R}^6$, and the equations of motion of state vector $\boldsymbol{X}(t) \in \mathbb{R}^6$ in BCRFBP are governed by Eq. (5). The LD is defined as

$$M(\boldsymbol{X}, t_0, t_f) = \int_{t_0}^{t_0+t_f} \left|\mathcal{F}(\boldsymbol{X}, t)\right|^q dt, \tag{12}$$

where $t_0$ indicates the initial epoch and $t_f$ denotes the forward integration interval. Substituting the modulus of angular momentum into Eq. (12), the integrand can be derived as

$$\mathcal{F}(\boldsymbol{X},t)=\sqrt{[y\dot{z}-z(\dot{y}+x)]^2+[z(\dot{x}-y)-x\dot{z}]^2+[x(\dot{y}+x)-y(\dot{x}-y)]^2}, \tag{13}$$

where the unit of $\mathcal{F}(\boldsymbol{X},t)$ is set as the S-B1 nondimensional unit.

The integrand $|\mathcal{F}(\boldsymbol{x},t)|^q$ in Eq. (12) is a bounded, positive quantity, while $q$ is the exponent defining the norm. In this study, $q$ is set as 1/2. The choice of integrand $|\mathcal{F}(\boldsymbol{x},t)|^q$ determines which dynamical features are emphasized. More precisely, quantities related to the dynamical state can be used as integrand, commonly including the modulus of velocity, modulus of acceleration, and the curvature [26,27]. In multibody astrodynamics, prior studies often adopt the velocity norm as the integrand [28,32]. In contrast, to better capture the characteristics of cislunar escape motion, one can select physically motivated quantity as the integrand [33]. The integrands built from the modulus of the orbital angular momentum w.r.t. the EMB are employed.

The rationale for selecting the modulus of the EMB-referenced orbital angular momentum as the integrand is detailed as follows. Fig. 4 illustrates representative cislunar escape and bounded trajectories simulated within the BCRFBP, alongside the time histories of their corresponding orbital angular momentum. Fig. 4a displays the angular momentum $H$ profiles in the S–B1 nondimensional unit, where the three colors indicate distinct evolution patterns. Fig. 4b projects the associated trajectories into the S–B1 rotating frame, utilizing an identical color code. Consistent with theoretical expectations, a successful departure from the cislunar environment strongly correlates with an increase in the EMB-referenced angular momentum. Conversely, failed escapes, such as quick returns or long-term bounded motions, exhibit no such increase. It is worth noting that all escape paths are generated via Lunar Gravity Assist (LGA), a well-established technique [34,35]. In Fig. 4b, the lunar orbit is denoted by a dashed black circle, while the dashed green circle defines the formal escape boundary of the cislunar region.

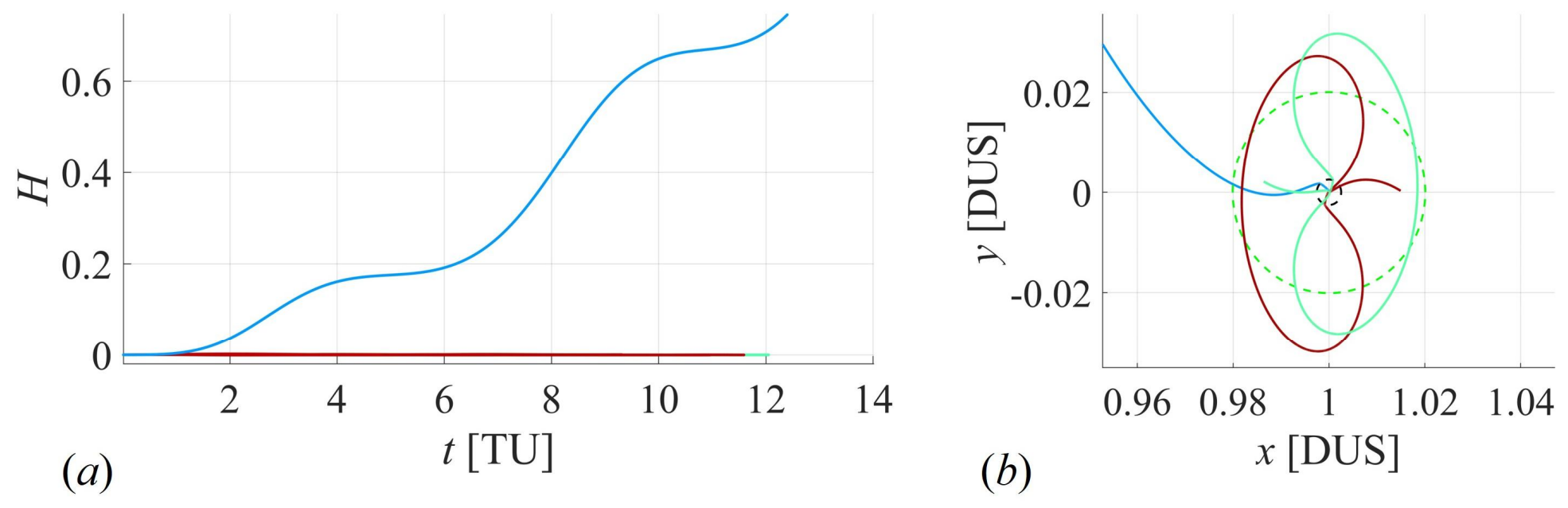

**Fig. 4 The comparison of successful escape and unsuccessful escape.**

The LD field is mathematically defined as $M(t_0,t_f):=\{M(\boldsymbol{X}_0,t_0,t_f),\ \boldsymbol{X}_0\in\Theta\}$, where $\Theta$ represents the domain of initial conditions. Because this computation aims to characterize cislunar escape trajectories, the integration time interval must satisfy $t_f > 0$. During numerical simulations, the integration stops automatically if the spacecraft collides with either the Earth or the Moon.

## 3. Departure Dynamics from DROs

During the cislunar operations, various assets, ranging from deployed CubeSats to end-of-life spacecraft, will inevitably require disposal maneuvers to depart from the nominal DROs. This section comprehensively investigates the departure dynamics from the DROs within the BCRFBP framework. The first Section establishes the fundamental characteristics of the DROs, detailing the orbital geometry and stability. To accurately assess successful departure from the nominal orbit, a new momentum integral (MI)-based metric is applied to evaluate the post-disposal trajectories. The maximum stretching direction of the phase flow is utilized for comparing the disposal maneuver direction of fast departure. Finally, this section systematically examines the time-to-depart characteristics, quantifying the influence of disposal maneuver magnitude, the disposal execution location, and the maneuver direction. Due to the gravitational perturbation of the Sun, the solar phase has a significant impact on the possibility of departure.

### 3.1 Characteristics of DROs

The DRO family consists of planar periodic orbits in the CRTBP, characterized by retrograde motion about the Moon in the Earth-Moon rotating frame. As the Jacobi constant increases, the average distance of the position along DRO w.r.t the moon gradually decreases, as illustrated in Fig. 5. The blue and red markers in Fig. 5 denote the Earth and the Moon, respectively; and the "+" symbols represent the L1 and L2 points, respectively.

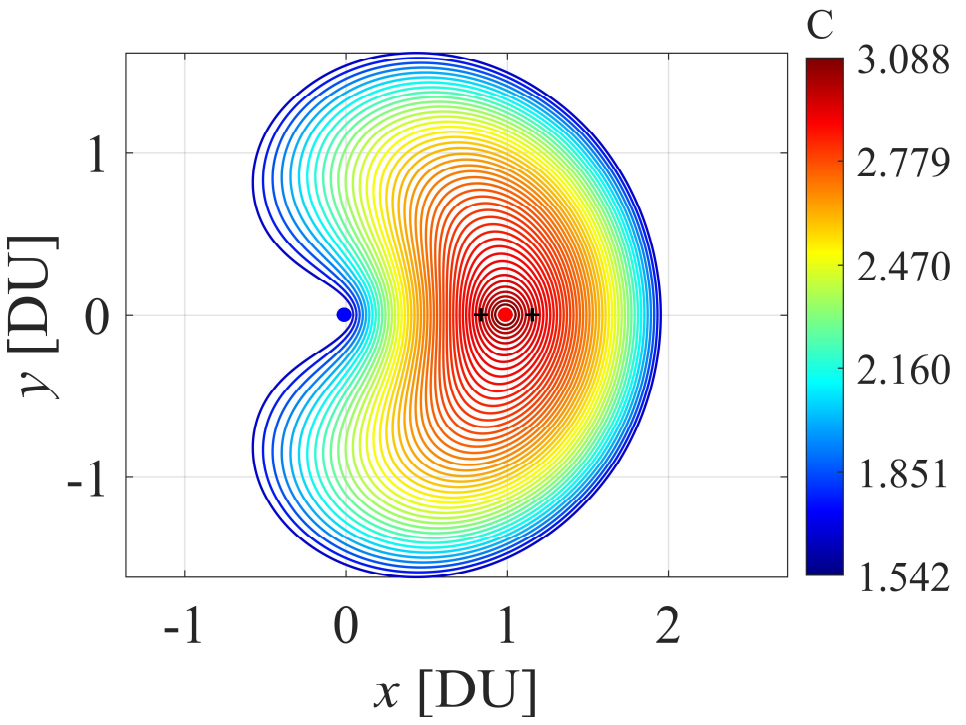

**Fig. 5 The DRO family with different Jacobi constant.**

In the BCRFBP, a strictly periodic orbit must satisfy a resonance condition with the synodic period of the system. The synodic period in the BCRFBP equals the solar orbital period surrounding the EMB, which is governed by $T_{syn} = |2\pi / \omega_S| = 6.79119$ [TU] and is approximately 29.5 days [36]. The generation of initial conditions for periodic solutions in BCR4BP is achieved by leveraging the synodic resonant orbits in the Earth-Moon CRTBP. Specifically, the solar mass ratio $\varepsilon$ is introduced as a homotopy parameter, so that a periodic solution in the CRTBP with $\varepsilon = 0$ is gradually corrected into BCRFBP with $\varepsilon = 1$. In the present study, the initial solar mass ratio is set to 0.001, the homotopy step size $\delta\varepsilon$ is taken as 0.001, and the initial solar phase $\alpha_{S0}$ is fixed at 0. To assess the stability of periodic orbits in the BCRFBP, the Lyapunov exponent $\varphi_i$ is defined as the real part of the Floquet exponent, i.e.,

$$\varphi_i = R\left(\frac{\ln \lambda_i}{T}\right), \tag{14}$$

where $T$ is the period of orbit, and $\lambda_i$ denotes the *i*-th eigenvalue of the monodromy matrix associated with the periodic solution in BCRFBP. The phase space in the BCRFBP contains six dimensions, and thus, the monodromy matrix corresponding to a periodic solution admits six eigenvalues, labeled $\lambda_1$ through $\lambda_6$. In the absence of perturbation, the strict linear stability of a periodic solution is achieved only when their Lyapunov exponents are all zero, i.e., the corresponding multipliers lie on the unit circle. Otherwise, the orbit is linearly unstable.

The two DROs exhibit slightly different stability properties, as shown in Table 2. 2:1 DRO is strictly linearly stable, whereas 3:1 DRO holds an unstable Lyapunov exponent pair of ±0.02132. The geometry of two DROs are illustrated in Fig. 6. The stronger stability of DRO is favorable for long-duration operations with reduced station-keeping demand, but it may be less advantageous for disposal and escape design. In addition, both 2:1 DRO and 3:1 DRO are more stable than the 9:2 NRHO, the Lyapunov exponent pair of the latter is ±0.5208 [15]. This comparison suggests that departure from DROs may be more challenging than from a 9:2 NRHO, as will be examined in the following Sections.

**Table 2 The Lyapunov exponent of DROs.**

| Orbit | $\varphi_{1,2}$ | $\varphi_{3,4}$ | $\varphi_{5,6}$ |
|---|---|---|---|
| 2:1 DRO | 0 | 0 | 0 |
| 3:1 DRO | ±0.02132 | 0 | 0 |

In the BCRFBP, the disposal location along the DRO is uniquely associated with the solar phase. As illustrated in Fig. 6a and Fig. 6d, the orbital phase along the DRO, denoted by $\theta_{DRO}$, is measured in a clockwise direction and also coupling with the solar phase $\alpha_{S0}$. Fig. 6b and Fig. 6e further illustrates the geometric relationship between the DRO state and the solar phase in the S-B1 rotating frame. For the position of $\theta_{DRO}=0$, the solar phase is $\alpha_{S0}=0$. Fig. 6d and Fig. 6e illustrate the 3:1 DRO in the Earth-Moon rotating frame and S-B1 rotating frame, respectively. The unit in Fig. 6b and Fig. 6e, i.e. DUS, is the distance from the Sun to EMB, which equals $1.49460947\times10^{8}$ km. Fig. 6c and Fig. 6f show 2:1 and 3:1 DROs, respectively, after transition to the ephemeris model, represented in the S–B1 rotating frame. The transition procedure follows the methods described in Refs. [37,38]. The DRO trajectories are shown in red, and the lunar orbit is shown in black. In the ephemeris model, the time-varying Earth–Moon and Sun–B1 distances introduce time-dependent perturbations, causing the DRO trajectories to exhibit quasi-periodic rather than strictly periodic behavior. Despite the loss of exact periodicity, their overall geometry remains qualitatively consistent with the corresponding periodic orbits in the BCRFBP.

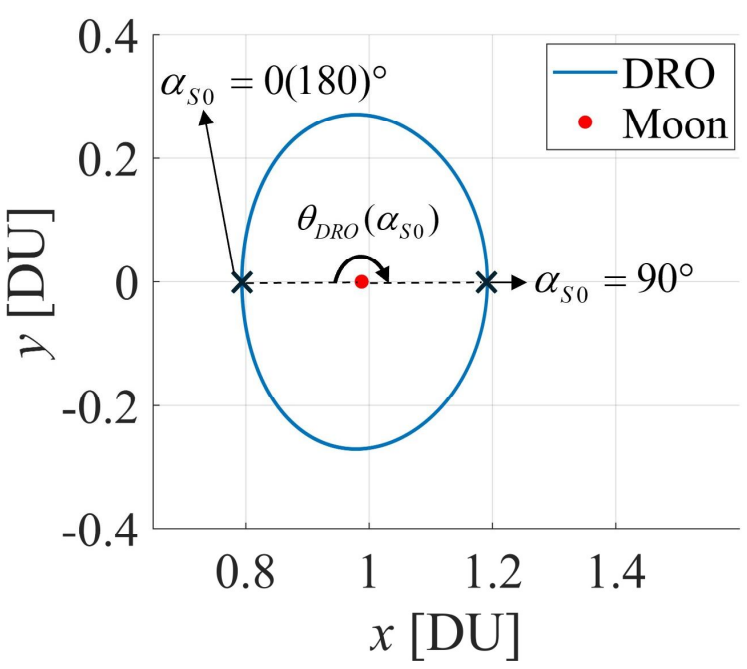

(*a*) Departure points of 2:1 DRO in EM rotating frame

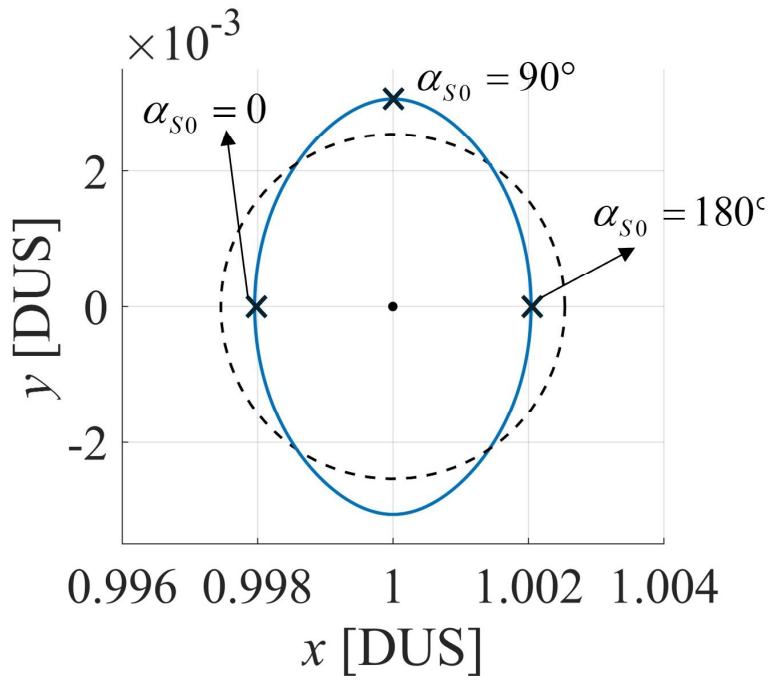

(*b*) Departure points of 2:1 DRO in S-B1 rotating frame

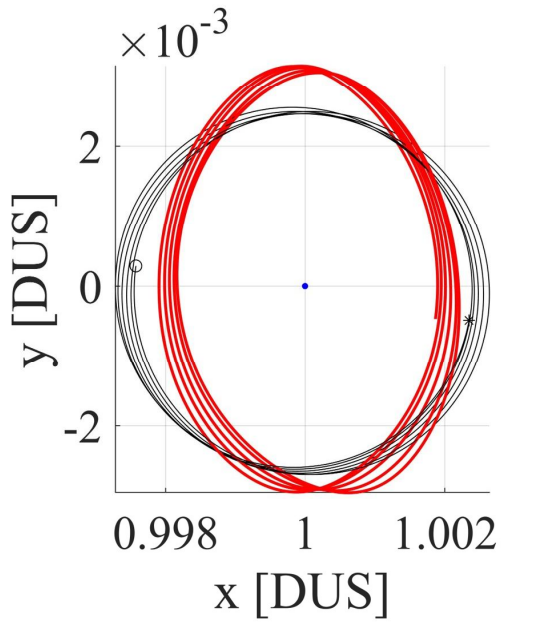

(*c*) 2:1 DRO in high-fidelity ephemeris model (S-B1 rotating frame)

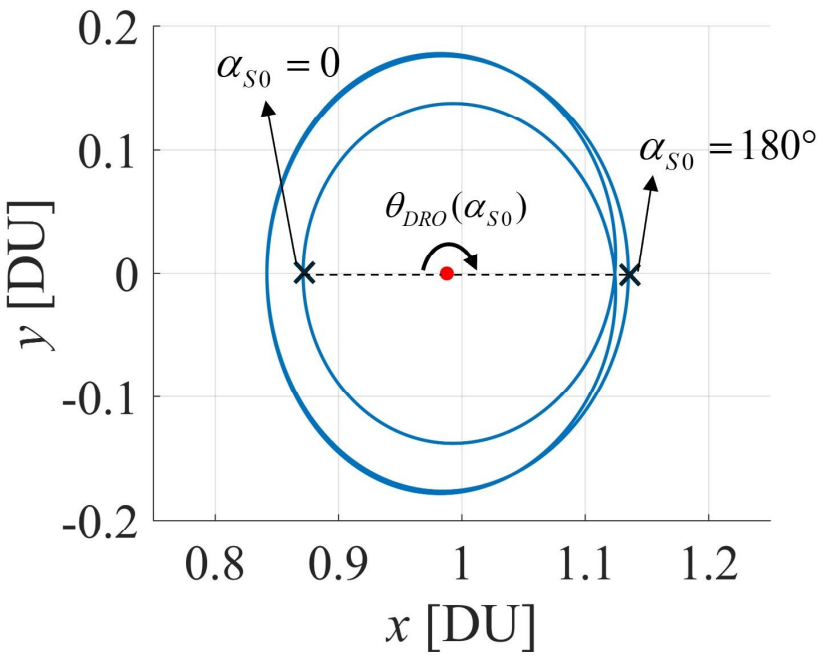

(*d*) Departure points of 3:1 DRO in EM rotating frame

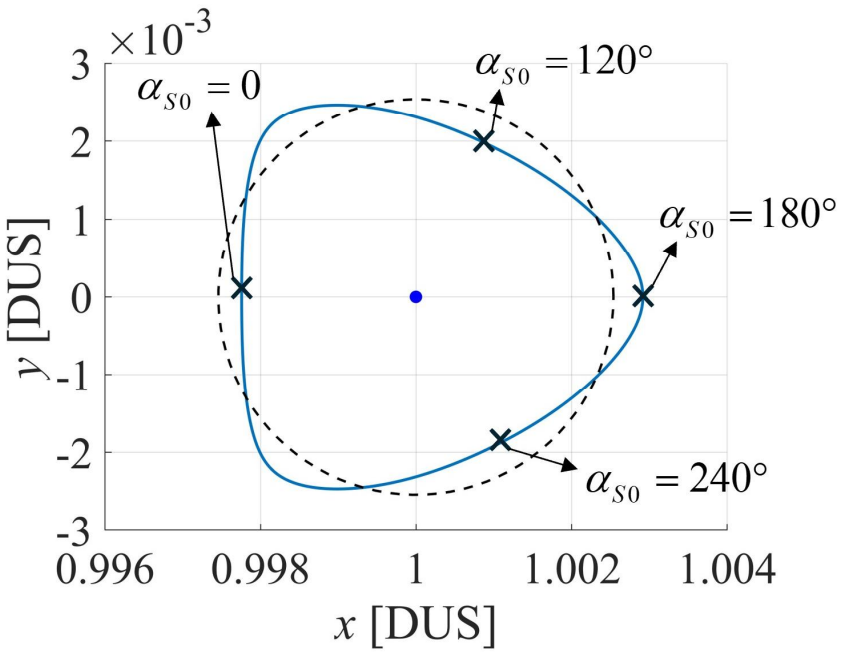

(*e*) Departure points of 3:1 DRO in S-B1 rotating frame

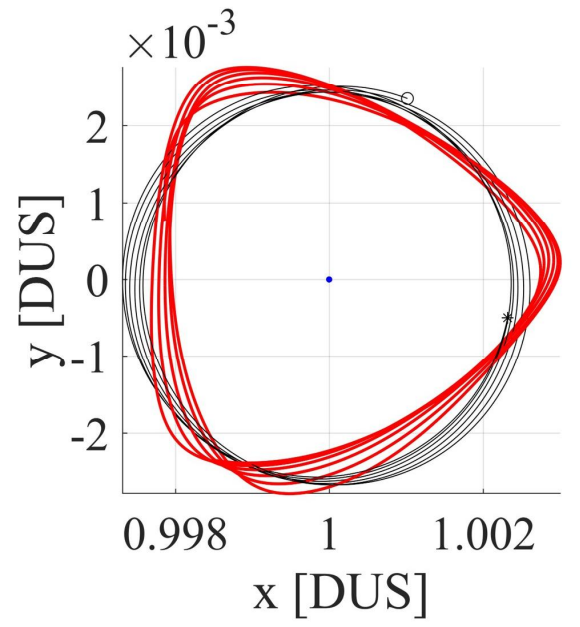

(*f*) 3:1 DRO in high-fidelity ephemeris model (S-B1 rotating frame)

**Fig. 6 The relationship between the position of the disposal and the solar phase for DROs.**

Before presenting the departure results, it is necessary to clarify both the direction of disposal maneuver and the position of disposal. Fig. 7 illustrates the definition of the disposal direction along the DRO and introduces the local velocity-binormal-normal (VBN) reference frame. The VBN reference frame is defined in the Moon-centered rotating frame, where *V*-axis is aligned with the spacecraft velocity vector $(\dot{x},\dot{y},\dot{z})$ in the Moon-centered rotating frame, the $N$-axis is aligned with the negative orbital normal direction which can be derived as $\hat{N} = -(\hat{x}, \hat{y}, \hat{z}) \times (\hat{\dot{x}}, \hat{\dot{y}}, \hat{\dot{z}})$, and the $B$-axis completes the right-handed coordinate system and $\hat{B} = \hat{N} \times \hat{V}$. The three-dimensional direction for the disposal maneuver is explored, with each direction parameterized by azimuth and elevation angles relative to the VBN reference frame, as depicted in Fig. 7. The disposal maneuver vector in the VBN reference frame is then expressed as

$$\Delta \boldsymbol{v}_{VBN} = \Delta v[\cos\phi\cos\theta,\ \cos\phi\sin\theta,\ \sin\phi]^T, \tag{15}$$

where $\theta \in [0,\ 360°]$, $\phi \in [-90°,\ 90°]$. The vector of disposal maneuver can be converted into Earth-Moon rotating frame as

$$\Delta \boldsymbol{v} = [\hat{V}, \hat{B}, \hat{N}] \cdot \Delta \boldsymbol{v}_{VBN} = \boldsymbol{T} \cdot \Delta \boldsymbol{v}_{VBN} \tag{16}$$

where $\boldsymbol{T}$ is the transition matrix from VBN reference frame to Earth-Moon rotating frame.

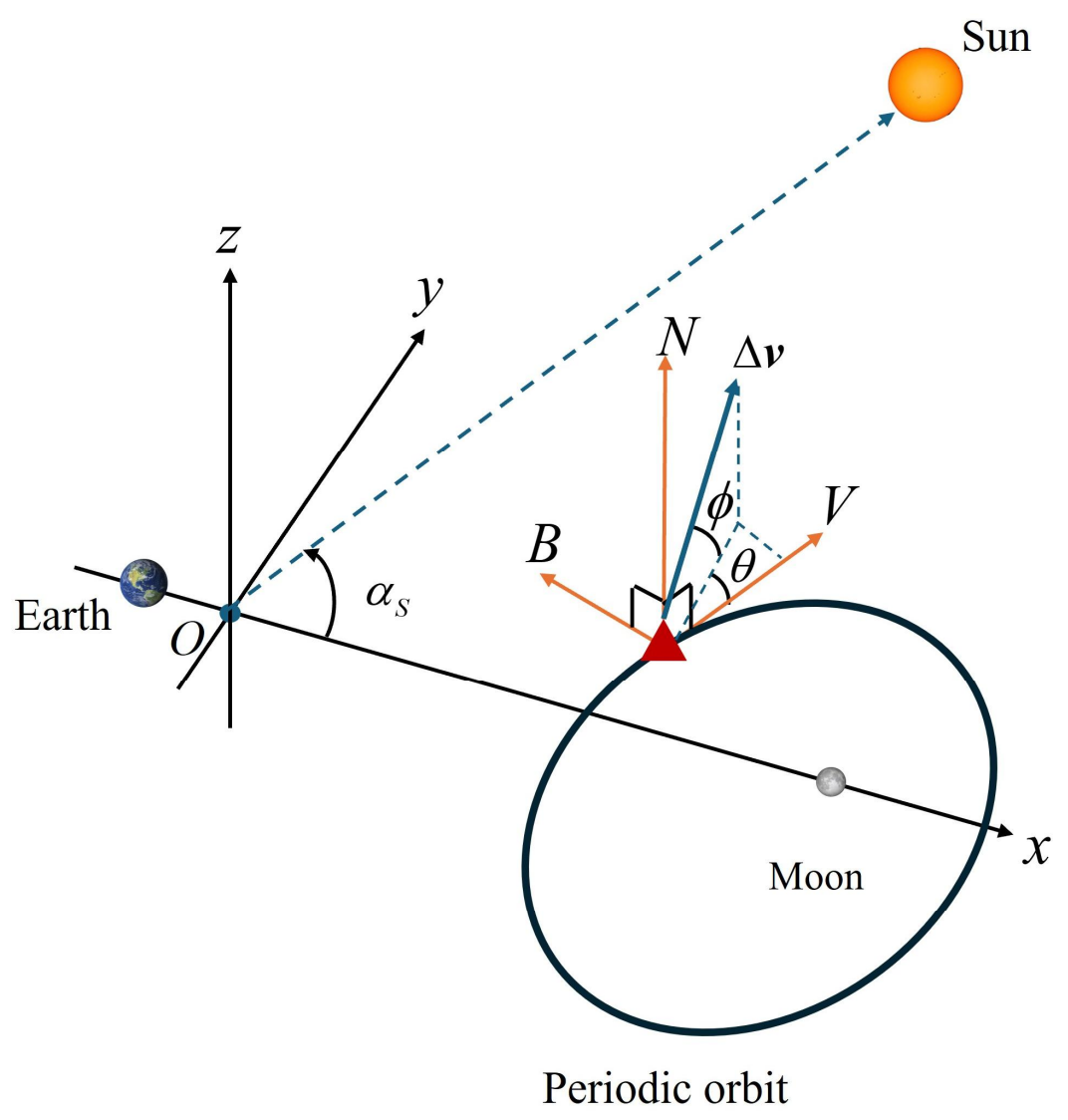

**Fig. 7 The direction of disposal maneuver of the position along DRO and the VBN reference frame.**

### 3.2 Departure Metric Based on Momentum Integral

Building on previous studies of departure dynamics from NRHO [15,17], this work adopts the momentum integral (MI) as a key diagnostic tool. The MI is defined as a line integral reflecting the position vector from the initial time $t_0$ to the current time $t$,

$$\mathrm{MI}(t) = \int_{t_0}^{t} \boldsymbol{r}(t)\cdot\boldsymbol{v}(t)dt = \int_{r(t_0)}^{r(t)} \boldsymbol{r}(t)d\boldsymbol{r}(t) = \int_{t_0}^{t} [x(t)\cdot\dot{x}(t) + y(t)\cdot\dot{y}(t) + z(t)\cdot\dot{z}(t)]dt, \tag{17}$$

where $x$, $y$, and $z$ are components of the position vector w.r.t. the EMB in the rotating frame; and $\dot{x}$, $\dot{y}$, and $\dot{z}$ are the corresponding velocity components. According to the Kelvin–Stokes theorem, over a complete period of a periodic orbit, the MI can be interpreted as the circulation of the velocity field, or equivalently, as the integral of the curl of the position vector over the area enclosed by the orbit [39]. Therefore, in the complete period $T$ of orbital motion,

$$\mathrm{MI}(T) = \oint_C \boldsymbol{r}(t) d\boldsymbol{r}(t) = \oiint_S \nabla \times \boldsymbol{r}(t) d\boldsymbol{s} = 0, \tag{18}$$

where $S$ is the area swept in the radial direction, and Eq. (18) holds for any closed loop in a gradient field. For strictly periodic motion, the MI over one or more orbital periods is zero, making it a sensitive indicator of long-term periodicity. If the MI of a disposal trajectory remains similar to that of the nominal periodic orbit, the motion retains quasi-periodic characteristics. Conversely, significant divergence in MI indicates that the trajectory has departed from the nominal orbit. Notably, the MI in Eq. (17) admits a closed-form:

$$\mathrm{MI}(t) = \frac{1}{2}\left(\left\|\boldsymbol{r}(t)\right\|^2 - \left\|\boldsymbol{r}(t_0)\right\|^2\right). \tag{19}$$

In previous work about departure dynamics from NRHO [15], a constant threshold $\mathrm{MI}_{\text{dep-NRHO}} = 0.1$ was used: a trajectory was deemed departed once the MI exceeded the threshold $\mathrm{MI}_{\text{dep-NRHO}}$. However, this approach is not directly transferable to other periodic orbits due to differences in MI scale. To enable a more general analysis across different orbital families, this study introduces a new normalized MI-based departure metric.

First, the difference between the post-disposal MI and the nominal periodic orbit's MI is defined as

$$\Delta\mathrm{MI}(t) = \mathrm{MI}(t) - \mathrm{MI}^*(t). \tag{20}$$

This difference $\Delta\mathrm{MI}(t)$ is then normalized by the amplitude of the MI for the nominal orbit

$$\Delta\mathrm{MI}_{\text{nd}}(t) = \Delta\mathrm{MI}(t) / [\max(\mathrm{MI}^*) - \min(\mathrm{MI}^*)]. \tag{21}$$

For the cases of 2:1 DRO and 3:1 DRO, the amplitude of the MI is notably larger than that for 9:2 NRHO, owing to the larger distance from the Moon. Finally, the new departure metric is defined as

$$\Delta\mathrm{MI}_{\text{nd}}(t) \ge K, \tag{22}$$

where $K$ is the predefined constant. In this research, $K = 1.6$.

The proposed metric in Eq. (22) features two crucial properties. When the trajectory reaches the threshold in Eq. (22), on the one hand, the trajectory has departed substantially from the nominal periodic orbit, indicating a loss of nominal periodicity; on the other hand, the trajectory is beyond the region near the nominal periodic orbit and headed towards the exterior cislunar region. The normalized MI is not intended as a vectorial departure metric: it quantifies the one-sided change in squared barycentric radius and is used here to identify the loss of periodic bounded motion after disposal. In addition, the setting of $K$ also needs to be selected correctly. If the value of $K$ is too small, it is possible that the spacecraft has not yet left the vicinity of DRO but has been determined to be the departure state mistakenly. If the value of $K$ is too large, the spacecraft may have already escape the cislunar space but has not yet been determined to be the departure state. The selection of $K=1.6$ is a heuristic value in this work which consistently avoids both of these issues for the range of scenarios considered.

Fig. 8 showcases three representative departure cases from the 2:1 DRO. In Fig. 8, the red curve in the right panel denotes the MI evolution of the nominal orbit; blue and orange curves correspond to the MI of disposal trajectories before and after the departure epoch, respectively; and the green curve corresponds to the $\Delta\text{MI}_{\text{nd}}$ of disposal trajectories. The red "*" symbol marks the departure event. The period of the 2:1 DRO, equivalent to the synodic period, is denoted as $T_{\text{DRO}}$. Specifically, Fig. 8a presents a fast departure, with the spacecraft leaving the DRO only 2.65 $T_{\text{DRO}}$ from disposal maneuver to departure. The case in Fig. 8b shows a much slower departure, requiring about 11.0 $T_{\text{DRO}}$. Fig. 8c demonstrates a case where the object, after 4.5 $T_{\text{DRO}}$, departs from the vicinity of the DRO but remains bounded within cislunar space and may eventually return near the original orbit. The case (c) in Fig. 8 illustrates the benefits of the new metric, as the previous departure metric required MI to exceed the constant threshold that is much larger than the nominal amplitude of MI. Although the MI in case (c) of Fig. 8 does not increase positively, it can still be considered as departure. If the constant threshold that is much larger than the nominal amplitude of MI is adopted in the departure situation of DRO, then cases (a) and (b) can be considered as departure, but case (c) in Fig. 8 cannot be considered as departure. However, the trajectory in case (c) has clearly left the vicinity of DRO after 4.5 $T_{\text{DRO}}$.

The new departure metric in Eq. (22) effectively distinguishes between departure and non-departure scenarios. However, it should be noted that departure is only the first phase following a disposal maneuver. As highlighted in Fig. 8c, the object may persist in cislunar space for extended durations without achieving escape. A more comprehensive analysis on how the magnitude, direction, and position along the DRO of the disposal maneuver influence the departure results will be discussed in the subsequent section.

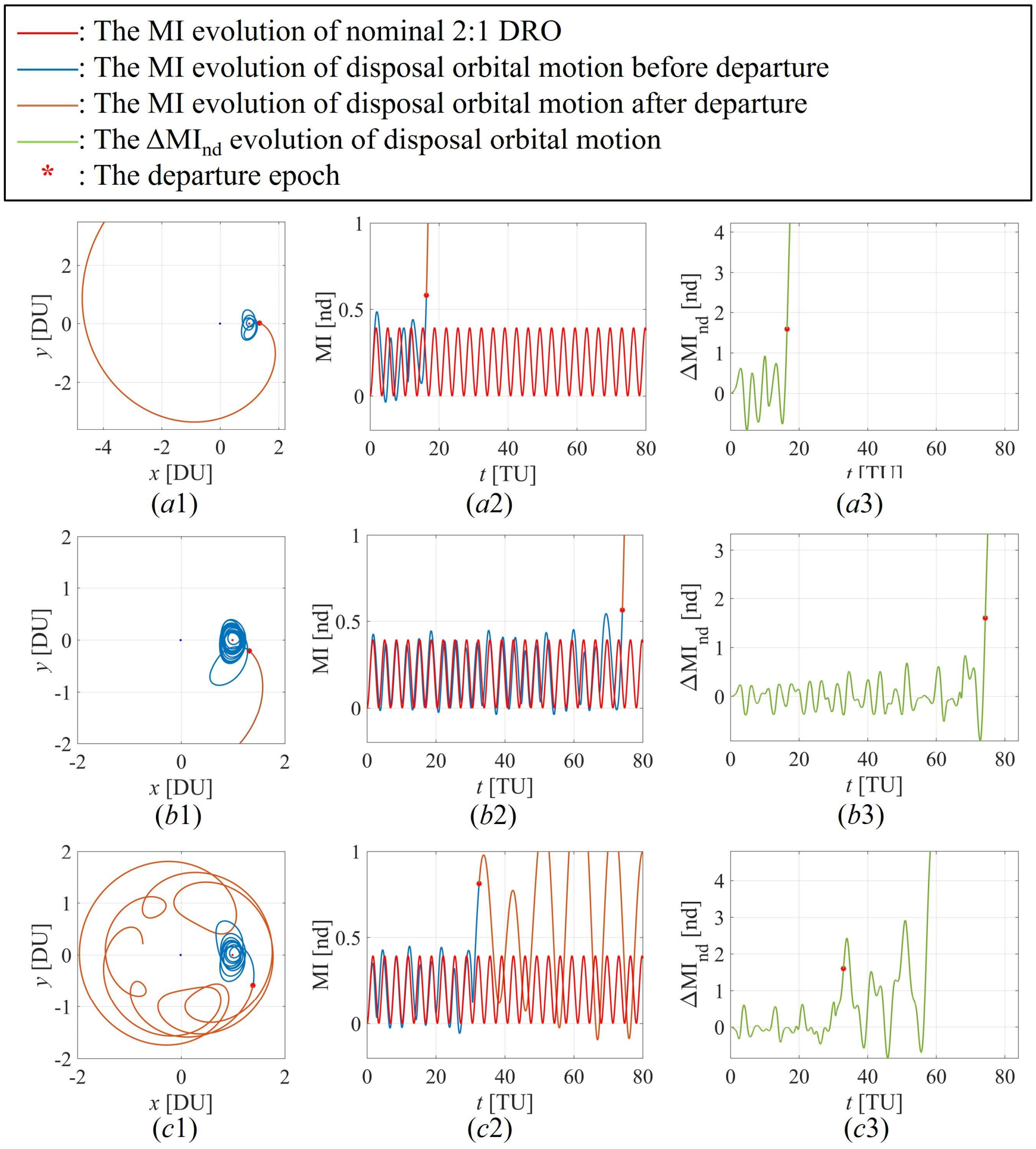

**Fig. 8 The disposal trajectories of 2:1 DRO and the related MI evolution.**

### 3.3 Time-to-Depart Analysis

Each disposal location along the DRO uniquely corresponds to a solar phase, thus the initial solar phase $\alpha_{S0}$ defines the disposal epoch. In addition, the disposal maneuver magnitude $\Delta v$ and the direction $\theta$ and $\phi$ form a three-dimensional parameter space. The parameter space for departure dynamics is therefore four-dimensional, consisting of the disposal location (initial solar phase), magnitude, and two direction angles of the maneuver. For clarity, the following analysis first fixes the initial solar phase and maneuver magnitude, and examines how time-to-depart varies with the maneuver direction. In this study, the orbital dynamical equations are propagated with the function ode45, with the 'RelTol' and 'AbsTol' set as $10^{-11}$. The step size of all the angular variables is set as 0.0314 [rad].

The time-to-depart is defined as the interval between the execution of the disposal maneuver and the epoch when the trajectory crosses the departure threshold introduced in Eq. (22). $\alpha_{S0} = 0$ and a disposal maneuver magnitude of $\Delta v = 25$ m/s for 2:1 and 3:1 DROs are chosen firstly. The time-to-depart map w.r.t $\theta$ and $\phi$ is illustrated in Fig. 9. Four representative cases with distinct time-to-depart behavior are highlighted for 2:1 DRO, and two cases are selected for 3:1 DRO. The color scale on the right side indicates time in the unit of $T_{DRO}$, with 3.5 $T_{DRO}$ corresponding to 103.35 days. The specific values of $\theta$ and $\phi$ for six cases in Fig. 9 are listed in Table 3. The corresponding orbital motion, MI evolution, and $\Delta \mathrm{MI}_{\mathrm{nd}}$ evolution in six cases selected in Fig. 9 are shown in Fig. 10. In the case of $\alpha_{S0} = 0$, the velocity direction corresponds to $\theta = 0°, \phi = 0°$, and the anti-velocity direction is $\theta = 180°, \phi = 0°$. Notably, for both 2:1 and 3:1 DROs, the maneuver direction of the fastest departure does not necessarily occur strictly along the velocity direction, as a sharp contrast to the NRHO case [15]. This result also reflects the complexity of dynamics in the vicinity of DRO. For NRHO, a 5 m/s disposal maneuver applied at perilune is typically sufficient for fast departure, due to the weaker stability of NRHO and higher sensitivity to velocity direction perturbations. By contrast, for the DRO, cases (a), (c), (d), and (e) in Fig. 10 achieve departure from the vicinity of the DRO but remain bound within cislunar space, because the last close lunar encounter does not provide sufficient energy increase for escape. Only cases (b) and (f) demonstrate both departure and successful escape, and the detailed escape dynamics will be analyzed in the next section. The most important fact is that the new orbit departure metric defined by Eq. (22) can be accurately applied to 2:1 and 3:1 DRO, not limited to one periodic orbit.

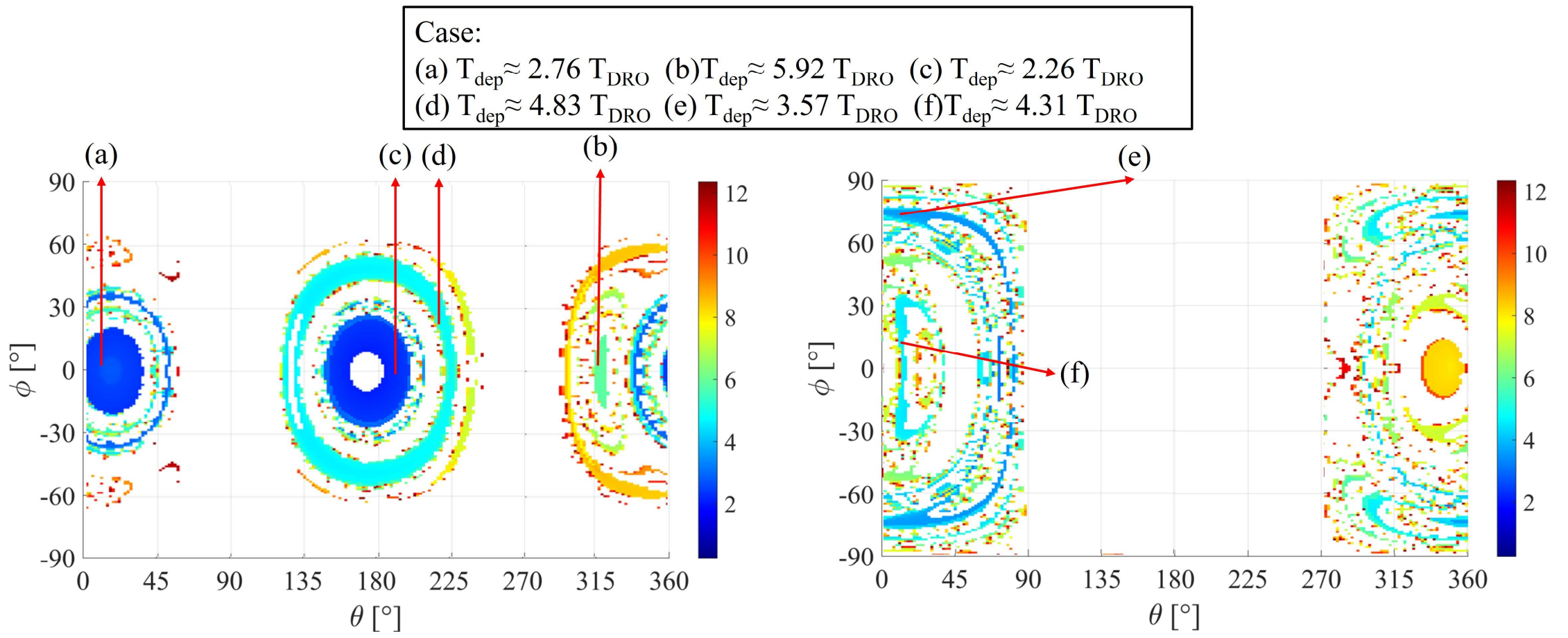

(*a*) 2:1 DRO (*b*) 3:1 DRO

**Fig. 9 The time-to-depart map with $\alpha_{S0} = 0$ for (a) 2:1 DRO and (b) 3:1 DRO.**

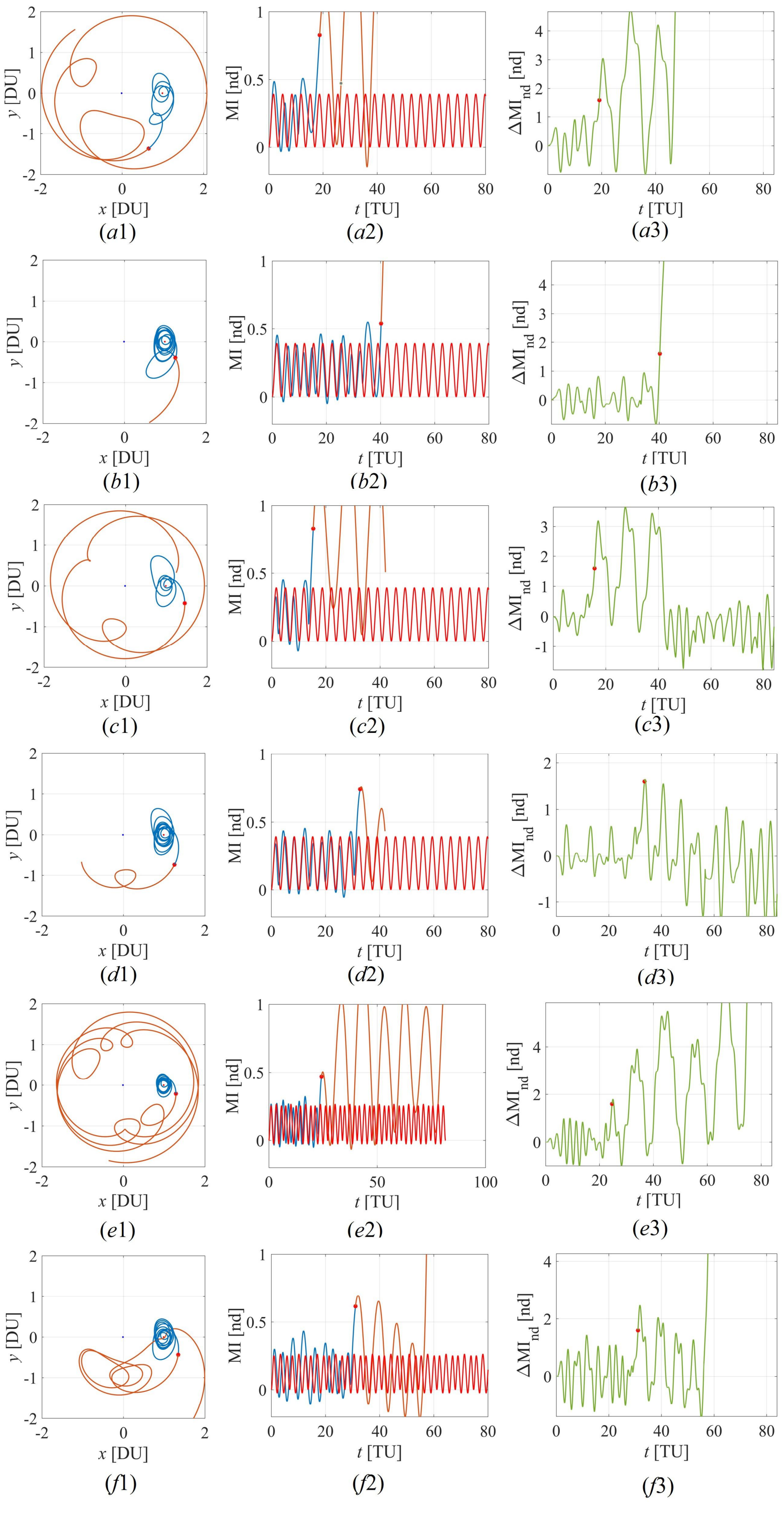

**Fig. 10 The orbital motion and MI evolution in six cases selected in Fig. 9.**

**Table 3 Six cases selected in Fig. 9.**

| Case | $\theta$ [°] | $\phi$ [°] | Time-to-depart [$T_{DRO}$] |
|---|---|---|---|
| (a) | 13.12 | 1.36 | 2.7602 |
| (b) | 318.84 | 0.45 | 5.9193 |
| (c) | 190.40 | 0.45 | 2.2580 |
| (d) | 219.35 | 23.07 | 4.8286 |
| (e) | 9.49 | 75.53 | 3.5732 |
| (f) | 9.49 | 12.85 | 4.3079 |

The time-to-depart is also highly sensitive to the maneuver magnitude. Fig. 11 shows the time-to-depart maps for smaller disposal magnitudes at the same initial solar phase for 2:1 DRO. It is evident that reducing the maneuver magnitude substantially increases the difficulty of successful departure. These results consistently demonstrate that departing from a 2:1 DRO is more challenging than from 9:2 NRHO, primarily due to the greater orbital stability of the former.

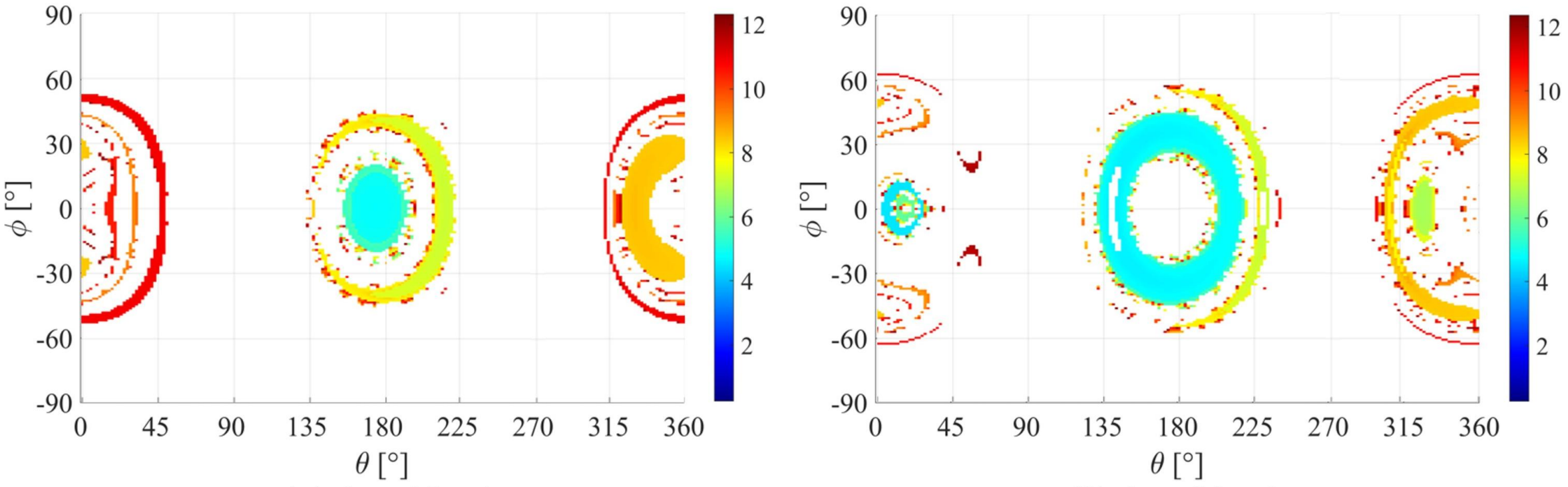

(*a*) $\Delta v = 15$ m/s (*b*) $\Delta v = 20$ m/s

**Fig. 11 The time-to-depart map for 2:1 DRO** $\alpha_{S0} = 0$ **with (a)** $\Delta v = 15$ $m/s$ **and (b)** $\Delta v = 20$ $m/s$**.**

The influence of the initial solar phase for 2:1 and 3:1 DROs is explored for values ranging from 30° to 90° with the step of 30°, as shown in Fig. 12. The time-to-depart displays dramatic variation with solar phase. For instance, when the initial solar phase is 30° for 2:1 DRO or 60° for 3:1 DRO, departure within 3.5 $T_{\mathrm{DRO}}$ is nearly impossible. These results stress that the departure process from the DRO depends on a complex interplay of parameters, and that the selection of disposal solar phase is a critical consideration.

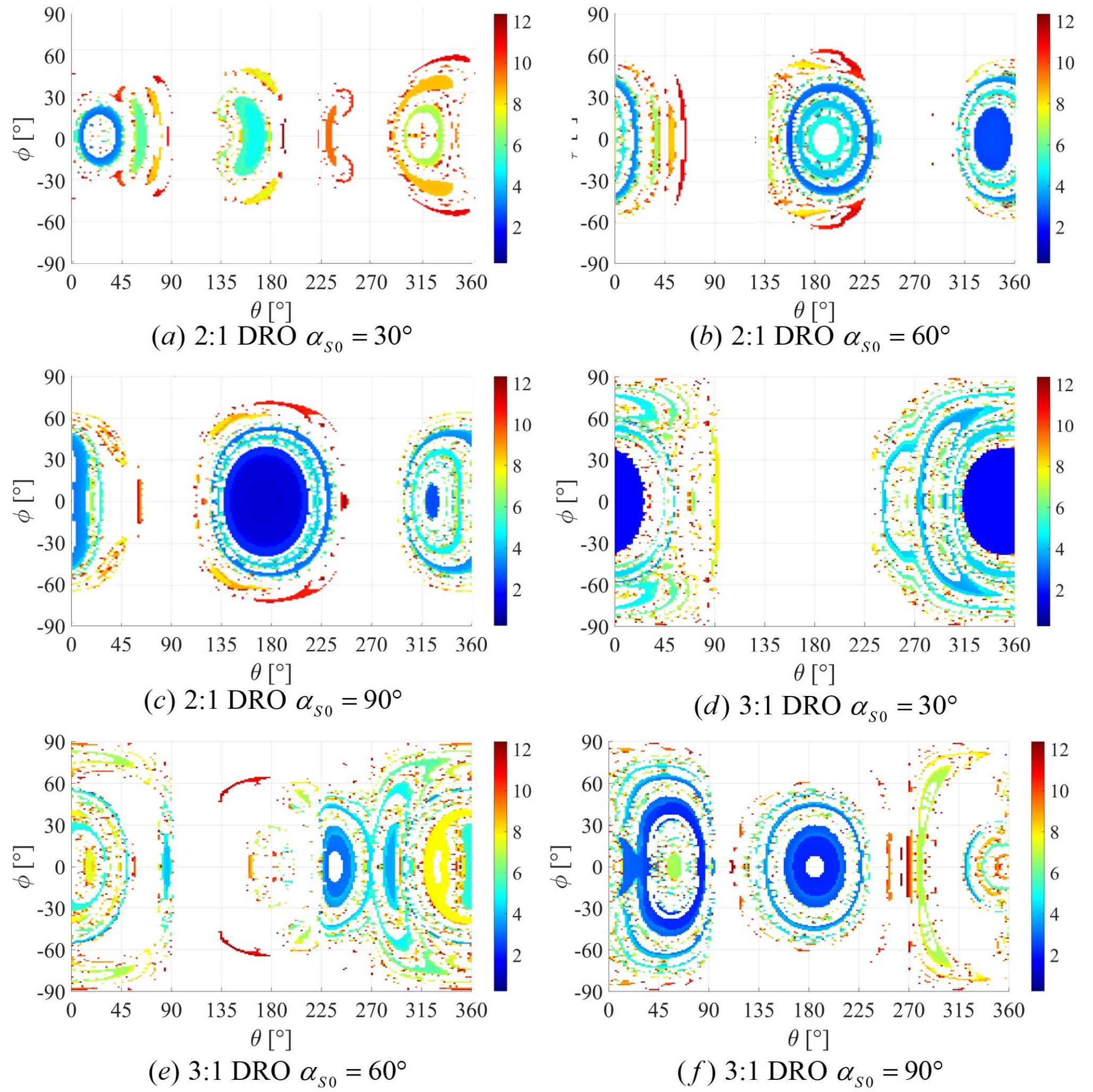

($a$) 2:1 DRO $\alpha_{S0} = 30°$ ($b$) 2:1 DRO $\alpha_{S0} = 60°$

($c$) 2:1 DRO $\alpha_{S0} = 90°$ ($d$) 3:1 DRO $\alpha_{S0} = 30°$

($e$) 3:1 DRO $\alpha_{S0} = 60°$ ($f$) 3:1 DRO $\alpha_{S0} = 90°$

**Fig. 12 The time-to-depart map for 2:1 and 3:1 DROs w.r.t. initial solar phases and $\Delta v = 25\ m/s$.**

A comparative analysis is conducted among the maximum stretching direction, the velocity (tangential) direction, and the disposal direction associated with fast departure. Here, fast departure refers to cases where the time-to-depart is less than 3.5 $T_{\mathrm{DRO}}$, evaluated across various initial solar phases. The maximum stretching direction for velocity perturbation is calculated based on prior research [39,40], which is determined by computing the submatrix $\Phi_{rv,v}$ of state transition matrix (STM) $\Phi_{rv,rv}$ over 1.5 orbital periods originating from each disposal position along the DRO. The singular value decomposition (SVD)of the submatrix $\Phi_{rv,v}$ is written as

$$U\Sigma V^T = \Phi_{rv,v}(t_0,\ t_0 + 1.5T_{\mathrm{DRO}}), \tag{23}$$

where $U$ and $V$ are the orthonormal matrices containing the final and initial principal stretching directions, respectively; and the quantity $\Sigma$ is a diagonal matrix with $\sigma_1 \geq \sigma_2 \geq ... \geq \sigma_n$, where $\sigma_i = \Sigma_{ii}$ $i$-th diagonal element. The matrix $V$ therefore specifies the principal stretching directions at the initial epoch along the phase flow. The maximum stretching direction is thus derived as the first column vector of $V$. Fig. 13 visualizes the relationship: red dots indicate the angular difference between the maximum stretching direction and the velocity direction along the positions in DRO, while blue dots correspond to the disposal directions that achieve fast departure for different initial solar phases. All results in Fig. 13 are obtained with $\Delta v = 25\ m/s$. For both 2:1 and 3:1 DROs, the maximum stretching direction is almost coinciding with the velocity direction. Besides, although the fast departure direction sometimes aligns closely with the maximum stretching direction for certain initial solar phases, there are also some phases, where the maximum stretching direction is far from fast departure direction. For 2:1 DRO, the initial solar phase at 150º results in none fast departure; while for 3:1 DRO, the initial solar phase at 0º and 60º results in none fast departure. By contrast, in the CRTBP, the evolution of the fast departure direction regularly matches the maximum stretching direction throughout the positions along DRO. These findings highlight the increased complexity of departure dynamics from DRO in the BCRFBP: the most effective disposal direction is not always along the velocity direction, and in some cases, the initial solar phase can significantly hinder efficient departure from the DRO. Therefore, detailed consideration is needed for the initial solar phase of disposal.

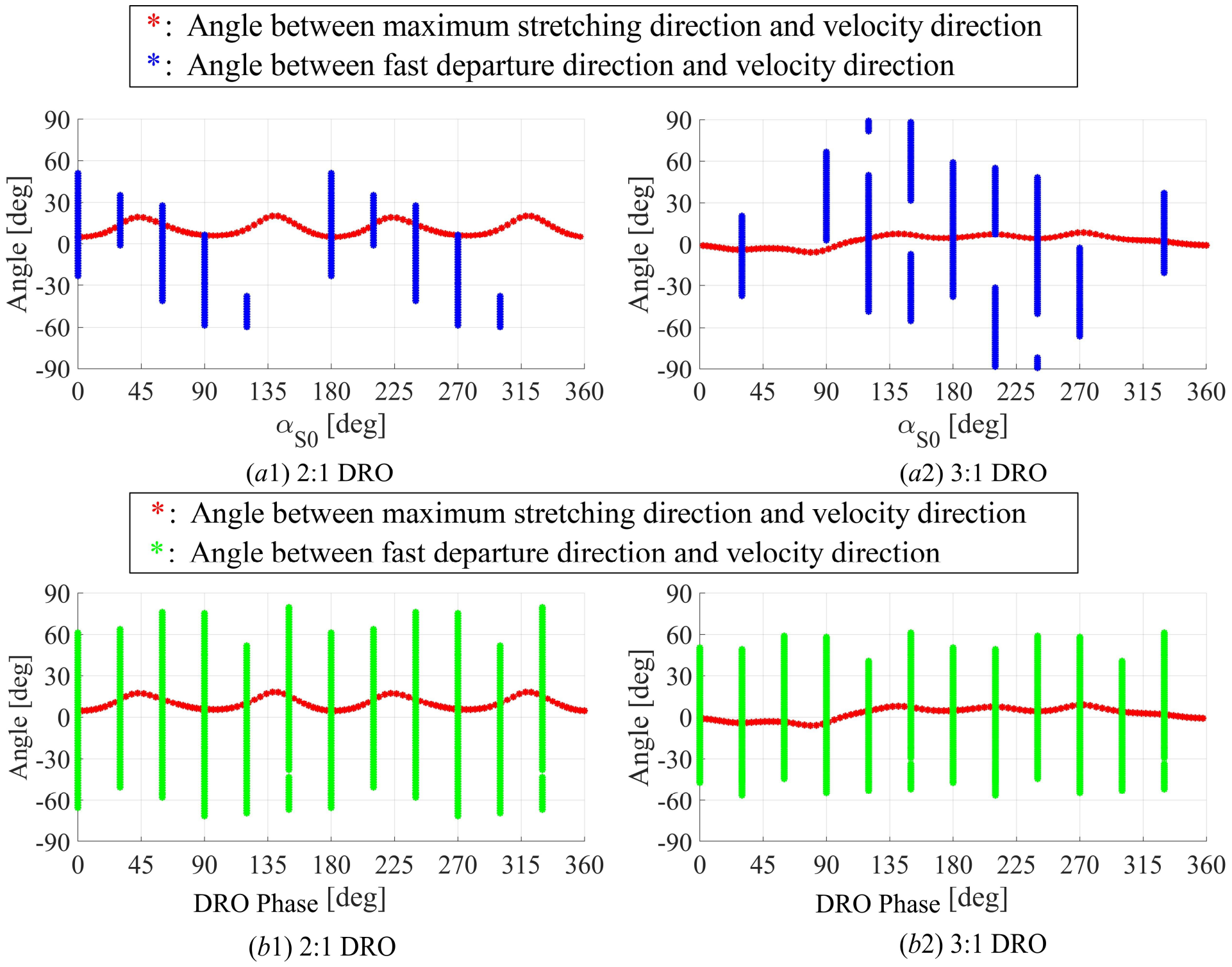

**Fig. 13 The difference between fast departure direction, the maximum stretching direction, and the velocity direction in (a) BCRFBP and (b) CRTBP.**

### 3.4 Statistical Analysis of The Threshold of Departure Metric

Because K is a heuristic threshold, a sensitivity analysis to evaluate its effect on departure detection is performed. At present, no unique definition of a departure event has been established for restricted three- and four-body orbital dynamics. An operational definition is therefore adopted:

1. At epoch $t_d$, the minimum distance between the object and any point on the nominal periodic orbit exceeds 1 DU;

2. During the remaining propagation interval $[t_d, t_f]$, the distance between the object and the Earth–Moon barycenter exceeds 1 DU for more than 50% of the time.

The first condition identifies departure from the nominal periodic orbit, whereas the second requirement is to avoid entering the interior cislunar region as much as possible. Three types of samples are considered: true positive (TP), false positives (FP), and false negatives (FN). A TP corresponds to a true departure that is correctly identified by $\Delta\mathrm{MI}_{\mathrm{nd}}$. An FP corresponds to a nondeparture incorrectly identified as a departure, whereas an FN corresponds to a true departure that is missed by $\Delta\mathrm{MI}_{\mathrm{nd}}$. The detection performance is quantified using recall and precision:

$$\mathrm{recall}=\mathrm{Num}(\mathrm{TP})\big/\left[\mathrm{Num}(\mathrm{TP})+\mathrm{Num}(\mathrm{FN})\right],\ \mathrm{precision}=\mathrm{Num}(\mathrm{TP})\big/\left[\mathrm{Num}(\mathrm{TP})+\mathrm{Num}(\mathrm{FP})\right], \tag{24}$$

where Num( ) represents the sample counting function. Values of recall and precision closer to 100% indicate more reliable departure identification.

The results of 2:1 and 3:1 DROs with $\Delta v = 25\ m/s$ over $\alpha_{S0} \in [0, 360°]$ are shown in Fig. 14. The step size of $\alpha_{S0}$ is 30º. The tested values of K were 1.4, 1.6, 1.8, and 2.0. The upper panels in Fig. 14 show the statistical analysis for the 2:1 DRO, whereas the lower panels correspond to the 3:1 DRO. As the K value decreases, more nondeparture cases are classified as departures since the relaxation of the criterion for Eq. (22). The increase of the number of FP samples also reduces the precision. Conversely, as K increases, more true departure cases are missed, increasing the number of FN samples and reducing the recall. For both DROs, K=1.6 yields recall and precision values above 0.92. Although K=1.8 also provides good overall performance, its recall is slightly lower than that obtained with K=1.6. Therefore, the threshold K can be selected for different periodic orbits through statistical sensitivity analysis to achieve an appropriate balance between recall and precision.

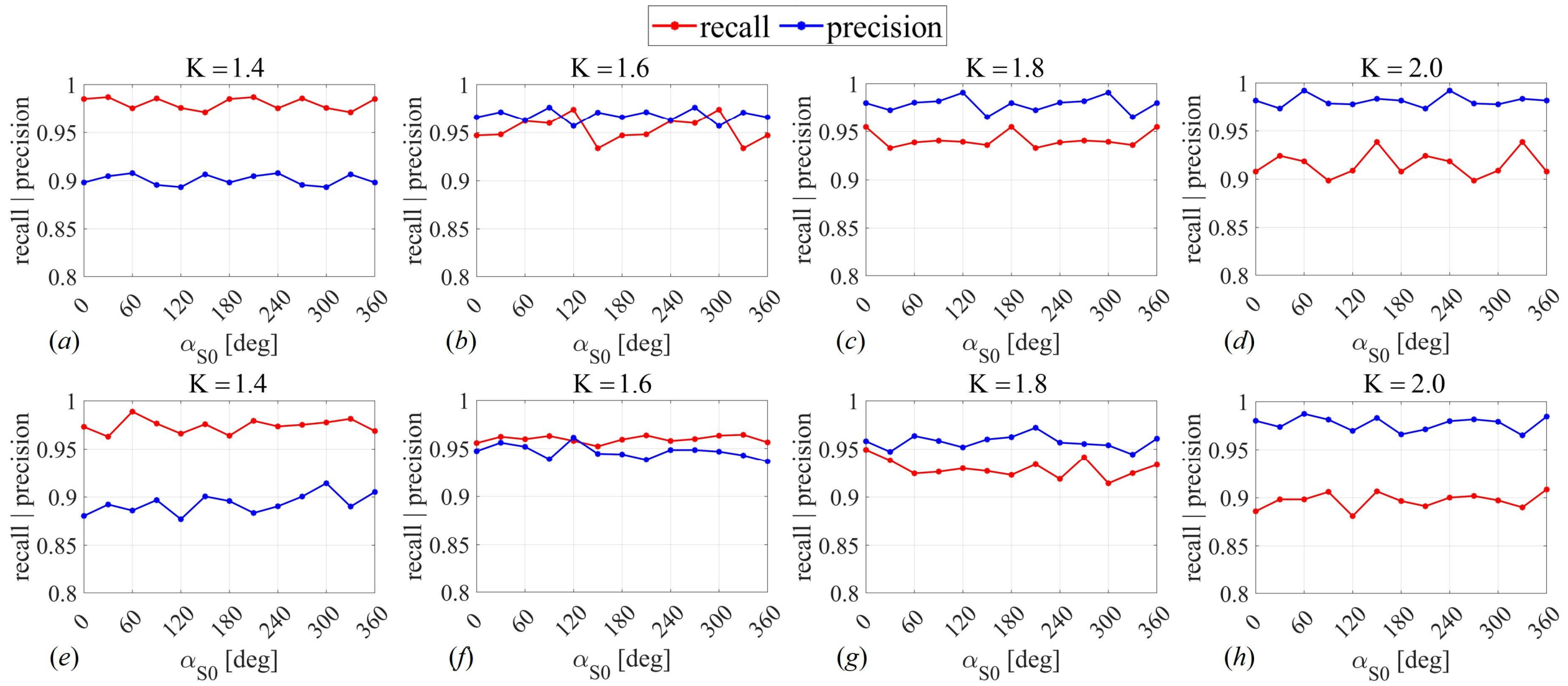

**Fig. 14 Statistical analysis of the selection of threshold K.**

## 4. Escape Dynamics from DROs

The previous section focused on the first phase of disposal dynamics, i.e., departure from the periodic orbit. The second, equally critical phase is cislunar escape dynamics, which ultimately determines the effectiveness of the disposal design. This section presents a detailed investigation of escape trajectories originating from the 2:1 and 3:1 DROs, emphasizing the dynamical structures revealed by Lagrangian Descriptor (LD) fields. To deepen understanding, the energy evolution of cislunar escape trajectories is analyzed. By combining LD field analysis with energy evolution, a more comprehensive interpretation of escape dynamics is achieved. Furthermore, the LD field enables precise classification of escaping orbital families from the DRO. The large-area and continuous region in the LD field can be selected as favorable disposal conditions. Finally, the influence of disposal maneuver magnitude and initial solar phase on escape outcomes is systematically compared.

### 4.1 The Energy Evolution of Cislunar Escape Trajectory

This Section presents a semi-analytical investigation of the energy evolution patterns underlying low-energy cislunar escape. The influence of SGP on the mechanical energy and $z$-component orbital angular momentum is examined, with a focus on mathematically elucidating the critical role of SGP in facilitating cislunar escape.

Before proceeding with the derivation, several points should be emphasized:

1. All studied escape trajectories originate from disposal maneuvers on the DRO.

2. The analysis focuses on the phase after the departure from DRO, and before the spacecraft approaches the cislunar escape boundary (or final integration epoch).

3. The forthcoming energy evolution analysis is applicable to both spatial and planar orbital dynamics.

As discussed in Section 2.1, the BCRFBP does not admit any true integral of motion. Nonetheless, the relationship among instantaneous Jacobi energy *C*, mechanical energy *E*, and z-component angular momentum $H_z$ in Eq. (10) remains valid. Firstly, the partial derivative of $\Omega_S$ w.r.t. time can be derived as [41]

$$\begin{aligned}\frac{\partial\Omega_S}{\partial t} &= \frac{-m_S\rho\omega_S(x\sin\alpha_S - y\cos\alpha_S)}{\left[(x-\rho\cos\alpha_S)^2+(y-\rho\sin\alpha_S)^2+z^2\right]^{3/2}} + \frac{m_S\omega_S}{\rho^2}(x\sin\alpha_S - y\cos\alpha_S)\\ &\approx \frac{-3m_S\omega_S r^2\sin 2\eta}{2\rho^3}.\end{aligned} \tag{25}$$

Eq. (25) is simplified via Taylor expansion of the denominator terms. Here, $\boldsymbol{r}=[x,\ y,\ z]^T$ and $\boldsymbol{\rho}=\rho[\cos\theta_S,\ \sin\theta_S,\ 0]^T$. The variable $\eta$ represents the angle between the spacecraft's planar position vector (with respect to the EMB) and the Sun's orientation vector, measured counterclockwise.

Integrating the derivative of instantaneous mechanical energy *E* from the initial epoch $t_i$ to the final epoch $t_f$, the net increase in mechanical energy can be obtained as

$$\begin{aligned}\Delta E &= \int_{t_i}^{t_f}\frac{dH_z}{dt}\cdot dt + \int_{t_i}^{t_f}\frac{d\Omega_S}{dt}\cdot dt - \int_{t_i}^{t_f}\frac{\partial\Omega_S}{\partial t}\cdot dt\\ &\approx \int_{t_i}^{t_f}(1-\mu)\mu y\left(\frac{1}{r_1(t)^3}-\frac{1}{r_2(t)^3}\right)dt + \int_{t_i}^{t_f}\frac{(1+\omega_S)3m_S r(t)^2\sin 2\eta(t)}{2\rho^3}dt + \int_{t_i}^{t_f}\frac{d\Omega_S}{dt}dt\\ &= \Delta E_{EM} + \Delta E_{S1} + \Delta E_{S2}.\end{aligned} \tag{26}$$

According to Eq. (26), the first term $\Delta E_{EM}$ captures the gravitational influences of the Earth and Moon, while the second and third terms are both due to SGP. More precisely, the second term $\Delta E_{S1}$ in $\Delta E$ reflects the variation of trajectory-dependent SGP energy, determined by the path of the spacecraft. In the EMB-referenced S-B1 frame, $\Delta E_{S1} > 0$ when the spacecraft mainly transits the second or fourth quadrants, while $\Delta E_{S1} < 0$ applies for the first or third quadrants. The third term $\Delta E_{S2}$ in $\Delta E$ can be reformulated as

$$\Delta E_{S2} = E_{S2}(t_f) - E_{S2}(t_i) = \int_{t_i}^{t_f}\frac{d\Omega_S}{dt}dt \approx \frac{m_S}{2\rho^3}\left[2x_{B1}(t)^2 - y_{B1}(t)^2 - z_{B1}(t)^2\right]\Big|_{t_i}^{t_f}. \tag{27}$$

The subscript "B1" in Eq. (27) indicates that the spacecraft's state is expressed in the S-B1 rotating frame. Thus, the term $\Delta E_{S2}$ is the variation of state-dependent SGP energy, which depends solely on the initial and terminal positions in this frame. During the escape process, the gravitational influences of the Earth and Moon on mechanical energy can be neglected, i.e., the value of $\left|\Delta E_{\mathrm{EM}}\right|$ is relatively small, as supported by Fig. 3. Therefore, $\Delta E_{S1}$ and $\Delta E_{S2}$ dominate the energy evolution.

Additionally, the evolution of angular momentum $H$ can also be derived according to Eq. (26). The three components of angular momentum are given by

$$\begin{cases} H_x = y\dot{z} - z(\dot{y} + x) \\ H_y = z(\dot{x} - y) - x\dot{z} \\ H_z = x(\dot{y} + x) - y(\dot{x} - y) \end{cases} . \tag{28}$$

Given that the post-disposal trajectories from the 2:1 DRO generally have small z-components, variations in $H$ are predominantly determined by changes in $H_z$. The variation of $H_z$ during the cislunar escape can be obtained as

$$\Delta H_z \approx \int_{t_i}^{t_f} (1-\mu)\mu y \left( \frac{1}{r_1(t)^3} - \frac{1}{r_2(t)^3} \right) dt + \int_{t_i}^{t_f} \frac{3 m_S r(t)^2 \sin 2\eta(t)}{2\rho^3} dt = \Delta H_{z-\mathrm{EM}} + \Delta H_{z-S1}. \tag{29}$$

Since trajectories typically exceed the escape boundary through the second or fourth quadrants of the S-B1 frame, $\Delta H_z > 0$ is valid. This analytical result accounts for the observation in Section 2.3 that the angular momentum of escaping orbits is significantly larger than that of non-escaping trajectories.

The analytical expression of $\Delta E_{S2}$ in Eq. (27) reveals the regional characteristics of the state-dependent SGP energy in the EMB-referenced S-B1 rotating frame. We define the state-dependent SGP energy as

$$\mathcal{U}(x_{B1}, y_{B1}, z_{B1}) = \frac{m_S}{2\rho^3}(2x_{B1}{}^2 - y_{B1}{}^2 - z_{B1}{}^2) \approx \frac{m_S}{2\rho^3}(2x_{B1}{}^2 - y_{B1}{}^2). \tag{30}$$

This approximation is justified by the fact that the *z*-axis component of the spacecraft trajectory, measured in S-B1 units, is within the order of magnitude of $10^{-3}$, and thus can be neglected in regional analyses, as depicted in Fig. 15b. Consequently, the right side of Eq. (30) describes a hyperbolic paraboloid in the $x_{B1} - y_{B1}$ plane. The condition for zero state-dependent SGP energy corresponds to

$$y_{B1} = \sqrt{2} x_{B1} \ \text{ or } \ y_{B1} = -\sqrt{2} x_{B1}. \tag{31}$$

That is, if both the initial and final positions of the spacecraft satisfy the conditions in (31), $\Delta E_{S2} \approx 0$. In practice, since the initial position is typically close to the DRO, and $\left|E_{S2}(t_i)\right|$ is sufficiently small compared with $\left|E_{S2}(t_f)\right|$.

Eq. (30) divides the $x_{B1} - y_{B1}$ plane into four regions, as shown in Fig. 15a. The light blue region indicates where state-dependent SGP energy decreases, while the white region indicates an increase. Intuitively, if $\Delta E$ is sufficiently large, particularly if it ensures $E_f = E_i + \Delta E > 0$ at the final epoch, then the trajectory becomes hyperbolic w.r.t. the EMB, which naturally results in cislunar escape. If the escape boundary is crossed within the region where $\mathcal{U}(x_{B1}, y_{B1}) < 0$ region (i.e., $\Delta E_{S2} < 0$), the Sun's perturbation tends to do negative work, opposing escape. Besides, if the spacecraft's trajectory exceeds the escape boundary and reaches apoapsis in the $\mathcal{U}(x_{B1}, y_{B1}) < 0$ region, the SGP will also continue to suppress cislunar escape. The apoapsides of the quick-return trajectories in Fig. 4b both locate at the $\mathcal{U}(x_{B1}, y_{B1}) < 0$ region and exceed the escape boundary, leading to unsuccessful escape. Conversely, if the spacecraft exits cislunar space through the region with $\mathcal{U}(x_{B1}, y_{B1}) > 0$ (i.e., $\Delta E_{S2} > 0$), the SGP contributes positive work, facilitating escape.

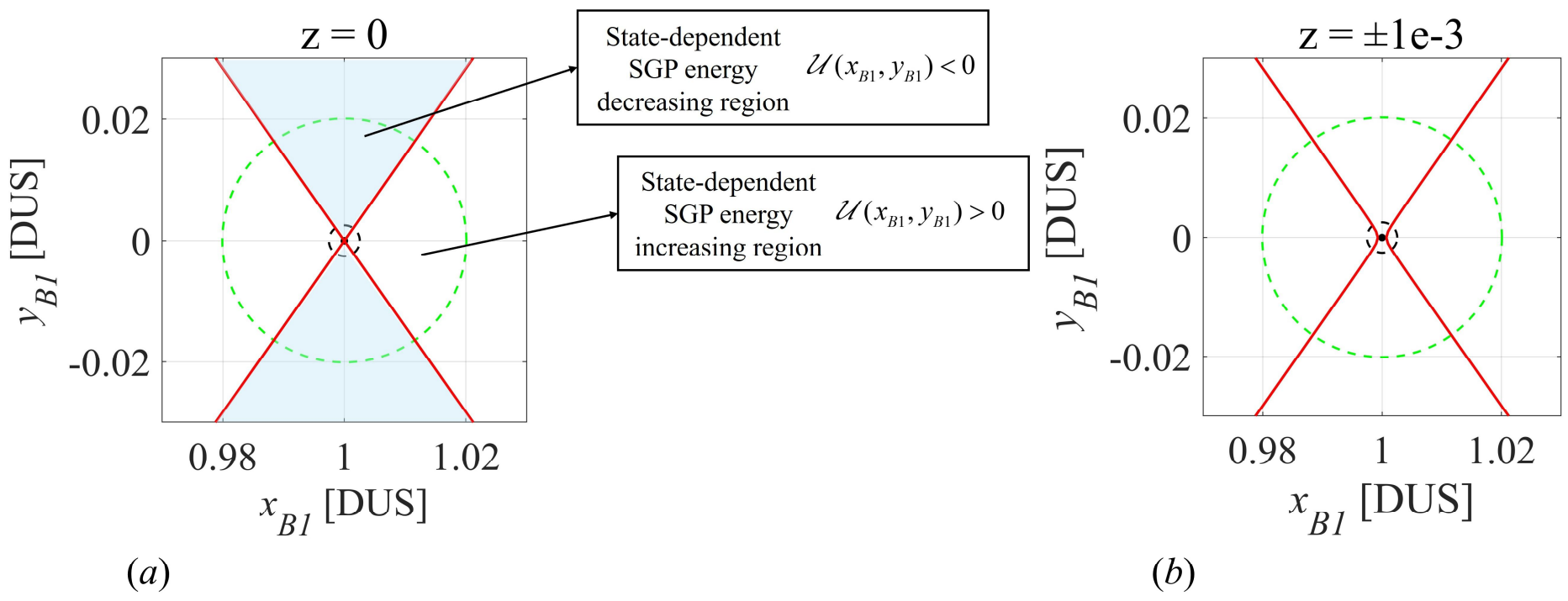

**Fig. 15 The geometry of state-dependent SGP energy.**

In the evolution of cislunar escape, accurate mechanical energy variations are caused by the combination of $\Delta E_{EM}$, $\Delta E_{S1}$, and $\Delta E_{S2}$, and the instantaneous Jacobi energy variations are also related to $\Delta E_{S1}$ and $\Delta E_{S2}$. The next Section will provide a detailed comparison on the combinations of $\Delta E_{EM}$, $\Delta E_{S1}$, and $\Delta E_{S2}$ in escape process. As a preliminary conclusion, the combinations of $\Delta E_{S1}$ and $\Delta E_{S2}$ w.r.t. geometry of SGP field are listed in Table 4.

**Table 4 The combinations of different terms in $\Delta E$ w.r.t. the geometry of SGP field.**

| Change of spacecraft's energy | $< 0$ | $> 0$ | $= 0$ ($\approx 0$) |
|---|---|---|---|
| $\Delta E_{S1}$ | Mainly moves in the first or third quadrants | Mainly moves in the second or fourth quadrants | Crosses different quadrants, ultimately resulting in no variation of trajectory-dependent SGP energy |
| $\Delta E_{S2}$ | Final position locates in the state-dependent SGP energy decreasing region | Final position locates in the state-dependent SGP energy increasing region | Initial and final positions both locate in the zero state-dependent SGP energy boundary |

### 4.2 The Lagrangian Descriptor Field

This Section examines cislunar escape dynamics from the 2:1 and 3:1 DROs, focusing on the low-energy escape phase space. While previous studies have commonly employed the finite-time Lyapunov exponent (FTLE) as a dynamical indicator [20,21], this work utilizes the Lagrangian Descriptor (LD), which offers reduced computational cost and more detailed information compared to FTLE. For the calculation of LD fields, the integration interval $t_f$ is set to one year. For the sake of simplicity, two LD fields generated from 2:1 DRO and one LD field from 3:1 DRO are shown in this Section. The LD fields are also constructed with two direction angles $\theta$ and $\phi$, which can be compared with the time-to-depart maps. Besides, the energy evolution patterns will be introduced to further interpret the cislunar escape behaviors. The 'RelTol' and 'AbsTol' in the propagation of dynamical equations are set as $10^{-11}$.

Consistent with the parameters in Fig. 9 of Section 3.3, the LD field with a disposal maneuver of $\Delta v = 25$ m/s from 2:1 DRO with $\alpha_{S0} = 0$ is shown in Fig. 16. Within the evaluated LD field, initial conditions that generate dynamically similar trajectories cluster into distinct regions. The LD values experience sharp gradients across the boundaries of these similar regions. Consequently, the transverse derivatives of the LD field with sharp gradients become discontinuous. In the context of cislunar escape dynamics, these abrupt variations boundary in the LD field effectively delineate the geometric projections of low-energy escape manifolds. The region inside the low-energy escape manifolds will be interpreted in the following.

The LD field in Fig. 16 can be divided into four main regions. Region (a) exhibits higher LD values and corresponds to faster escape, aligning with the outer parts of region (a) in Fig. 9. The cyan region (b) forms a strip with longer escape times than region (a), and similarly corresponds with the outer area of region (b) in Fig. 9. The light blue region (c) is associated with even longer escape time. The central dark blue area of region (c) is due to lunar collisions and is not further analyzed. Finally, the majority of the LD field falls within the blue region (d), indicating trajectories that remain bound within cislunar space for at least one year. Representative orbital evolutions for each region are illustrated in Fig. 17. The left and right subpanels in each panel of Fig. 17 represent the orbital evolution in Earth-Moon rotating frame and EMB-centered S-B1 rotating frame, respectively. Both regions (a) and (b) feature escape via LGA, with the number of revolutions surrounding the Moon inversely correlating with LD value; more revolutions yield lower LD. Region (c) is characterized by a pronounced apogee increase, analogous to WSB transfers, where the apogee location in the fourth quadrant in S-B1 frame enhances orbital energy. Fig. 18 displays the evolution of mechanical and Jacobi energy for cases (a) and (d) after applying the disposal maneuver. The orange trajectory in panel (a) marks the phase between the departure from DRO and the escape boundary crossing. Notably, region (d) trajectories remain within cislunar space, maintaining negative mechanical energy throughout. In contrast, region (a) leverages favorable SGP to rapidly increase orbital mechanical energy and achieve escape.

In summary, analysis of the LD field in Fig. 16 demonstrates that higher LD values correspond to faster escape, and most successful escape results are achieved via LGA. At the unit scale considered in this study, “LD value > 4.5” is treated as an indicator of successful escape.

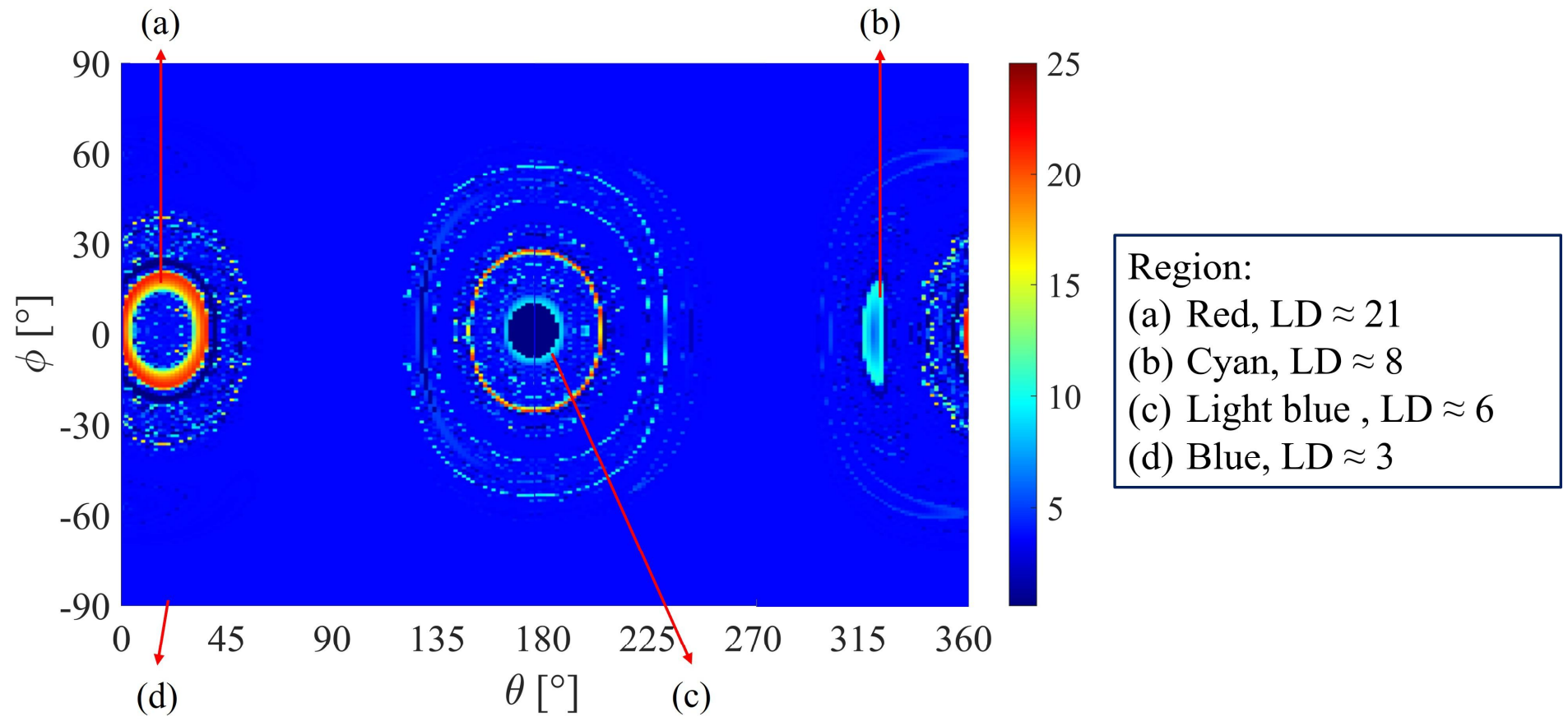

**Fig. 16 The LD field with $\alpha_{S0} = 0$ and $\Delta v = 25\ m/s$ for 2:1 DRO.**

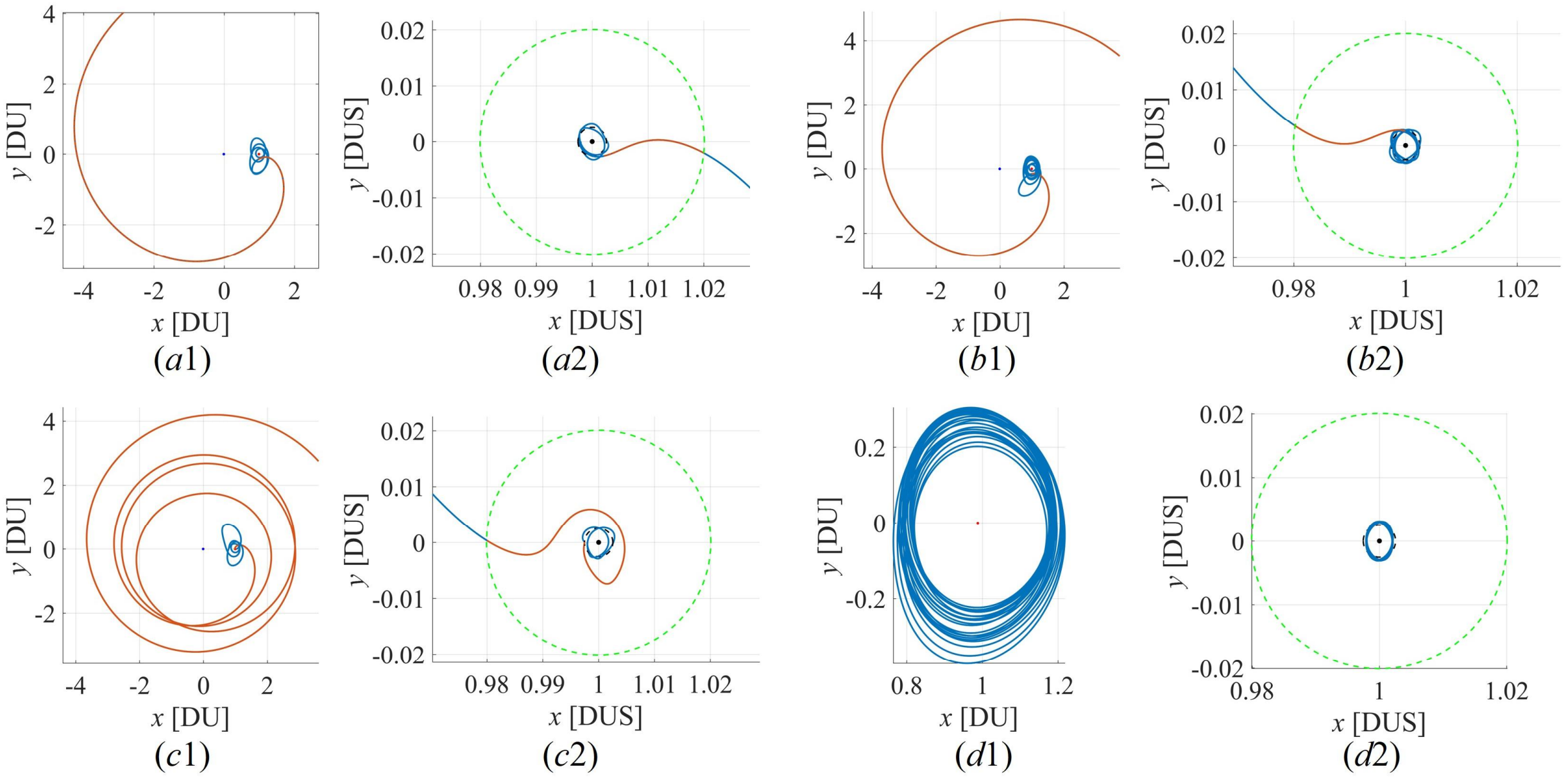

**Fig. 17 Four representative cases of cislunar escape evolution in Fig. 16.**

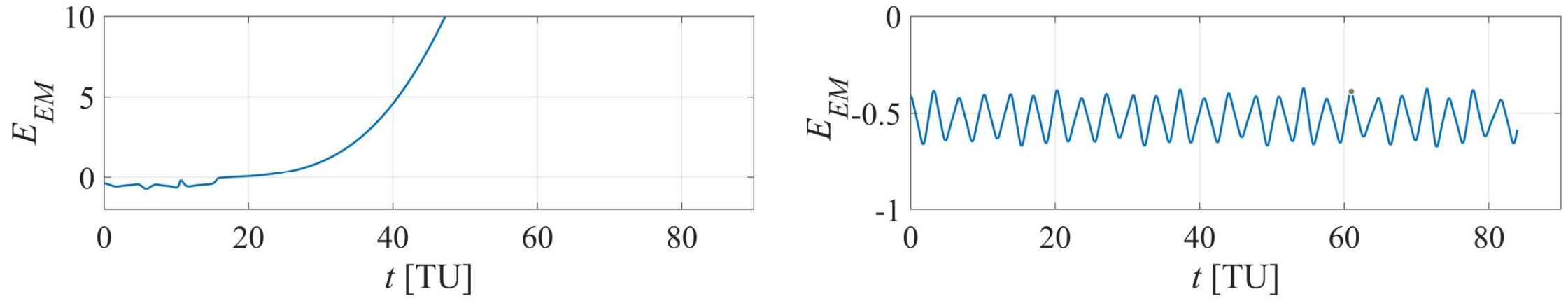

(*a*) Region (a) (*b*) Region (d)

**Fig. 18 The evolution of mechanical energy and Jacobi energy in cases (a) and (d) of Fig. 16.**

To distinguish genuine chaotic structures in LD field from integration noise, the integration tolerances should be discussed. 'RelTol' and 'AbsTol' are set as $10^{-10}$ and $10^{-12}$, respectively, for comparison in Fig. 16. The LD values in Fig. 16 are selected as the reference values, and the differences between LD values with tolerance ('RelTol' and 'AbsTol') set as $10^{-10}$ and $10^{-12}$ and the reference values are shown in Fig. 19. It is obvious that most regions have almost zero differences, with only some chaotic points around the annular region in Fig. 16 showing relatively large differences. The chaos points in Fig. 19 are sensitive to numerical integration and will be eliminated through algorithms in the following Sections, focusing only on the continuous-distributed large-area LD regions.

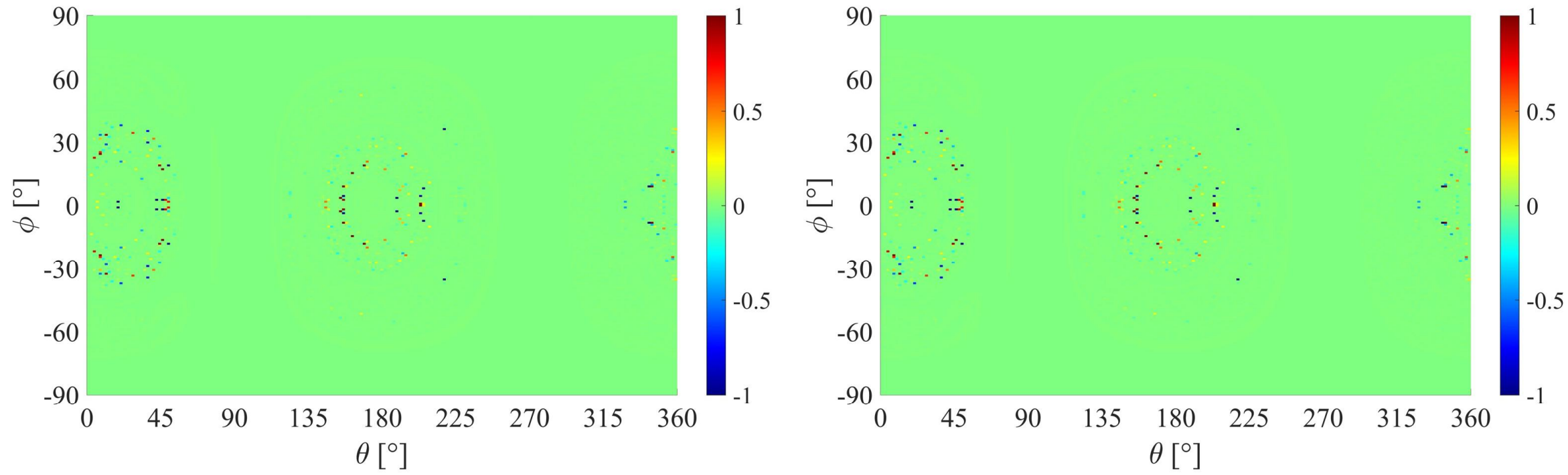

(*a*) Tolerance with $10^{-10}$ (*b*) Tolerance with $10^{-12}$

**Fig. 19 The differences between LD values with different tolerance and the reference LD values.**

The escape possibility with the disposal maneuver magnitude at $\Delta v = 25\ m/s$ is relatively low. Fig. 20 shows the LD field with $\Delta v = 55\ m/s$ and $\alpha_{S0} = 0$. Six distinct regions are selected, with corresponding representative orbital evolutions shown in Fig. 21. The evolution of the three components of mechanical energy before escape boundary crossing for cases (a), (b), (e), and (f) is provided in Fig. 22.

The escape dynamics associated with the LD field in Fig. 20 are notably more complex. In particular, the trajectory in region (a) escapes from the third quadrant of the EMB-referenced S-B1 rotating frame. As confirmed by Fig. 22a, although $\Delta E_{S1} < 0$, the mechanical energy after the LGA is sufficiently positive to enable escape. In contrast, the escape process of region (b) is more successful since $\Delta E_{S1} > 0$ and $\Delta E_{S2} > 0$ result in the large increase of mechanical energy. Region (d) also exhibits features of a WSB transfer, but unlike case (c) in Fig. 17, the apogee is located in the third quadrant of the EMB-referenced S-B1 frame. The mechanical energy decreases as the spacecraft approaches apogee; however, upon re-entering the second quadrant, the SGP gradually increases the energy, eventually enabling escape. For region (f), the trajectory returns to cislunar space after initially reaching the escape boundary. This result corresponds to the state-dependent SGP energy decreasing region described in Section 4.1. The result in Fig. 22d further demonstrates that $\Delta E_{S2} < 0$ during the escape with initial conditions generated from region (f), suppressing successful escape. However, the unfavorable SGP does not lead to long-term bounded motion within the escape boundary, and case (f) ultimately achieved escape from cislunar region. Thus, the LD values in region (f) are still obviously larger than 4.5.

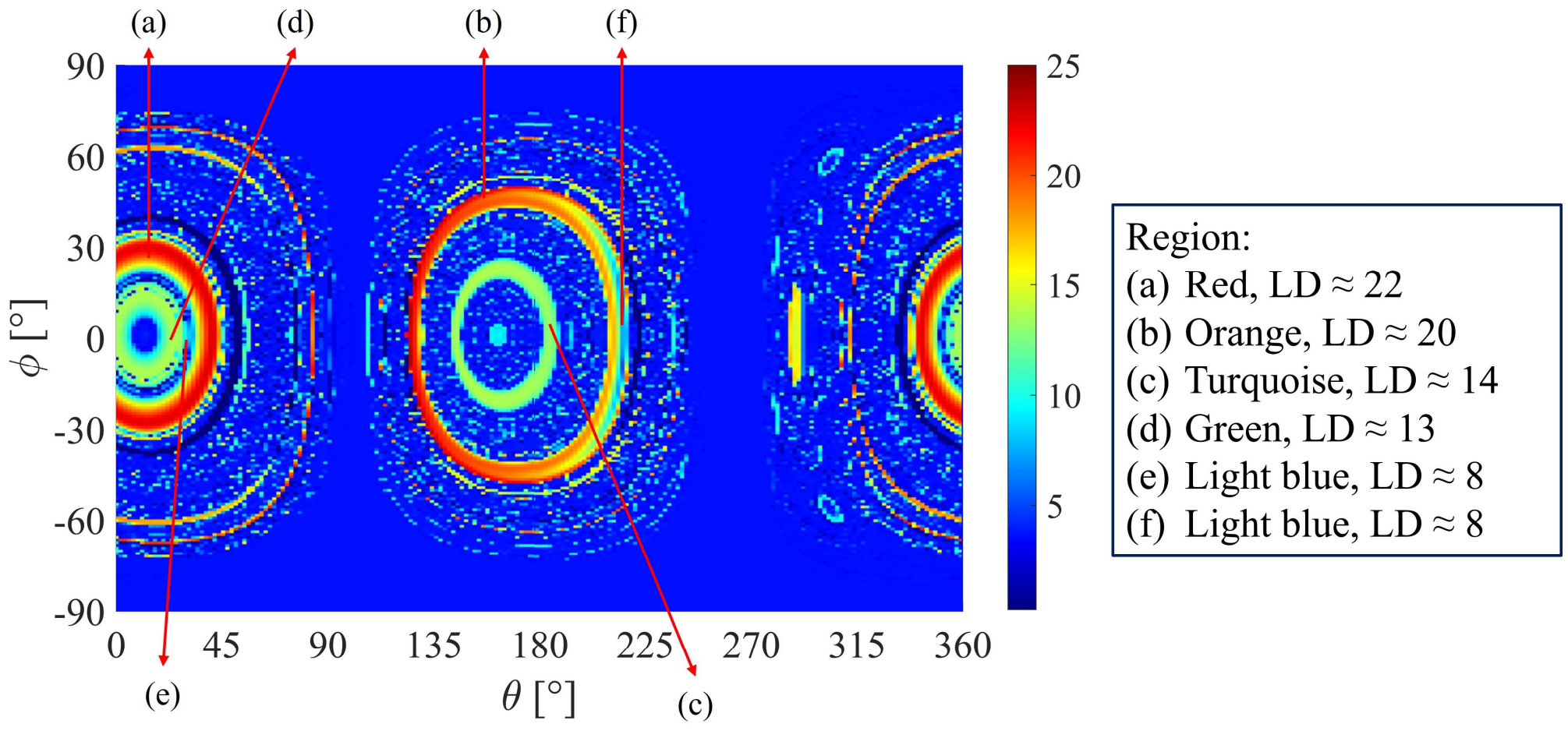

**Fig. 20 The LD field with $\alpha_{S0} = 0$ and $\Delta v = 55\ m/s$ for 2:1 DRO.**

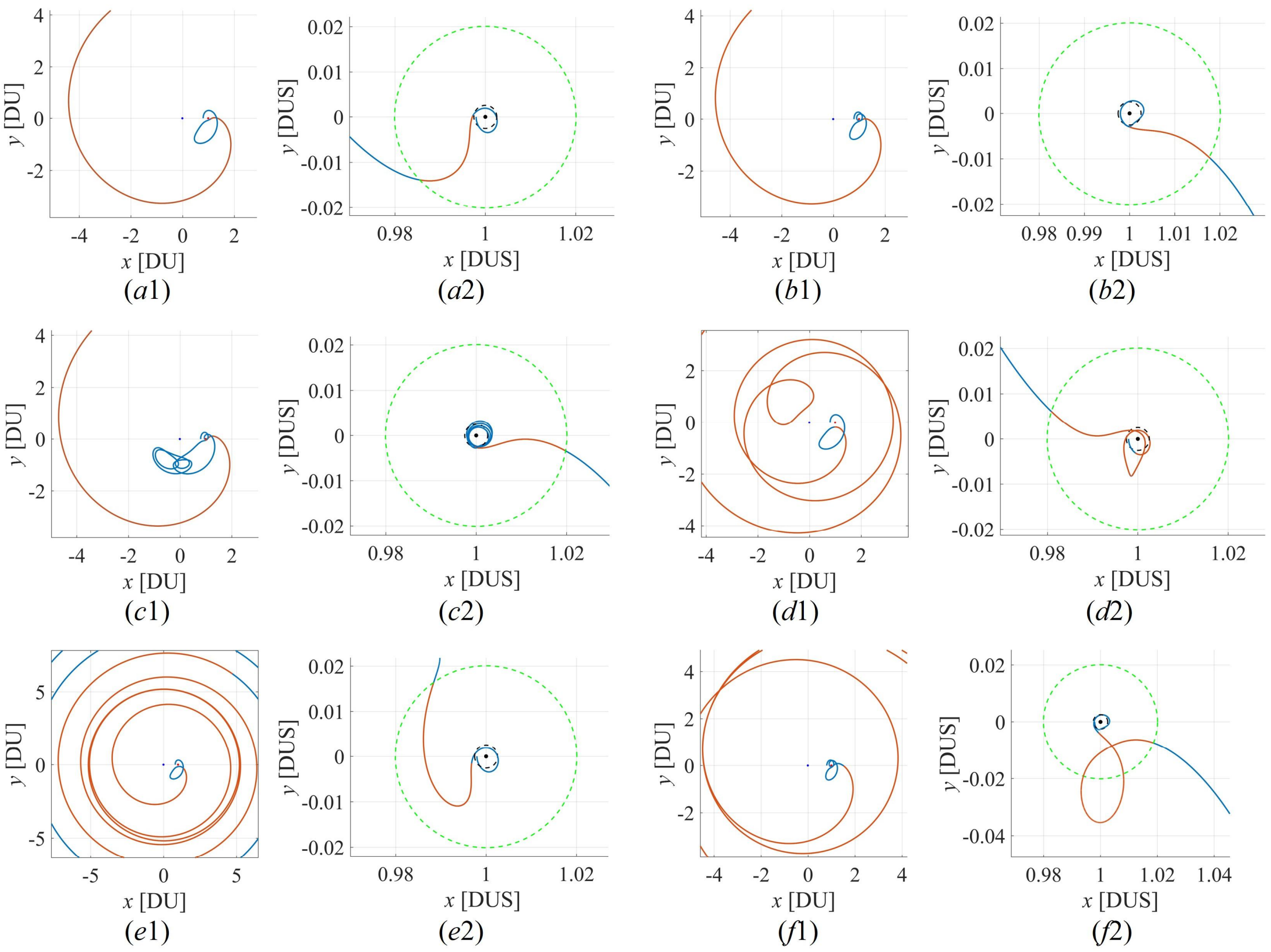

**Fig. 21 Six representative cases of cislunar escape evolution in Fig. 20.**

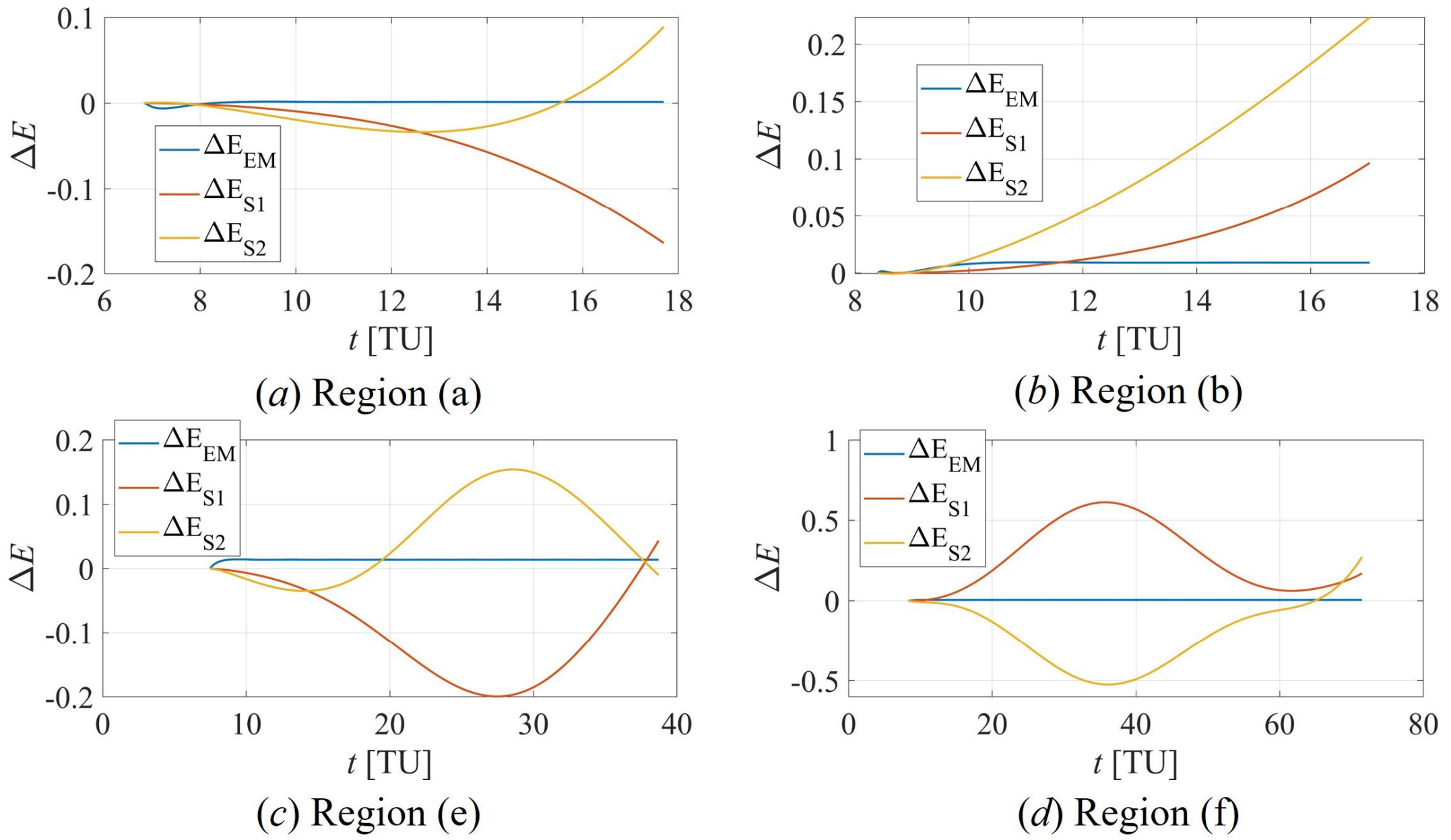

**Fig. 22 The three terms of mechanical energy evolution in cases (a), (b), (e) and (f) of Fig. 20.**

Fig. 23 shows the LD field with $\Delta v = 25\ m/s$ and $\alpha_{S0} = 90°$ from 3:1 DRO. Two distinct regions are selected, with corresponding representative orbital evolutions shown in Fig. 24. Region (a) in Fig. 23 corresponds to the fast escape result; while region (b) in Fig. 23 exhibits the apogee in the state-dependent SGP energy decreasing region, and thus the escape process is suppressed. The regional distribution in Fig. 23 is obviously different from Fig. 16, which indicates that the geometry of the disposal location has a significant impact on the escape conditions. Both for 2:1 and 3:1 DRO, smaller disposal maneuver magnitude results in fewer escape opportunities.

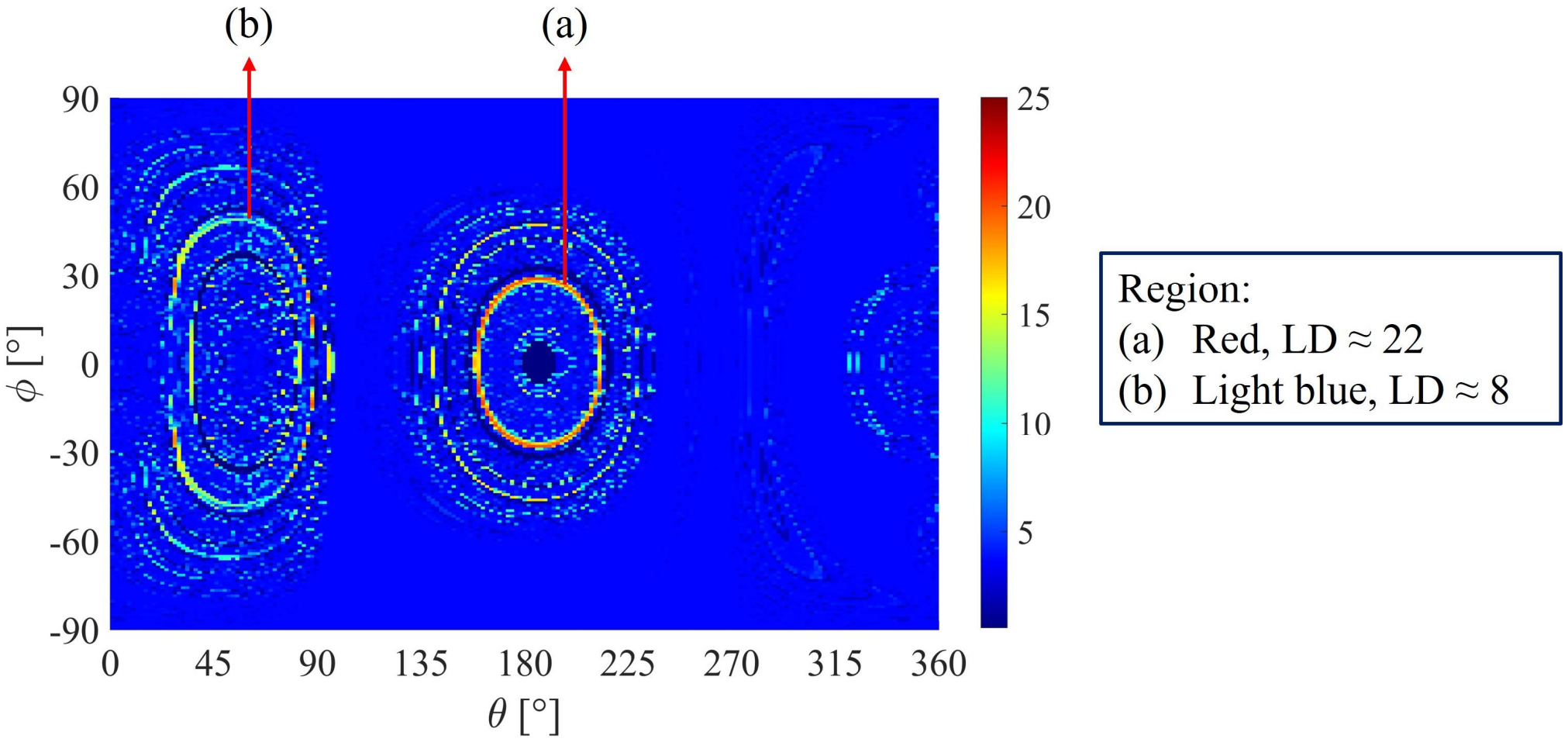

**Fig. 23 The LD field with $\alpha_{S0} = 90°$ and $\Delta v = 25\ m/s$ for 3:1 DRO.**

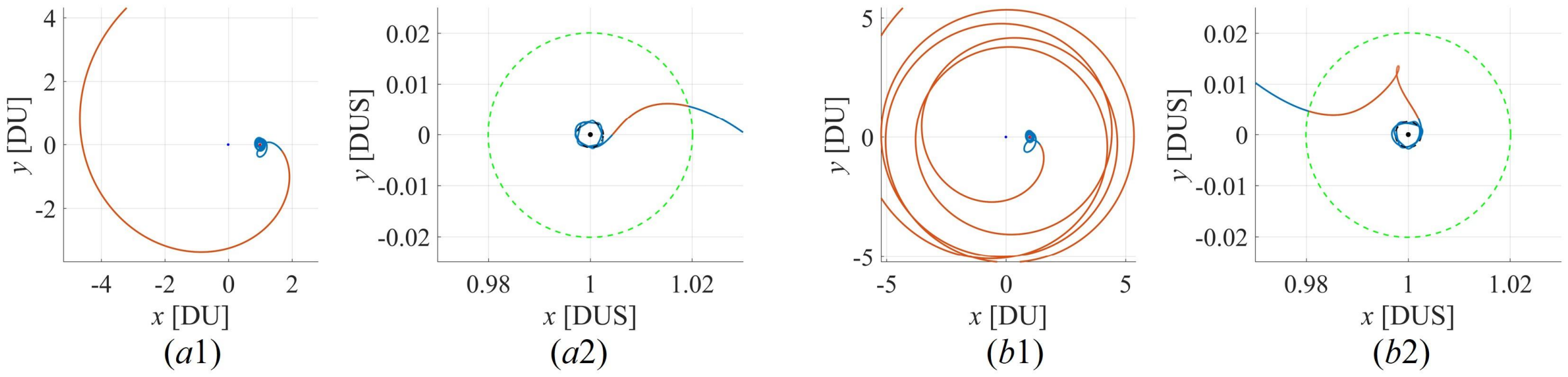

**Fig. 24 Two representative cases of cislunar escape evolution in Fig. 23.**

In summary, initial conditions in the LD field with higher LD values correspond to faster escape, and the overall escape probability increases with larger disposal maneuver magnitude. Additionally, when the trajectory contains at least one apogee passage, the influence of SGP becomes more intricate, and the apogee's location in different quadrants of EMB-centered S-B1 rotating frame plays a significant role in the escape process.

Across all LD fields, "LD value > 4.5" reliably serves as an effective escape criterion. Nevertheless, LD fields exhibit complex distributions, with regions of varying LD values forming irregular clusters. Besides, there are also many chaotic points in LD field, as shown in Fig. 20. Thus, it is necessary to directly extract conditions for disposal maneuver direction and classify orbital families with similar escape characteristics from the LD field.

The LD field construction clearly demonstrates the influence of SGP on orbital evolution for given initial conditions. The main findings on cislunar escape dynamics from this framework can be summarized as follows:

1. Similar orbital evolution results are organized into distinct regions within the LD field; boundaries correspond to projections of low-energy cislunar escape manifolds.

2. Regions with relatively high LD values pinpoint the most favorable combinations of disposal maneuver direction and initial solar phase for escape.

3. LD serves as an effective indicator: the larger values correlate with shorter escape time intervals, and a threshold can separate successful and unsuccessful escape outcomes (in this research, the threshold is taken as 4.5).

4. Most escape trajectories leverage LGA to increase orbital mechanical energy, and some disposal maneuver direction conditions result in more apogee passages analogous to WSB transfer.

5. The LD field can be evaluated together with time-to-depart map, and the region favorable for escape in LD field are also favorable for departure.

6. There are many chaotic points in the LD field, and large-area continuous regions with LD $\geq 4.5$ are more favorable as disposal conditions.

### 4.3 Escape Orbits Clustering Method

Section 4.2 demonstrates that the LD field distribution with respect to disposal maneuver direction is highly complex. This section presents an automated method for extracting escape conditions and efficiently clustering LD field samples to classify disposal scenarios exhibiting similar escape behaviors.

Firstly, three main types of escape trajectories from DROs are defined, as explained in Table 5. The type $\mathcal{L}$ represents the escape behaviors primarily via LGA, and the types $\mathcal{W}_{\mathcal{A}}$ and $\mathcal{W}_{\mathcal{D}}$ mainly use the effect of WSB. The trajectories belong to $\mathcal{L}$, $\mathcal{W}_{\mathcal{A}}$ or $\mathcal{W}_{\mathcal{D}}$ all experience the escape boundary crossing. More specifically, the farthest apogee in the $\mathcal{W}_{\mathcal{A}}$ type is located in the 2nd/4th quadrant of the EMB-centered S-B1 rotating frame, thereby increasing the orbital mechanical energy during the apogee passage. The farthest apogee in the $\mathcal{W}_{\mathcal{D}}$ type is located in the 1st/3rd quadrant of the EMB-centered S-B1 rotating frame, thereby decreasing the orbital mechanical energy during the apogee passage. Trajectories not matching these criteria are considered as unsuccessful escape. If the total number of samples in a LD field is $M$, then the proportion of successful escape conditions is

$$\eta_{escape} = \frac{I(\mathcal{L} \cup \mathcal{W}_{\mathcal{A}} \cup \mathcal{W}_{\mathcal{D}})}{M}, \tag{32}$$

where the function $I(\bullet)$ represents the number of elements in the set. The proportion of specific type can also be obtained by Eq. (32). For example, the proportion of escape type $\mathcal{L}$ can be obtained by $I(\mathcal{L})/M$.

**Table 5 Characteristics of three main escape categories.**

| Type | Characteristics |
|---|---|
| $\mathcal{L}$ | After applying the disposal maneuver, the spacecraft experience LGA and the apogee radius w.r.t. EMB does not exceed 2 [DU] |
| $\mathcal{W}_{\mathcal{A}}$ | The apogee radius w.r.t. EMB exceeds 2 [DU], the farthest apogee is located in the 2nd/4th quadrant of the EMB-centered S-B1 rotating frame, and there is at least one perigee after the apogee |
| $\mathcal{W}_{\mathcal{D}}$ | The apogee radius w.r.t. EMB exceeds 2 [DU], the farthest apogee is located in the 1st/3rd quadrant of the EMB-centered S-B1 rotating frame, and there is at least one perigee after the apogee |

Considering that the LD field contains a large number of chaotic points (like the chaotic points between the region (b) and region (c)), as shown in Fig. 20. These discrete chaotic samples are sensitive to parameters and not suitable for practical use. Therefore, the following extraction and clustering method is proposed to identify continuous, large-area regions of disposal maneuver direction. The first layer of the algorithm consists of two steps:

1. Binarization: The samples in LD field are binarized by retaining only samples with LD values larger than 4.5. Samples below this threshold are discarded, leaving only successful escape conditions.

2. Regional Connectivity Filtering: Using the 'regionprops' function [42] which measures the attributes of each connected component in the LD field, regions are screened for connectivity. Only regions where the proportion of continuously distributed samples exceeds the threshold $\varepsilon$ are retained ($\varepsilon = 0.01$ in this study). This step removes isolated small-area regions.

The second layer of the algorithm is to perform clustering analysis. For each sample in the LD field, a specific classification information vector is constructed as

$$\mathcal{Y}_j = \left[LD_j,\ \cos\phi_j \cos\theta_j,\ \cos\phi_j \sin\theta_j,\ \sin\phi_j\right],\ j = 1,2,\ldots M. \tag{33}$$

The classification information vector includes LD value, disposal maneuver direction ($\theta$ and $\phi$) for each sample. The dataset $\mathcal{Y}$ are used for coarse classification via 'K-means' function [43]. The coarse classification mainly using LD values to classify dataset $\mathcal{Y}$ into several major categories. To capture escape features more thoroughly, characteristic points are augmented in $\mathcal{Y}_j$. The epoch and state of characteristic points include the perilune, perigee, and apogee passages and the escape duration; if the perilune, perigee, and apogee passages are absent, this part of vector is set to zero. Each vector in dataset $\boldsymbol{Y}$ are obtained as

$$\boldsymbol{Y}_j = \left[ LD_j,\ t_{perilune},\ \boldsymbol{X}_{perilune},\ t_{apoapsis},\ \boldsymbol{X}_{apoapsis},\ t_{periapsis},\ \boldsymbol{X}_{periapsis},\ t_{escape} \right],\ j = 1, 2, \ldots M. \tag{34}$$

Then, combined with information vector of characteristic points and escape time interval, DBSCAN algorithm [44,45] is used for fine classification. Subsequently, the major categories from dataset $\mathcal{Y}$ can be further subdivided into more detailed families.

The above algorithm procedure can be summarized as shown in Fig. 25. The initial binarization and connectivity screening effectively remove most discrete chaotic points, though small isolated chaotic points may remain near large-area continuous regions. These small isolated chaotic regions are further filtered out by DBSCAN, yielding ideal LD field samples. In addition, DBSCAN requires setting clustering parameters. The clustering parameters are the minimum neighborhood radius $\varepsilon_{\text{LD-radius}}$ and the minimum number of points within the family $min_{sap}$. In this study, $\varepsilon_{\text{LD-radius}}$ is set to 1.5 and $min_{sap}$ is set to 10, which ensure accurate noise points filtering and families partitioning. Following the rationale used in previous DBSCAN-based orbital classification studies [46], $min_{sap} = 10$ was selected to suppress isolated points while preserving large connected LD regions; thus, each cluster contains at least ten neighboring initial conditions. The value of $\varepsilon_{\text{LD-radius}}$ was calibrated using more than 100 LD fields by sweeping the parameter from 0.1 to 3.0 with a step size of 0.1. Smaller values caused excessive fragmentation and insufficient noise filtering, whereas larger values removed or merged small connected LD regions and resulted in overly coarse classifications. The density-based clustering validation index (VI) which varies in ranges [-1, 1] and holds better effect when closer to 1, is used to reflect the effect of $\varepsilon_{\text{LD-radius}}$. The specific calculation of VI can be obtained from previous research [45,47], In all tested cases, $\varepsilon_{\text{LD-radius}} = 1.5$ holds a mean VI of 0.767 and is the maximum value. Accordingly, $\varepsilon_{\text{LD-radius}} = 1.5$ was adopted as a balanced value for clustering considered in this study. The time required for running the above algorithm averagely on the CPU used (Intel (R) Core (TM) Ultra 9 285H, 32 GB of RAM) is 0.13 s.

Some results of the orbital family classification are illustrated in Fig. 26. The top, middle, and bottom panels correspond to parameters from Fig. 16, Fig. 20, and Fig. 23 respectively. Each panel is subdivided into two parts: (1) the original LD field, (2) the LD field color-coded by cluster index. It is obvious that Fig. 26 retains the main parts of the LD field with LD values larger than 4.5, and most of the chaotic points have been removed, leaving feasible disposal maneuver direction conditions. In the panel (a) of Fig. 26, 12 distinct escape orbital families are identified; in panel (b), 44 escape orbital families are identified; while for the panel (c), 8 escape orbital families are identified. The proportion of successful escape samples $\eta_{escape}$ for the three panels in Fig. 26 are 9.60%, 26.36%, and 9.19%, respectively. Overall, the larger magnitude of the disposal maneuver leads to more diversity of escape orbit families.

To validate the clustering accuracy, representative orbital families are displayed in Fig. 27, Fig. 28, and Fig. 29. Fig. 27 shows two families of type $\mathcal{L}$, represented by $\mathcal{L}_1$ and $\mathcal{L}_2$, respectively, as well as one family each of type $\mathcal{W}_{\mathcal{A}}$ and $\mathcal{W}_{\mathcal{D}}$. Fig. 28 also presents two additional $\mathcal{L}$ families, and two families of type $\mathcal{W}_{\mathcal{D}}$. Fig. 29 shows two families, denoted by $\mathcal{L}_1$ and $\mathcal{W}_{\mathcal{D}1}$. The clustering results demonstrate high accuracy, effectively distinguishing the orbital families. In practical disposal design, the LD field enables evaluation of disposal maneuver direction distributions for different orbital families. For instance, the $\mathcal{L}_2$ group in Fig. 27 corresponds to a relatively large-area, continuous region of $(\theta, \phi)$ with lower parameter sensitivity. Furthermore, the LD-based clustering algorithm developed in this study is also potentially able to analyze low-energy-transfer parameter fields.

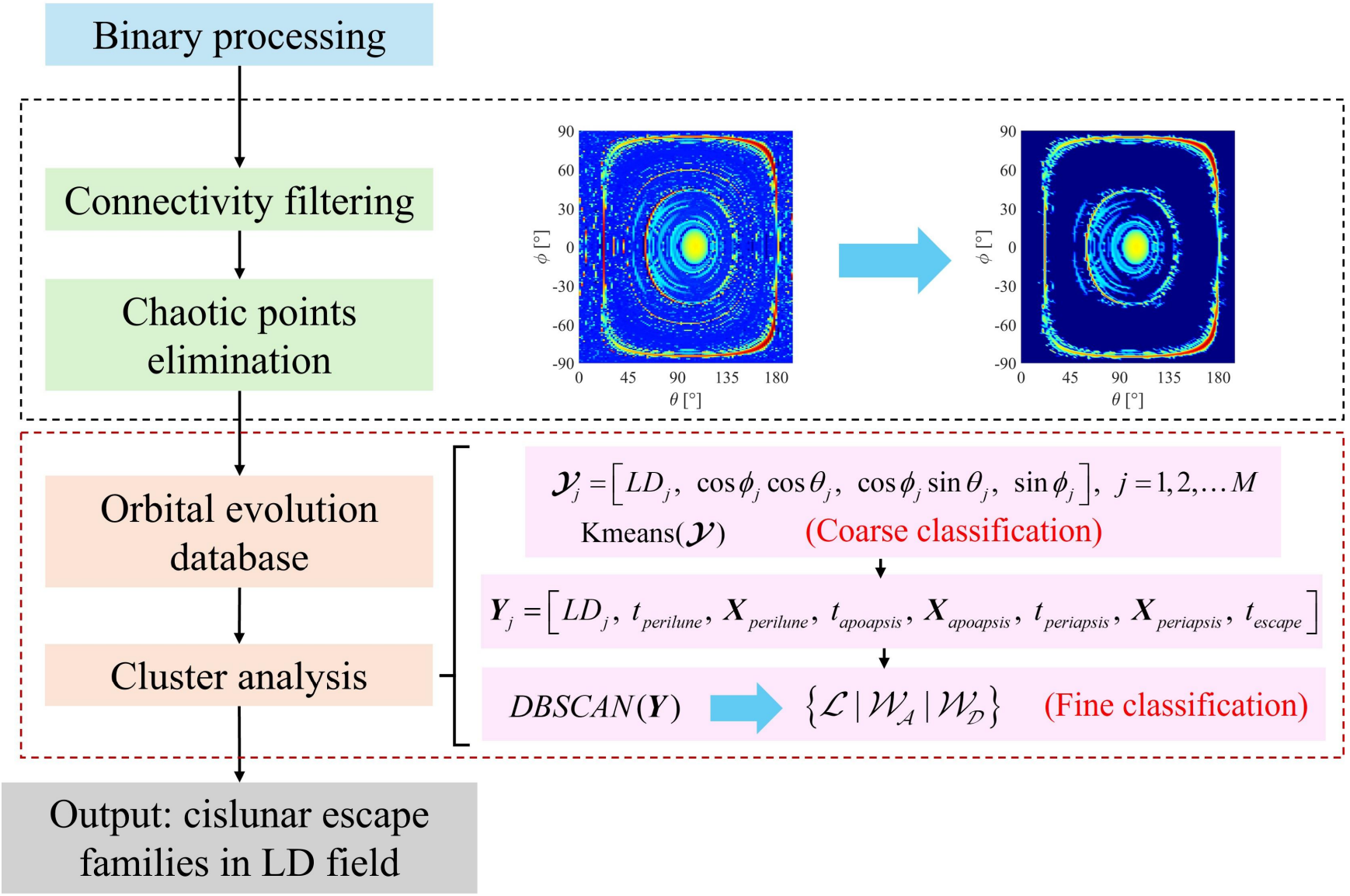

**Fig. 25 The two-layer escape orbital families clustering method.**

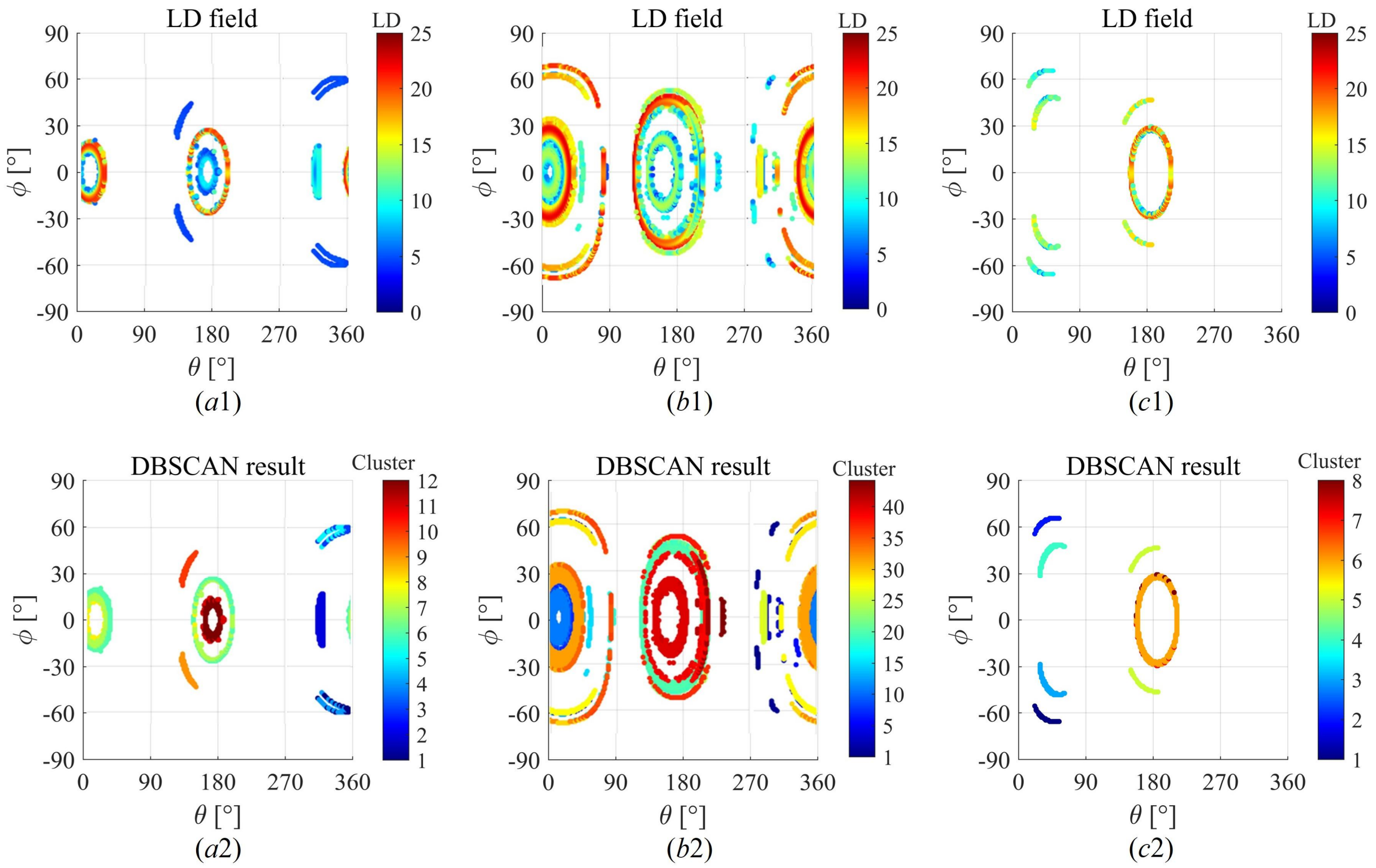

**Fig. 26 The clustering results for LD field from (a) 2:1 DRO $\Delta v = 25\,m/s, \alpha_{S0} = 0$, (b) 2:1 DRO $\Delta v = 55\,m/s, \alpha_{S0} = 0$, (c) 3:1 DRO $\Delta v = 25\,m/s, \alpha_{S0} = 90°$.**

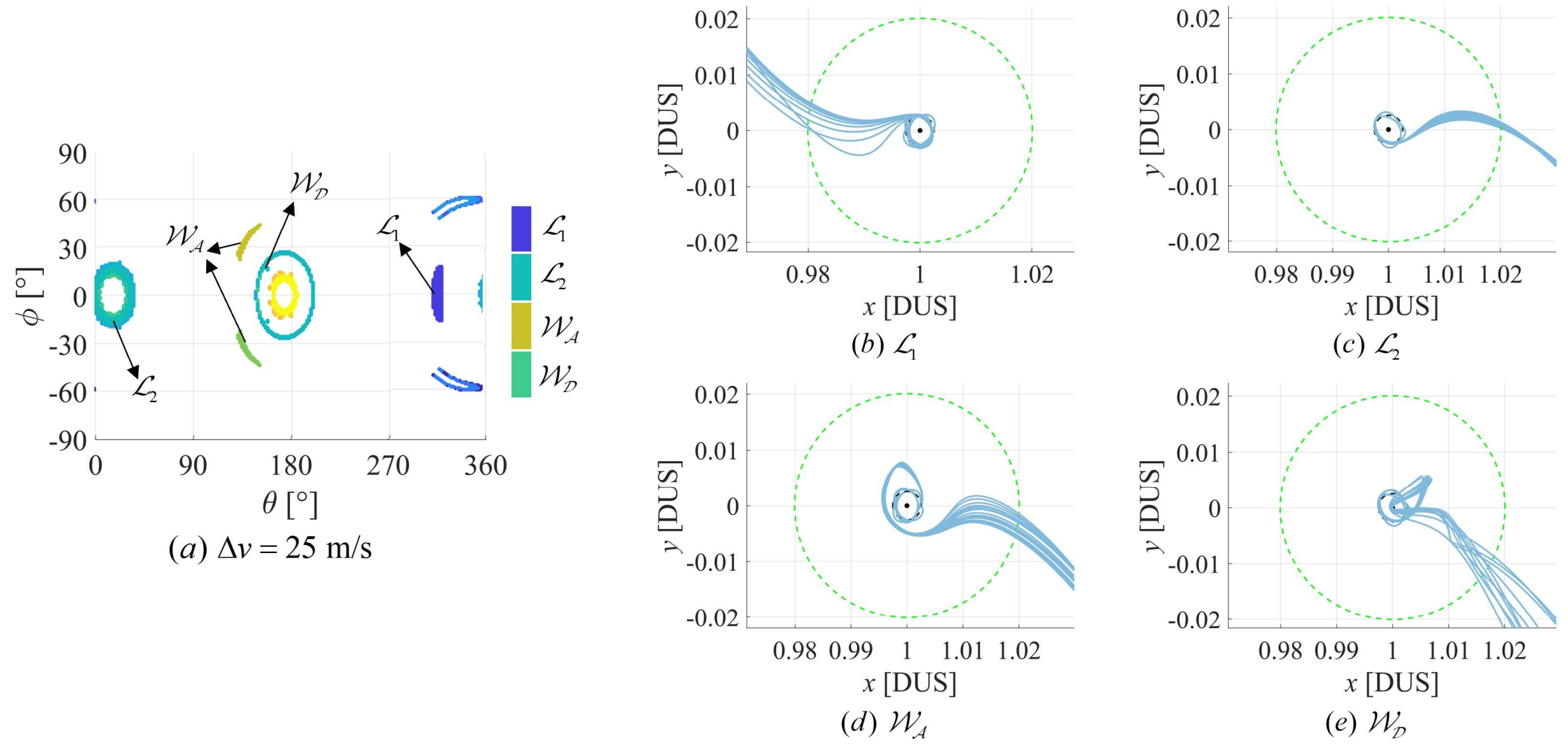

**Fig. 27 Four representative families of LD field for 2:1 DRO with $\Delta v = 25\,m/s, \alpha_{S0} = 0$.**

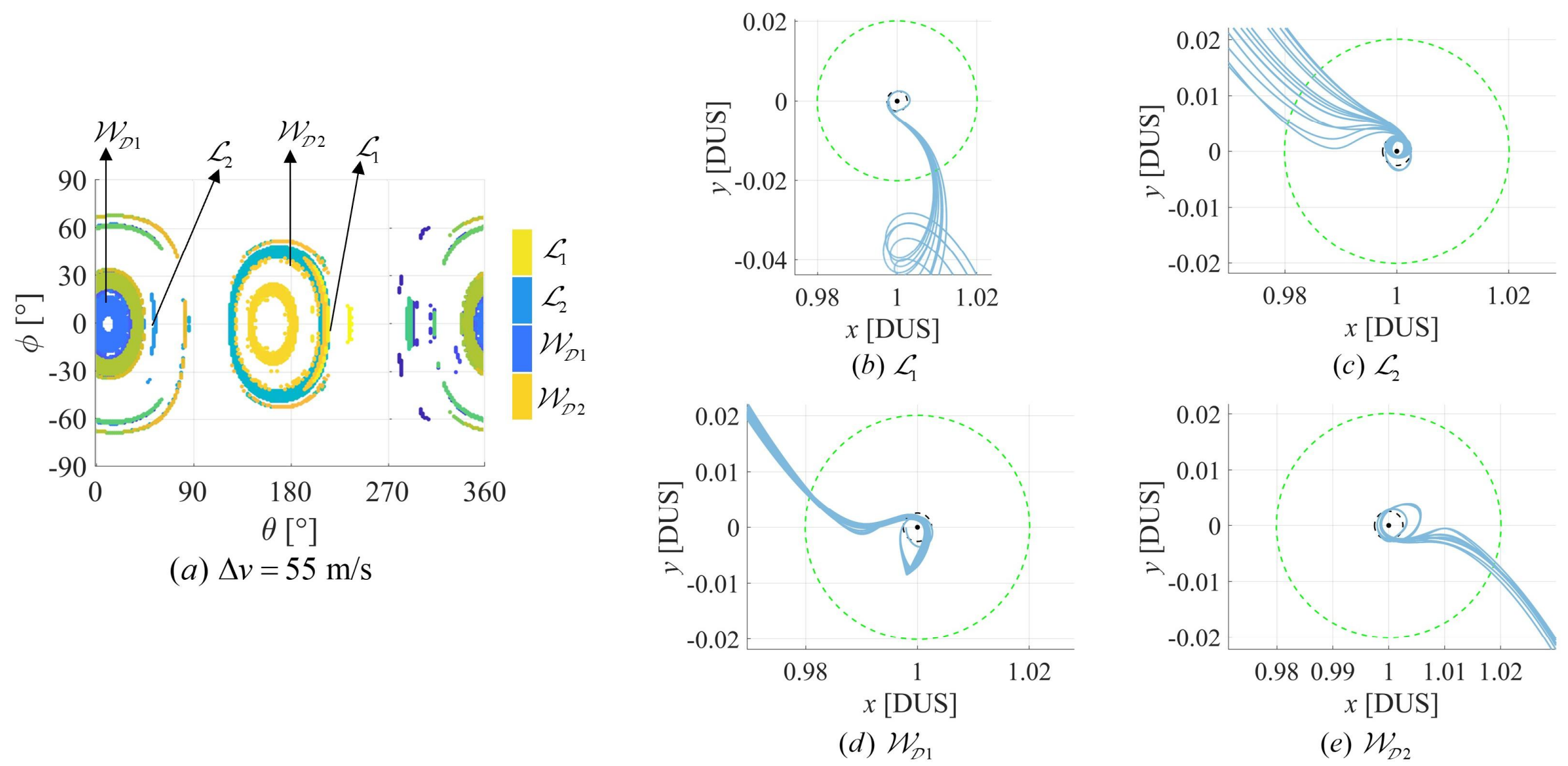

**Fig. 28 Four representative families of LD field for 2:1 DRO with $\Delta v = 55\, m/s, \alpha_{S0} = 0$.**

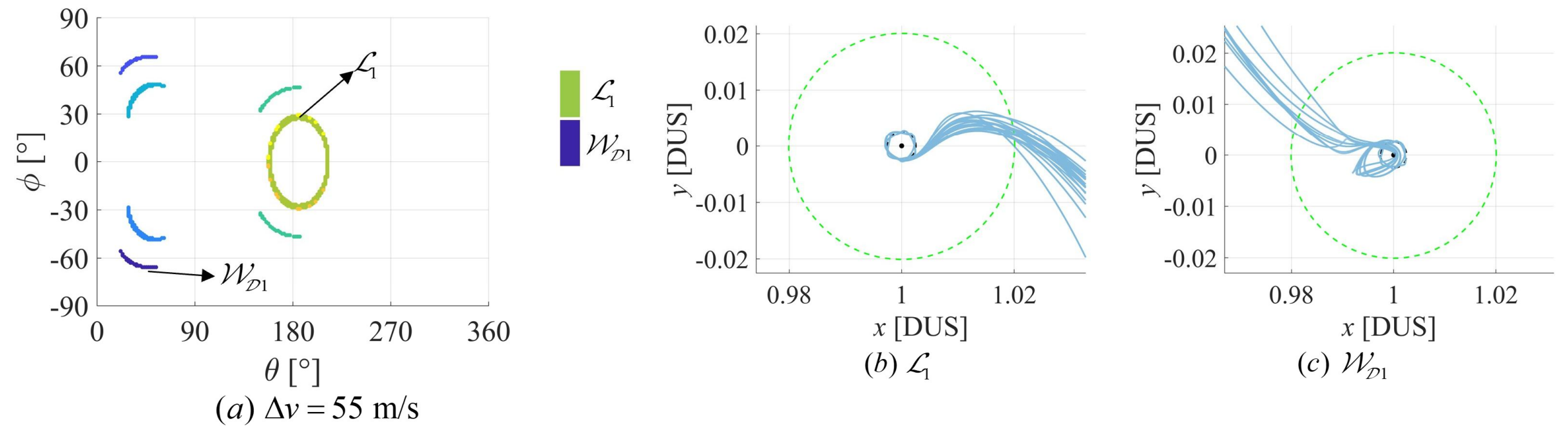

**Fig. 29 Two representative families of LD field for 3:1 DRO with $\Delta v = 25\, m/s, \alpha_{S0} = 90°$.**

### 4.4 The Influence of Disposal Maneuver Magnitude and Initial Solar Phase

By varying the initial solar phase at different disposal epochs and comparing LD fields, the qualitative influence of both disposal maneuver magnitude and initial solar phase can be assessed. Fig. 30 presents the proportion of escape conditions for both 2:1 and 3:1 DROs, which is calculated by the procedure in Section 4.3. The LD fields for 2:1 DRO with different initial solar phases are shown in Fig. 31 and Fig. 32, and the same cases for 3:1 DRO are shown in Fig. 33 and Fig. 34. The disposal maneuver magnitudes for the results in Fig. 31 (Fig. 33) and Fig. 32 (Fig. 34) are $\Delta v = 25\ m/s$ and $\Delta v = 55\ m/s$, respectively. The LD fields of $\alpha_{S0} = 0$ and $\alpha_{S0} = 180°$ differ by 180° in $\theta$, reflecting the dynamical symmetry inherent in the BCRFBP. As the disposal maneuver magnitude increases, the overall effect of solar gravity on escape probability diminishes. However, the initial solar phase still significantly influences the spatial distribution of escape favorable regions in the LD field. For example, for disposal from 2:1 DRO with $\Delta v = 55\ m/s$, the distributions of escape favorable regions in the LD field with $\alpha_{S0} = 60°$ are chaotic; while when $\alpha_{S0} = 15°$ or $\alpha_{S0} = 165°$, there are a large number of continuously distributed regions with LD value larger than 20, corresponding to the condition of fast escape. In addition, the escape possibility is also related to orbital stability. 3:1 DRO has the weak unstable component, resulting in a higher proportion of overall escape ratio.

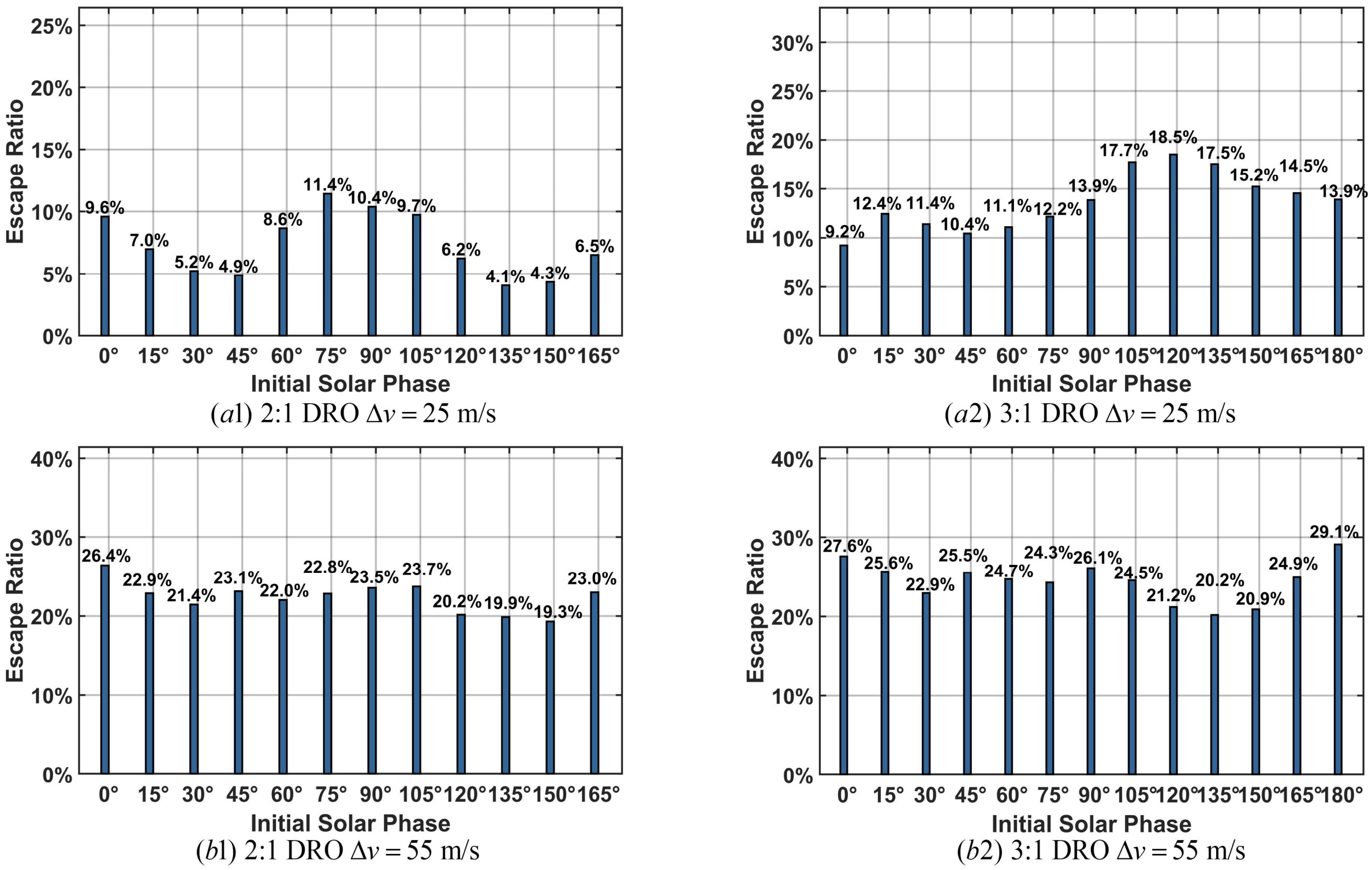

**Fig. 30 The escape ratio in different LD fields for 2:1 and 3:1 DROs.**

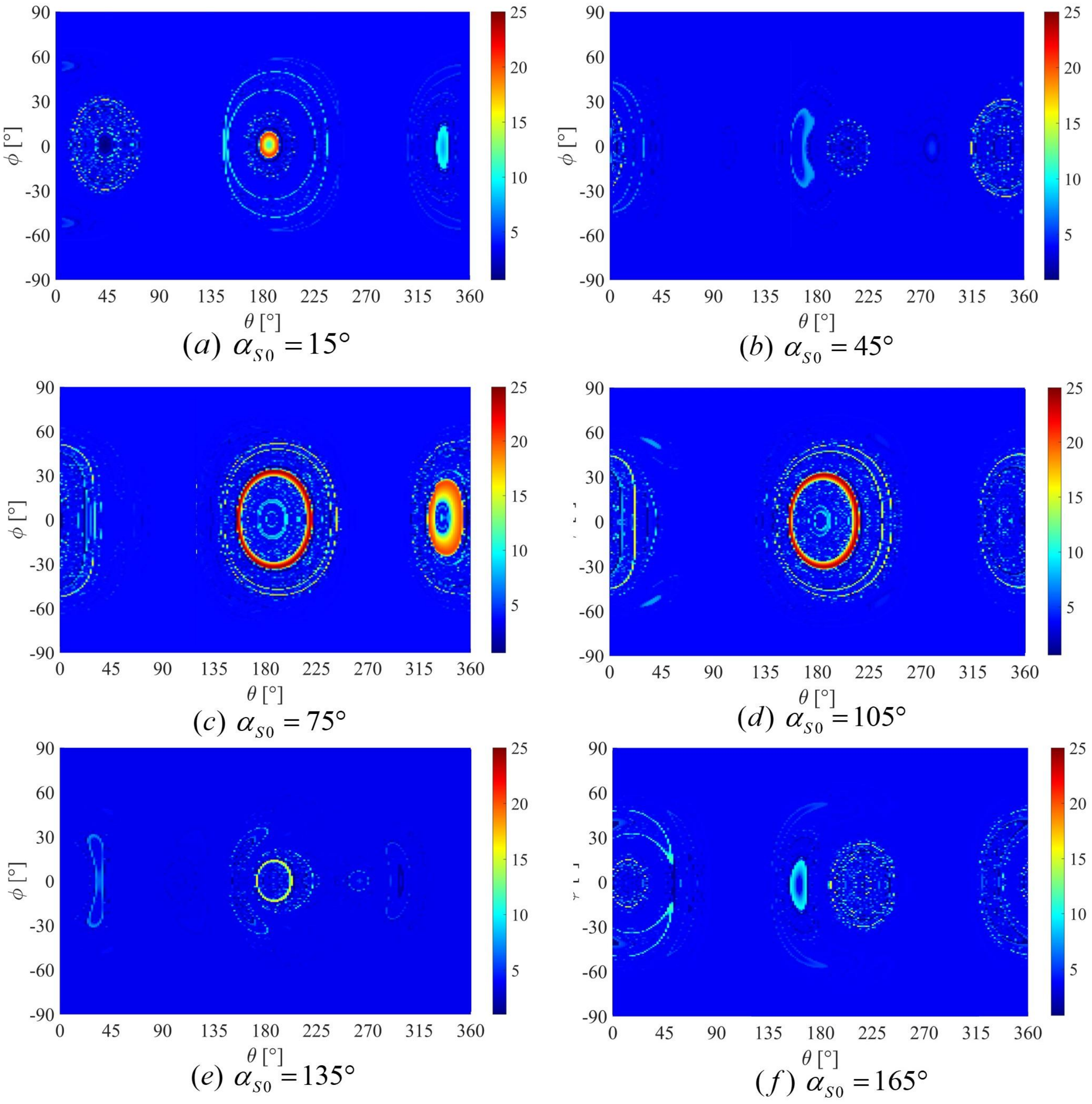

**Fig. 31 The LD fields with different initial solar phase when $\Delta v = 25\ m/s$ for 2:1 DRO.**

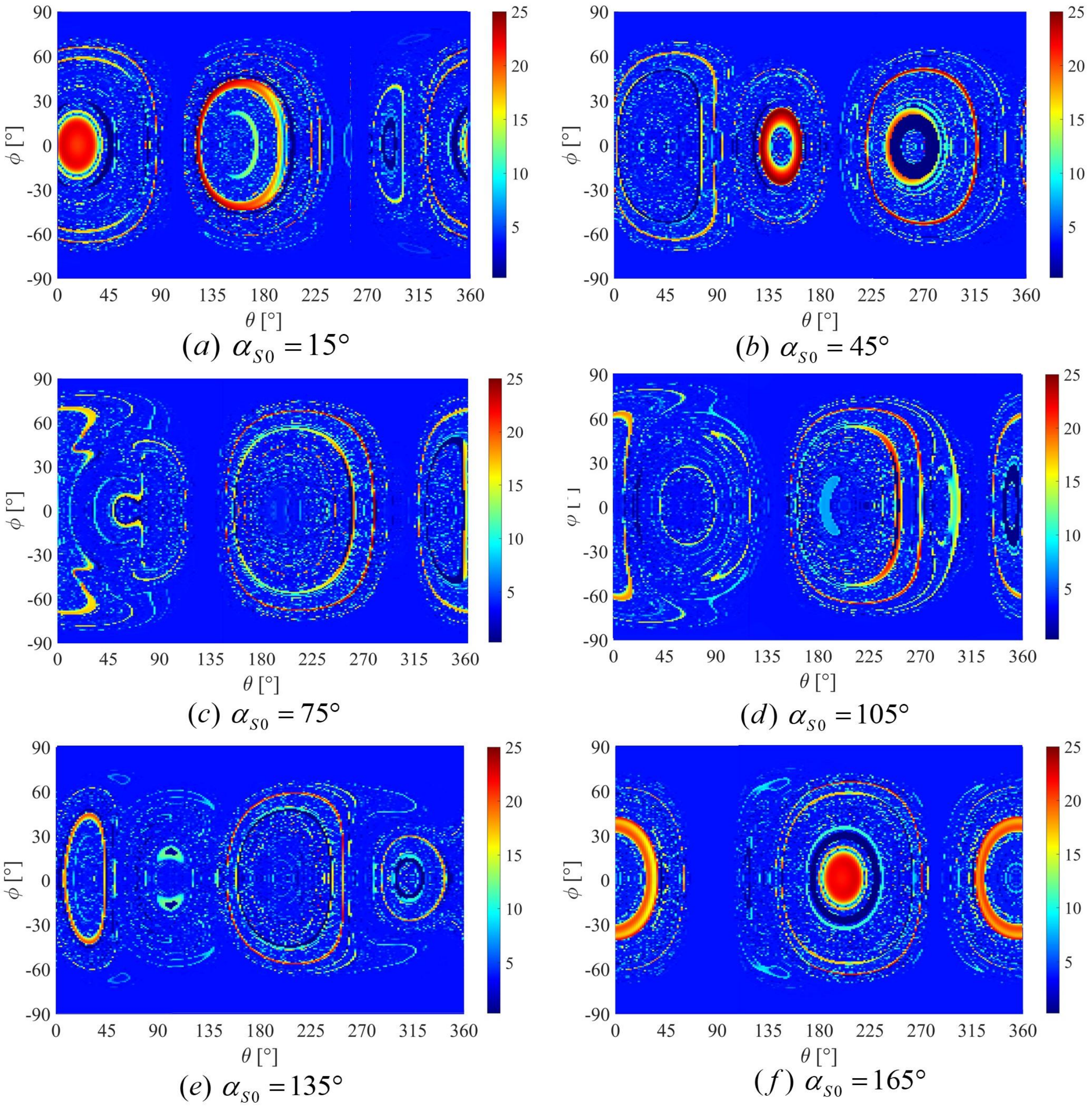

**Fig. 32 The LD fields with different initial solar phase when $\Delta v = 55\ m/s$ for 2:1 DRO.**

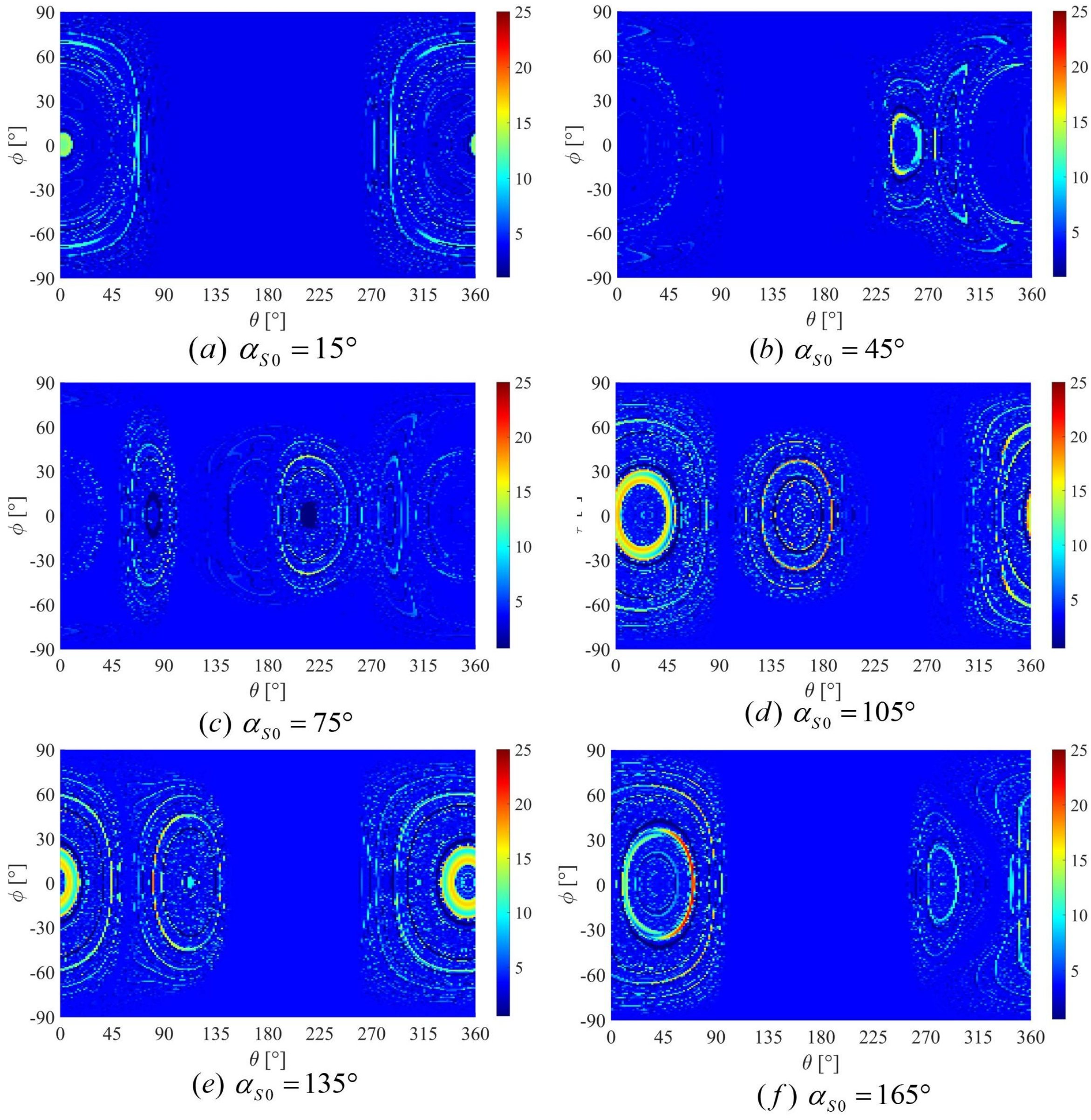

**Fig. 33 The LD fields with different initial solar phase when $\Delta v = 25\ m/s$ for 3:1 DRO.**

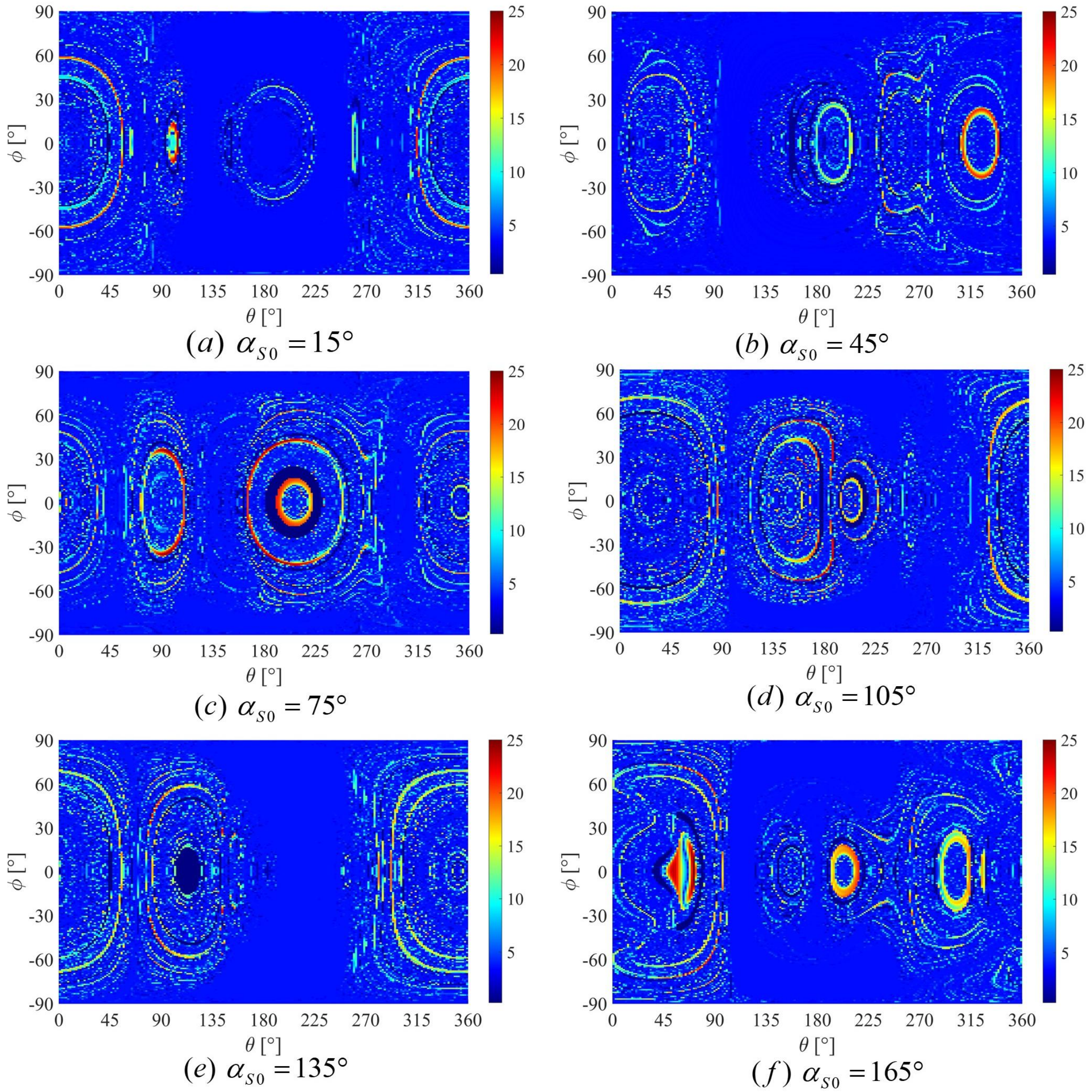

**Fig. 34 The LD fields with different initial solar phase when $\Delta v = 55\ m/s$ for 3:1 DRO.**

The results show that disposal-maneuver magnitude is the primary control parameter governing escape feasibility, whereas the initial solar phase determines the timing-dependent geometry of favorable escape channels. At low maneuver magnitudes, the post-disposal state remains near the bounded DRO region, making escape strongly dependent on feasible SGPs and local low-energy phase-space structures; consequently, the LD field is characterized by fragmented and narrow favorable regions. Increasing the maneuver magnitude enlarges the escape basin and produces broader, more continuous regions associated with fast escape. The solar phase also changes the relative Sun–Earth–Moon geometry, so an identical disposal direction may yield energy gain, energy loss, or weak SGP assistance. Accordingly, favorable LD regions may shift, split, merge, or become continuous, while phase pairs separated by $\theta$=180° exhibit the expected symmetry of the BCRFBP.

These results identify two disposal-design regimes. In the low-magnitude regime, escape is strongly phase-selective: favorable phases can open continuous fast-escape channels, whereas unfavorable phases yield scattered and chaotic regions. At higher magnitudes, the escape possibility becomes less phase-dependent, although the solar phase continues to determine the location and robustness of favorable disposal directions. Therefore, disposal design should consider both the scalar escape possibility and the topology of the LD field. Robust opportunities are associated with large, connected regions satisfying "LD≥4.5", which provide greater tolerance to execution errors. This framework supports the selection of disposal epochs and directions under propellant or solar-phase constraints and can be extended to other cislunar periodic orbits.

### 4.5 Statistical Analysis of LD Threshold and Integration Time for Escape

The proposed method for identifying successful escape depends on the LD threshold $LD^*$ and integration time $t_f$. This Section presents a statistical assessment of the escape-classification precision with respect to these two parameters. Without loss of generality, a trajectory is classified as a successful escape if the distance $r_{end}$ between the position and EMB at $t_f$ exceeds the escape boundary $R_{escape}$ .Accordingly, the minimum terminal distance $r_{end-\min}$ among the trajectories selected from the LD fields remains above this threshold, indicating that the escape criterion based on $LD^*$ provides high precision. Some initial conditions with LD value less than $LD^*$ but still successfully escape, however, in these cases, escape epochs are close to one year, making these trajectories unsuitable for practical disposal strategies.

Disposal conditions for 2:1 and 3:1 DROs consist of $\alpha_{S0} \in [0,180°]$ and $\Delta v \in [25,55]$ $m/s$, with the step size of 15° in $\alpha_{S0}$ and 5 m/s in $\Delta v$ . A total of 182 LD fields were therefore computed, and the corresponding $r_{end-\min}$ values were recorded. The influence of $LD^*$ on the precision of escape criterion is shown in Fig. 35. For any $\alpha_{S0}$ , all $r_{end-\min}$ with $\Delta v \in [25,55]$ $m/s$ are presented using box plots. As shown in panels (a)~(c) of Fig. 35, LD*=4 produces several FP samples since the accumulation of orbital angular momentum is not large enough, while $LD^*$=4.5 or $LD^*$=5 eliminate FP samples.

With $LD^*$=4.5, three integration times, $t_f$ = 0.5, 1.25, and 1.5 years, are considered. The corresponding box plots are illustrated in panels (d)~(f) of Fig. 35, and the black dashed line in each panel represents the $R_{escape}$ . Shorter integration times produce no FP samples among the trajectories satisfying "LD≥4.5". However, when $t_f$=1.5 years, some FP samples appear. The selection of LD* and $t_f$ can therefore be selected through a statistical analysis tailored to the intended application. Within the nondimensionalization adopted in this study, "$t_f$=1.0 year" and "LD$^*$=4.5" is an appropriate choice.

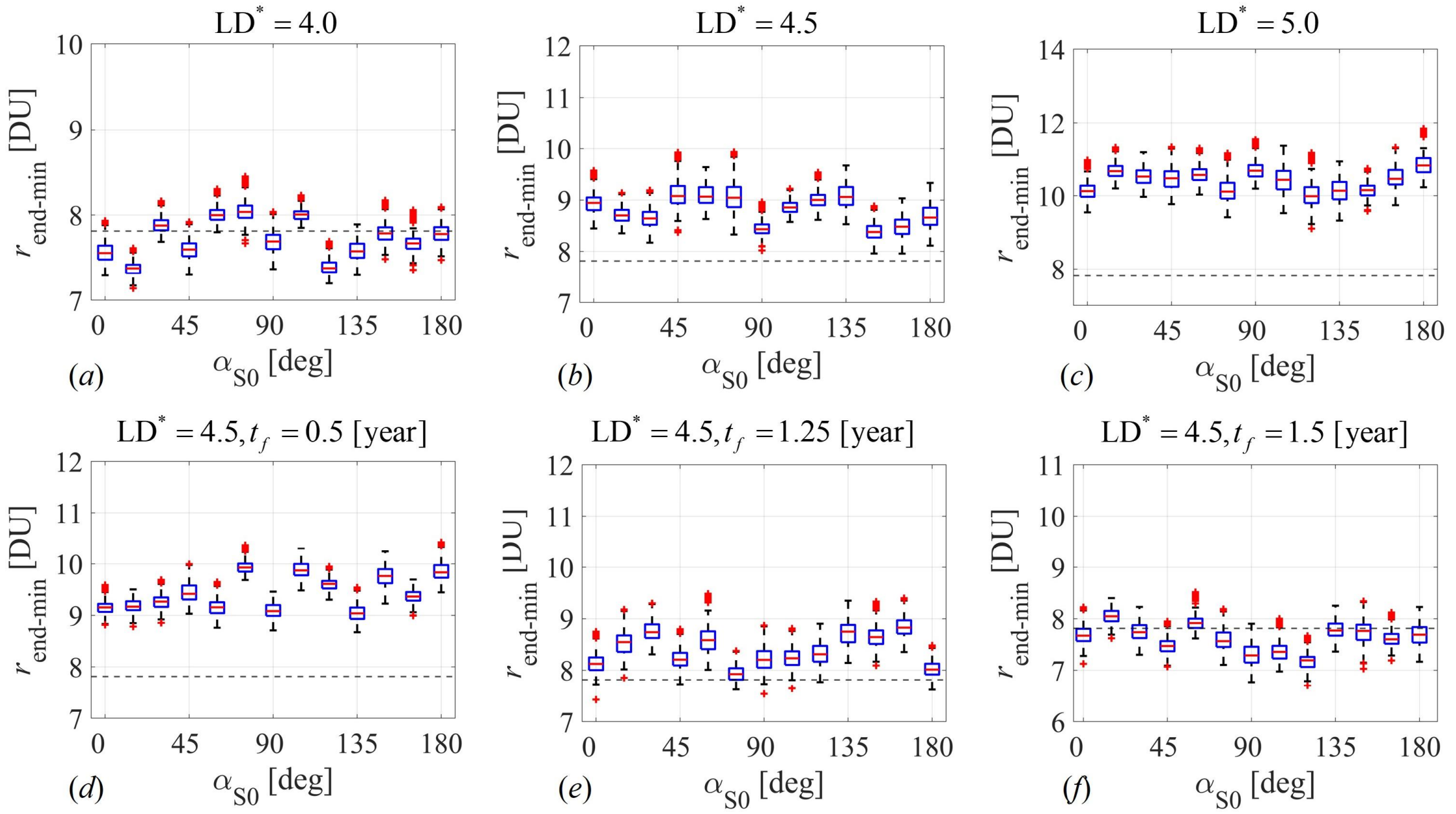

**Fig. 35 Statistical results of LD* and $t_f$ for escape criterion.**

## 5. Extension to Different Periodic Orbits and High-Fidelity Ephemeris Model

This section extends the proposed framework to other periodic orbits and to a high-fidelity ephemeris model to assess its generality. The southern 9:2 NRHO is selected as a representative orbit beyond the DRO family. In addition, representative escape and bounded-motion cases identified from LD fields are propagated in the ephemeris model and evaluated through Monte Carlo simulations.

### 5.1 The Disposal Dynamics of NRHO

The southern 9:2 NRHO in the BCRFBP is shown in Fig. 36. Nine revolutions of the 9:2 NRHO span exactly two Earth–Moon synodic periods, corresponding to approximately 59 days. The nine perilune passages are clearly visible in the S–B1 rotating frame, as shown in Fig. 36b and Fig. 36c.

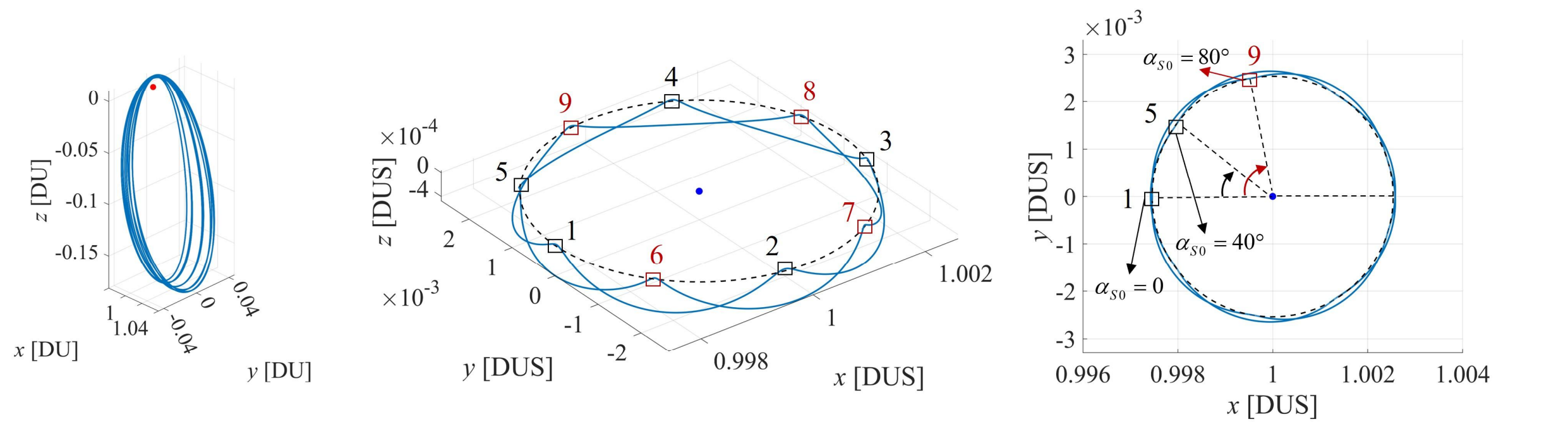

**Fig. 36 The cislunar southern 9:2 NRHO of L2 point in BCRFBP.**

For brevity, the first perilune is selected as the disposal position, and the maneuver magnitude is $\Delta v = 10\, m/s$. The time-to-depart map and LD field are shown in Fig. 37a and Fig. 37b, respectively. Two disposal direction conditions are selected, where case (a) is $\theta = 7.23°, \varphi = 6.78°$ and case (b) is $\theta = 43.41, \varphi = -38.44°$. The evolution of MI and trajectories for cases (a) and (b) are shown in panels (c) and (d) in Fig. 37. Both cases result in departure from the NRHO, indicating the applicability of criterion "K=1.6" for different periodic orbits. Case (a), however, does not lead to cislunar escape, and the corresponding LD value is 3.67. Case (b) leads to cislunar escape, with a corresponding LD value of 8.67.

Fig. 37 highlights a clear contrast with the disposal results for DROs. The greater intrinsic instability of the NRHO facilitates both departure and cislunar escape. Because the proposed framework is based on the dynamical characteristics of cislunar space, it is not restricted to a particular periodic-orbit family. The results for the 9:2 NRHO therefore support the applicability of the framework to other periodic orbits.

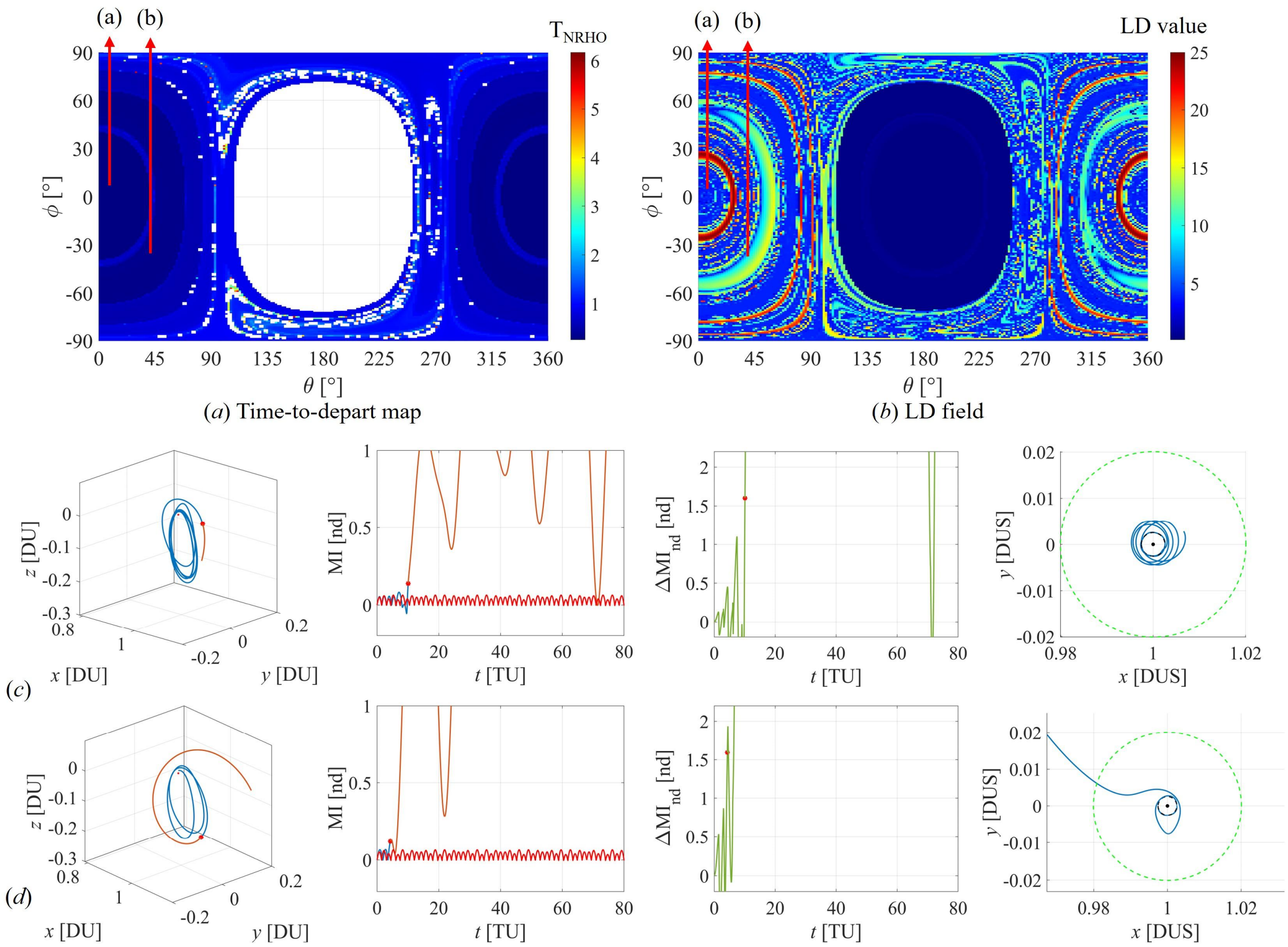

**Fig. 37 The validation of departure and escape phases for 9:2 NRHO.**

### 5.2 The Disposal Dynamics in High-Fidelity Ephemeris Model

In a realistic dynamical environment, post-disposal orbital evolution is strongly influenced by the time-varying Earth–Moon distance, the long-term accumulation of solar-radiation-pressure effects, and other perturbations. This Section assesses whether the disposal conditions extracted from the BCRFBP LD fields remain valid in a high-fidelity ephemeris model. The 2:1 and 3:1 DROs in the ephemeris model are introduced in Section 3.1. Three examples selected form LD fields are shown in Fig. 38, and the corresponding disposal conditions are listed in Table 6. In all simulations, disposal-execution errors following a standard normal distribution are included, consistent with previous research [15]. The standard deviation of $\Delta v$ error is 1.5% of the nominal magnitude. The standard deviations of errors for $\theta$ and $\varphi$ are 1°. The red trajectory denotes the nominal post-disposal trajectory in the S–B1 rotating frame, whereas the blue trajectory denotes the nominal trajectory in the Earth–Moon rotating frame. The gray clusters represent the trajectories with execution errors. Fig. 38 confirms that the connected regions identified from the BCRFBP LD fields retain robustness in the ephemeris model. Case (a) corresponds to fast escape, whereas case (c) achieves escape after a relatively long-duration lunar encounter. In case (b), the object remains captured in the vicinity of DRO for an extended period because of an unfavorable disposal direction. Future work will calculate LD fields directly in the ephemeris model to design robust disposal strategies that can be patched to stable heliocentric resonant orbits.

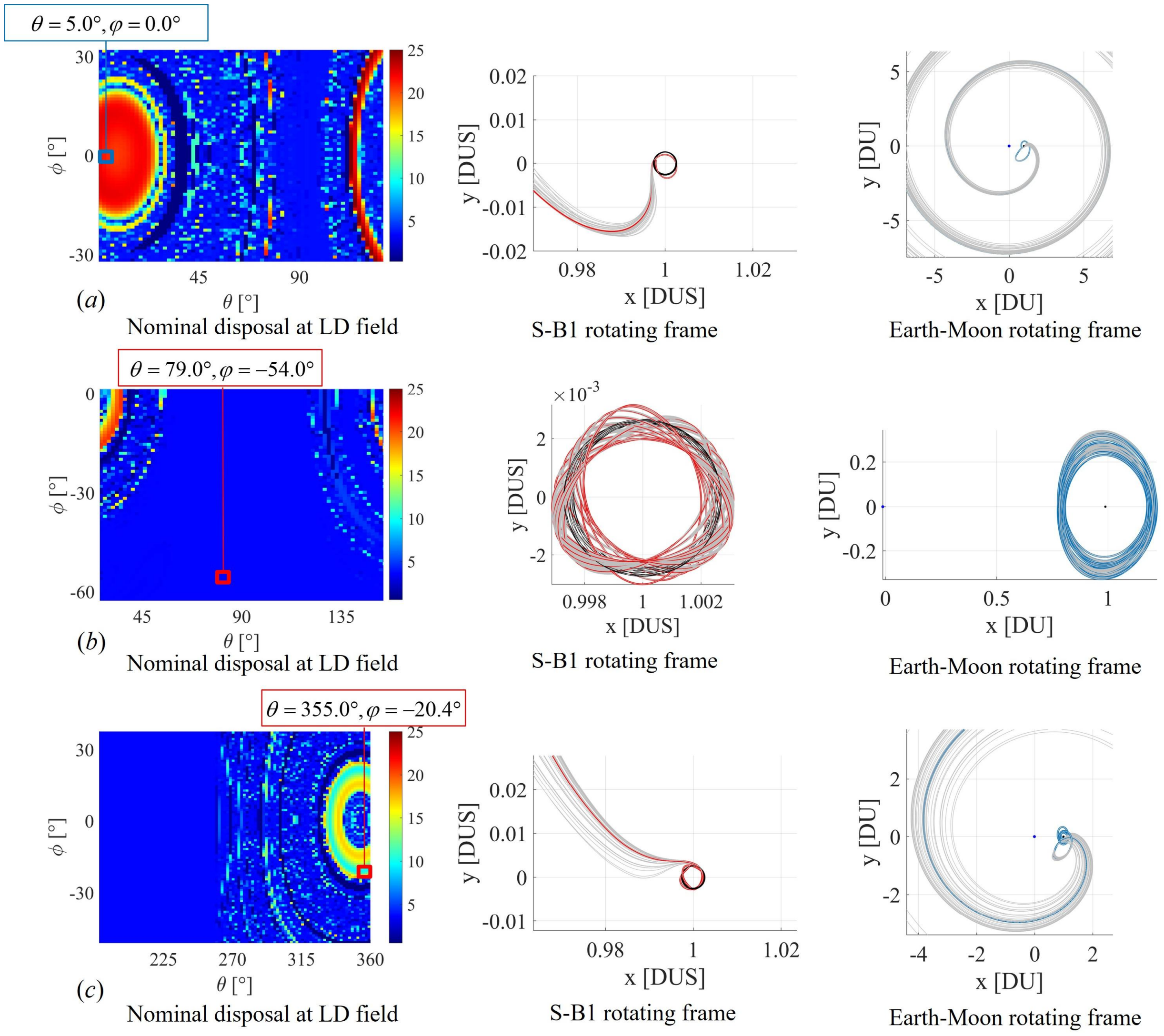

**Fig. 38 Verification of escape and bounded motion disposal results in ephemeris model.**

**Table 6 Disposal conditions for the three cases shown in Fig. 38.**

| Type | $\theta$ [°] | $\varphi$ [°] | $\alpha_{S0}$ [°] | $\Delta v$ $[m/s]$ | Date |
|---|---|---|---|---|---|
| ($a$) Escape | 5.0 | 0.0 | 15.0 | 55 | 2028-04-23 12:39:05 |
| ($b$) Bounded motion | 79.0 | -54.0 | 0.0 | 25 | 2028-04-24 19:42:06 |
| ($c$) Escape | 355.0 | -20.4 | 135.0 | 25 | 2028-05-12 10:28:26 |

## 6. Conclusion

This study developed a framework for assessing disposal dynamics from cislunar periodic orbits, with lunar distant retrograde orbits (DROs) selected as representative cases. Spatial disposal dynamics from 2:1 and 3:1 DROs were systematically investigated, covering both local departure from the nominal orbit and subsequent escape from the cislunar region.

Disposal from DROs is strongly affected by maneuver magnitude, maneuver direction, orbital phase, and initial solar phase. Owing to the inherent stability of DROs, departure from these orbits is more challenging than from near rectilinear halo orbits and generally requires larger disposal maneuvers. An improved momentum-integral-based metric was proposed, providing a dimensionless departure criterion that is independent of orbital geometry and thus applicable to other cislunar periodic orbits.

Cislunar escape was further analyzed from a low-energy transfer perspective incorporating Lagrangian descriptor and semi-analytical energy evolution. The results clarify the role of solar gravitational perturbations in shaping escape dynamics and show that more stable periodic orbits tend to exhibit lower escape probabilities. The clustering method based on Lagrangian descriptor values enables accurate extraction and classification of escape families, facilitating evaluation of disposal parameter influences on escape outcomes. The large-area and continuous region in the Lagrangian descriptor field can be can be selected as favorable disposal conditions. Furthermore, the proposed framework is examined in different periodic orbits and the high-fidelity ephemeris model.

This study advances understanding of disposal dynamics from cislunar DROs and provides a transferable methodology for evaluating disposal options from other cislunar periodic orbits. The Lagrangian descriptor analysis can also contribute to the broader study of low-energy transfer mechanisms in cislunar trajectory design.

## Acknowledgments

This work has been supported by the National Natural Science Foundation of China (12422214) and Fundamental Research Funds for the Central Universities.